\documentclass[fleqn,usenatbib]{mnras}

\usepackage{newtxtext,newtxmath}

\usepackage[T1]{fontenc}

\DeclareRobustCommand{\VAN}[3]{#2}
\let\VANthebibliography\thebibliography
\def\thebibliography{\DeclareRobustCommand{\VAN}[3]{##3}\VANthebibliography}

\usepackage{graphicx}	
\usepackage{amsmath}	
\usepackage{caption}
\usepackage{subcaption}
\usepackage{enumitem}

\DeclareCaptionLabelFormat{subfigure}{(#2)}
\title[SNe in photoionized GMCs]{Supernova feedback in porous photoionized Giant Molecular Clouds}

\author[C. S. C. Lau et al.]{
Cheryl S. C. Lau$^{1}$\thanks{E-mail: cheryl.lau@phys.ncts.ntu.edu.tw (CSCL)},
Ian A. Bonnell$^{2}$,
Yueh-Ning Lee$^{3,1}$,
and Ke-Jung Chen$^{4,5}$
\\
$^{1}$Physics division, National Center for Theoretical Sciences, National Taiwan University, Taipei 106319, Taiwan\\ 
$^{2}$School of Physics and Astronomy, University of St Andrews, St Andrews KY16 9SS, UK\\
$^{3}$Center of Astronomy and Gravitation, Department of Earth Sciences, National Taiwan Normal University, Taipei 116, Taiwan\\
$^{4}$Institute of Astronomy and Astrophysics, Academia Sinica, Taipei 10617, Taiwan\\
$^5$Heidelberger Institut für Theoretische Studien, Schloss-Wolfsbrunnenweg 35, 69118 Heidelberg, Germany
}

\date{Accepted XXX. Received YYY; in original form ZZZ}

\pubyear{\the\year{}}

\begin{document}
\label{firstpage}
\pagerange{\pageref{firstpage}--\pageref{lastpage}}
\maketitle

\begin{abstract}
We present a new suite of numerical simulations of Type II supernovae (SNe) detonating in Giant Molecular Clouds with a variety of density structures shaped by photoionization feedback. Ionizing radiation sculpts cavities and channels that guide SN energy to emerge from the cloud as shock-driven blowouts, rather than as a coherent spherically expanding shell as assumed in most sub-grid SN models adopted in galaxy or cosmological simulations. We investigate how such outflows differ to the 1-D descriptions, and whether or not the perturbations induced by the blowouts are sensitive to the host cloud's structure. A channelling parameter $P_\mathrm{chnl}$ is introduced to characterise the cloud's porosity and boundness using the morphology of the ionized channels. 
Our results reveal that the outflow velocities, whilst consistently higher than that of the spherical blasts, are in fact rather independent of the porosity of its local environment. The total kinetic energy and momentum deposited also appear similar across all runs. 
What is most sensitive to $P_\mathrm{chnl}$ is the mass of the materials carried in the outflows and their migration distances. 
It implies that SNe exploding in compact clouds with distinctive channel structures may have more confined metal injection radii and shortened turbulent driving scales, which consequently lead to a clumpier interstellar medium with higher density and metallicity fluctuations. 
We argue that molecular cloud structures play an equally important role to SN rates and energy budgets in stellar feedback sub-grid modelling.
\end{abstract}

\begin{keywords}
ISM: clouds -- ISM: supernova remnants -- ISM: bubbles -- HII regions -- methods: numerical
\end{keywords}



\section{Introduction} \label{sec:introduction}

Type II supernova (SN) feedback produces the volume-filling hot phase of the interstellar medium (ISM) \citep[][]{mckeeostriker77}. They drive galactic outflows \citep[e.g.][]{girichidis16,gatto16,afruni26}, distribute $\alpha$ elements throughout the galactic discs \citep[e.g.][]{kolborg22,zhang25}, and regulate star formation by keeping the discs turbulent \citep[e.g.][]{ostrikershetty11,padoan16,bacchini20}, rendering it a crucial ingredient in galaxy evolution. 

The challenge, however, is that the average spatial resolutions of Milky Way-like galaxies in cosmological simulations currently reach only down to $\gtrsim$100 pc \citep[e.g.][]{pillepich19}, or $\gtrsim$10 pc at best for dwarf galaxies \citep[e.g.][]{wetzel16}. Even in isolated disc galaxy simulations, we reach few tens of parsecs by furthest \citep[e.g.][]{girmateyssier24}. We very often lack the computational capacity to capture the dissipation of feedback energy that occurs ubiquitously at subparsec-scales\footnote{Exceptions are the dwarf galaxies in cosmological zoom simulations \citep[e.g.][]{rey25,tung25} or the isolated dwarf galaxy simulations \citep[e.g.][]{hu23}. These can reach extremely high resolutions down to a few parsecs, which are arguably sufficient for directly resolving the SNe.}. It is for this reason that we have remained reliant on sub-grid models \citep[][]{rosdahl17} in order to circumvent the numerical overcooling problem \citep[e.g.][]{katz92,thackercouchman00,dallavecchiaschaye08}. 

Early sub-grid models employ analytical methods to describe the energy and momentum input from spherical SNe in uniform medium \citep[e.g.][]{woltjer72,chevalier74,cioffi88,blondin98}. But needless to say, the ISM is highly inhomogeneous. Past studies have factored this into account by considering a medium that is dominated by warm low-density gas, scattered with numerous cold dense cloudlets developed from thermal instability \citep[e.g.][]{mckeeostriker77,cowie81,kim11}. These clouds occupy only a few percent of the volume \citep[e.g.][]{spitzer78} and they merely manifest as small obstacles that impede the SN's expansion \citep[e.g.][]{kimostriker15,guo25}. In this picture, the collision and evaporation of clouds are simply treated as a mass injection source in the 1-D SN equations \citep[e.g.][]{whitelong91,dyson02,pittard19}. 

Recent 1-D models have also begun to incorporate the influence from pre-SN feedback \citep[e.g.][]{rahner17,rahner18,fichtner24}. Ionizing radiation and stellar winds collectively sculpt bubbles and cavities in the environment within which the first SN explodes \citep[e.g.][]{dale14,mcleod21,chevance21}. For instance, \citet{dwarkadas05} and \citet{haid16} found that SNe shocks transition to snowplough phases upon collision with the shell around wind-blown cavities \citep[][]{walchnaab15}, leading to rapid cooling\footnote{Meanwhile, the presence of low-density cavities could in fact allow more energy to be retained as cooling is suppressed within these bubbles \citep[e.g.][]{bending22,fichtner24}.}. Factors such as metallicity \citep[e.g.][]{vink01} and binaries \citep[e.g.][]{cournoyercloutier25} equally contribute to shaping the circumstellar environment. In view of the large number of parameters involved, SN sub-grid models are indeed best handled with fast 1-D computations \citep[e.g.][]{karpov20}. 

However, 3-D simulations of feedback in Giant Molecular Clouds (GMCs) have revealed that SN outflows are often highly asymmetrical, and may behave far more complex than the 1-D descriptions. This is particularly the case if, unlike \citet{kimostriker15} or \citet{guo25}, the progenitor remained embedded within gravitationally-bound clouds that withstood the destruction by pre-SN feedback \citep[e.g.][]{dale12b,kortgen16}. Such scenario is likely common in large galaxies such as M51, where clouds are subjected to external pressure exerted by the diffuse molecular `fog' \citep[][]{pety13}. Similar outcomes may be achieved if the progenitor is embedded within hierarchically-structured dense clusters \citep[][]{bonnell03}, which are very resilient to gas removal by the early feedback outflows \citep[e.g.][]{krumholz19,dinnbier20}. 

In such compact star-forming environments, both mechanical and radiative feedback are found to be preferentially travelling through the low-density regions in the clumpy GMCs \citep[e.g.][]{walch12,walchnaab15,wareing16}. These paths of least resistance are commonly termed `channels' \citep[e.g.][]{rogerspittard13,wareing17,lucas20} or `chimneys' \citep[e.g.][]{garrattsmithson18} in cloud-scale studies\footnote{Large-scale `chimneys' are also observed in disc galaxies \citep[e.g.][]{fragile04,zhang21}, whereby the overpressured superbubbles rise above the disc scale-height and break out as mass-loaded hot winds \citep[e.g.][]{maclowmccray88,basu99,kim17}.}. They allow the majority of the feedback energy to swiftly escape the GMCs without imposing significant damage to the shielded dense molecular regions \citep[e.g.][]{rogerspittard13,dale14,ali18,lucas20}, allowing star formation to continue. The channelling effect is enhanced with the inclusion of pre-SN feedback, which serves to remove gas from existing channels and thicken the cavity walls \citep{lucas20}, facilitating the transfer of SN energy to the large-scale galactic environment. This phenomenon is hence one of the largest limitations to the 1-D SN models. 

In \citet[hereafter Paper~I]{laubonnell25}, we investigated how SNe detonating within cavities and channels behave differently to those in a uniform medium (which we termed `free-field models'). We presented an idealised `semi-confined model', which depicts a SN centred in a spherical ionized cavity bordered by a thin swept-up shell. The H {\scshape ii} region is embedded inside a GMC, and is connected to the external warm ISM only via cone-shaped channels. We analytically derived the evolution of SN outflow velocity in this scenario. The results show that semi-confined SNe have significantly smaller impact radii, yet impose a higher dynamical perturbation onto their local environments by producing faster outflows. We also found that the terminal momentum is slighted boosted. Of course, these models were based off simplified assumptions, thus motivates a new study to test these findings in realistic turbulent environments. 

The purpose of this paper is to examine whether or not the SN feedback outputs are sensitive to the porosity of their parent GMCs. This is crucial for sub-grid modelling, since SN outflows are responsible for distributing the metal-enriched gas into the ISM and drive the large-scale chemical evolution. The distances to which the materials migrate can affect the inhomogeneity of metals in the disc \citep[e.g.][]{krumholzting18}, which in turn influences how we interpret the galaxy assembly histories \citep[e.g.][]{iza25}. It is necessary to understand how the ISM mixing timescales would vary with the GMC properties. 

In Section~\ref{sec:numerical_methods} we describe the numerical methods employed in our simulations. Section~\ref{sec:calculations} details the initial condition of the GMCs and their subsequent evolution after injecting ionizing radiation and SNe. Section~\ref{sec:results} presents the analysis on the SN outflows from each GMC model. Finally, we discuss and conclude in Section~\ref{sec:conclusion}.

\section{Numerical methods} \label{sec:numerical_methods}

\subsection{Smoothed particle hydrodynamics simulations} \label{sec:sph}

We use the Smoothed Particle Hydrodynamics (SPH) code {\scshape phantom} developed by \citet{phantom18} to perform the simulations. Particles have equal masses and variable smoothing lengths that scale inversely with local density, enabling them to naturally resolve the dense star-forming regions. The code solves long-range gravity with a \textit{k}d-tree \citep{gaftonrosswog11} via multipole expansion up to quadrupole terms \citep[e.g.][]{aarseth67}, whereas short-range gravity \citep[][]{bateburkert97} is computed via direct summation. Particles are evolved on individual timesteps with leapfrog integrators \citep{verlet67,tuckerman92}. Shock capturing is achieved by introducing artificial viscosity and artificial thermal conductivity \citep[e.g.][]{monaghan92,monaghan97}. 

Star formation is followed with dynamically-created sink particles \citet{bate95}. We set the density threshold $\rho_\mathrm{crit}$ for sink formation to be $10^{-16}\ \mathrm{g\ cm^{-3}}$, and the critical radius $r_\mathrm{crit}$ is $0.01\ \mathrm{pc}$ within which no new sinks can be created. Accretion radii of new sinks are set to $1000\ \mathrm{au}$. All sinks are allowed to merge unconditionally within a radius of around $200\ \mathrm{au}$. 

We also emphasise that the sinks here only determine the location of the feedback sources; their accretion and merging activity are of less concern. Provided that the sinks are dynamically formed, we reproduce the massive star environment that feedback experiences. That includes the dense clumps and cores, the inward accretion flows, and the neighbouring massive stars, all of which are obstacles to the feedback outflows.

\subsection{Radiation hydrodynamics} \label{sec:rhd_scheme} 

Photoionization is incorporated by coupling the SPH code {\scshape phantom} \citep{phantom18} to a grid-based Monte Carlo Radiative Transfer (MCRT) code {\scshape cmacionize} \citep{vandenbrouckewood18,vandenbrouckecamps20}. This radiation hydrodynamics (RHD) scheme was first introduced by \citet{petkova21}, and was later further developed by \citet{lau25} to implement tree-based algorithms and the heating and cooling calculations. 

The scheme works as follows. At each timestep, the particles' positions, masses and smoothing lengths, along with the properties of the ionizing sources, are passed from {\scshape phantom} to {\scshape cmacionize} to run a MCRT simulation. There, $10^6$ photon packets are ray-traced across the density field. The optical depth of each cell is computed using the local density, the opacity and the photon's path length within it. This process is repeated for 10 times until the ionization and temperature structures converge \citep[][]{vandenbrouckewood18}. Upon completion, the ionic fractions are returned to {\scshape phantom}, which then heats up the particles accordingly. 

To reconcile between SPH and grid-based models, the particle-interpolated fluid densities are transferred onto Voronoi grids (and vice-versa) using the Exact mapping method developed by \citet{petkova18}. Lloyd iterations \citep{lloyd82} are employed to regularize the Voronoi grids which is necessary for highly irregular density fields \citep[][]{petkova21}. We also introduced pseudo-particles in \citet{lau25} which can temporarily de-refine the neutral regions during the mapping. 

In the current version, we assume a pure hydrogen composition, which is an appropriate approximation as far as gas dynamics is concerned. Ionic fractions are solved under the steady-state assumption, since GMCs are sufficiently dense for the recombination timescales to remain significantly shorter than the simulation timesteps. The re-emitted ionizing photons in optically thick GMCs are also extremely likely to be absorbed locally (i.e. that the `on-the-spot' approximation applies), which justifies using a case-B recombination coefficient $\alpha_B = 2.7 \times 10^{-13}\ \mathrm{cm^3\ s^{-1}}$ \citep[][]{osterbrock74}. We set the photoionization cross-section to be $\sigma_{\nu} = 6.3\times10^{-18}\ \mathrm{cm^2}$ \citep[][]{bisbas15}. 

Photoionization heating is likewise computed with the steady-state assumption, making it predominantly dependent only on the particle ionic fraction. Additionally, a background heating rate of $\Gamma = 2 \times 10^{-26}\ \mathrm{erg\ s^{-1}}$ \citep{koyamainutsuka02} is applied to all particles to include the photoelectric heating from dust grains, heating by cosmic rays, and the heating from $H_2$ formation and destruction \citep{koyamainutsuka00}. 

The heating terms are then balanced with radiative cooling rates to obtain a net temperature change for each particle. We adopted one of the cooling curves tabulated in \citet{derijcke13} for pure hydrogen composition (i.e. zero metallicity); the curve is akin to that of \citet{joungmaclow06}, with the high-temperature part constructed based on the work of \citet{sutherlanddopita93}. Finally, the internal energies are updated implicitly \citep[][]{vazquezsemadeni07} to alleviate the constraints on the timesteps, especially after SN injection. The fully-ionized particles are typically heated to around $10^4-10^5\ \mathrm{K}$. This RHD scheme has been verified against the {\scshape starbench} benchmark models \citep{bisbas15}.

\subsection{Type II supernova injection method} \label{sec:sn_method}

We inject $10^{51}\ \mathrm{erg}$ of SN energy purely in kinetic form based on the findings of \citet{fichtner24}. Before injection, we first delete all particles within 0.1 pc radius of the detonating sink. We also assume $25\%$ of the sink's mass becomes ejecta mass \citep{lucas20}. The emptied sphere is then, again, populated with new particles whose total mass corresponds to that of the deleted particles plus the ejecta. These particles are placed randomly but symmetrical across the Cartesian axes. This method ensures an equal redistribution of SN energy across ejecta particles, and that the explosion is initially spherically symmetrical. 

We set the ejecta particles' radial velocity profile to be an inward-skewed Gaussian, so that they can rapidly create a swept-up shell\footnote{Not to be confused with the shell formed during snowplough phases when cooling becomes significant.} whilst allowing some particles to remain in the SN cavity, which otherwise can be error-prone in SPH. The shock-heated particles are subjected to the same background heating and radiative cooling as implemented in the RHD scheme, even when photoionization is switched off. In cloud scales, the numerical overcooling problem should be insignificant, hence no special treatments are required.

\section{Simulation setup and evolution} \label{sec:calculations}

The following sections detail the setup (Section~\ref{sec:initial_conditions}) and the multi-stage evolution of our GMCs. First, the clouds are let to collapse under turbulence and self-gravity (Section~\ref{sec:cloud_collapse}). We then select the dynamically-formed massive sinks to be ionizing sources, and activate the RHD scheme to simulate photoionization (Section~\ref{sec:injecting_rad}). As the H {\scshape ii} regions evolve and expand, we arbitrarily select snapshots from their evolution that display a variety of ionized gas morphologies (i.e. different shapes of sculpted regions). For each selected snapshot, we detonate the ionizing source as SN, and showcase the evolution of the remnant in each scenario (Section~\ref{sec:injecting_sn}). 

Since we do not follow stellar evolution, the SN injection timings may not be consistent with the growth of the sinks. However, the evolution from the moment the OB stars reach zero-age main sequence to the formation of steady-state H {\scshape ii} regions, in fact, takes place very rapidly. Presumably the cavities and channels emerge within similar timescales, otherwise they would not have become diffuse enough to be exposed to the ionizing radiation. There is hence no harm in injecting SNe at earlier times, when the density fields around the progenitors have already been sculpted by photoionization.

\subsection{Initial conditions} \label{sec:initial_conditions}

Turbulent GMC models with different initial densities and virial ratios are created. Here, we define $\alpha = E_k/|E_p|$ \citep[][]{dale12b}, where $E_k$ is the total kinetic energy and $E_p$ is the total gravitational potential energy of the cloud. The parameter space is chosen such that the clouds are bound enough to withstand the disruption by ionizing feedback, at least during the first ten thousand years. We also require that the initial ionization front is within the clouds' domains, hence diffuse or unbound GMCs are not included in this study. All clouds have a total mass of $10^4\ \mathrm{M_\odot}$ and a mass resolution of $10^{-2}\ \mathrm{M_\odot}$. 

The clouds are initially set as ellipsoids, as GMCs are more often seen to be elongated rather than spherical \citep[e.g.][]{bonnell11}. We initialise the clouds at two different (uniform) densities -- $\rho = 10^{-21}\ \mathrm{g\ cm^{-3}}$ and $\rho = 10^{-20}\ \mathrm{g\ cm^{-3}}$. With a fixed total mass, the cloud sizes depend on their assigned density, but the ratios of semi-major to semi-minor axes are fixed at $2:1:1$ in all models. If we approximate the clouds as spheres, their free-fall times $t_\mathrm{ff} \equiv \sqrt{{3 \pi / 32 G \rho}}$ are $2.11\ \mathrm{Myr}$ and $0.67\ \mathrm{Myr}$ respectively. 

Similar to \citet{lucas17a} and Paper~I, a low-density warm envelope with $\rho_\mathrm{env} = 4\times10^{-25}\ \mathrm{g\ cm^{-3}}$ is placed around each cloud. This envelope mimics the photodissociation regions around the cold molecular gas, as well as the warm phase of the ISM in which the GMCs embed. Their volumes are set such that the low-density gas fill around 20 pc beyond the boundaries of the GMC, covering the spatial extents of the SN outflows that escape the cloud. 

Internal energies are evolved with an adiabatic equation of state (EOS). With cooling implemented (cf. Section~\ref{sec:rhd_scheme}), the gas temperatures are predominantly governed by their local densities. Prior to feedback injection, the GMCs remain at approximately $10\ \mathrm{K}$, whereas the envelopes are at around $10^3\ \mathrm{K}$. The mean molecular weight $\mu$ is set to $1.29$ and heat capacity ratio $\gamma$ is fixed at $5/3$; the gas is assumed to be monatomic since we are mostly interested in the ionized hot outflows in this study. Sound speed is defined as $c_s = \sqrt{\gamma k_\mathrm{b} T/\mu m_\mathrm{p}}$, where $k_\mathrm{b}$ is the Boltzmann constant, $m_\mathrm{p}$ is proton mass, and $T$ is gas temperature. Shock heating and $P\mathrm{d}V$ heating terms are applied. 

We add supersonic turbulence by imposing a pre-computed velocity field of dimension $32^3$ onto the particles during setup. The velocities are generated using the method adopted in \citet{bate03}. The code produces a randomly-generated divergence-free Gaussian velocity field that follows a power spectrum $P_v(k) \propto k^{-11/3}$ \citep{dubinski95}, where $k$ is the wavenumber of the velocity perturbation. Here in our simulations, we stick to one particular realisation of the turbulent velocity field and apply the same cube to all clouds. We do not manually drive turbulence during runtime. 

The initial boundness of the GMCs are varied by changing the root-mean-square Mach number $M_\mathrm{rms} \equiv \sqrt{1/n \sum_i v_i^2/c_s^2}$, where $n$ is the number of particles, $v_i$ denotes the individual particle velocity and $c_s$ is assumed to be initially uniform. We construct GMCs with $\alpha = 0.7$ \citep{dale12b} and $\alpha = 1.0$, with the latter being marginally bound. Note also that the $\alpha$ ratios are computed only with the particles that initially belong to the dense cloud, excluding those in the envelope. Hence, the `true' virial ratios of our simulated GMCs are likely lower, as they are further subjected to the external thermal pressure exerted by the warm envelope. No turbulence is set inside the envelope.

\subsection{Cloud collapse} \label{sec:cloud_collapse}

\begin{figure}
    \vspace{0.2in}
    \centering
    \includegraphics[width=\linewidth]{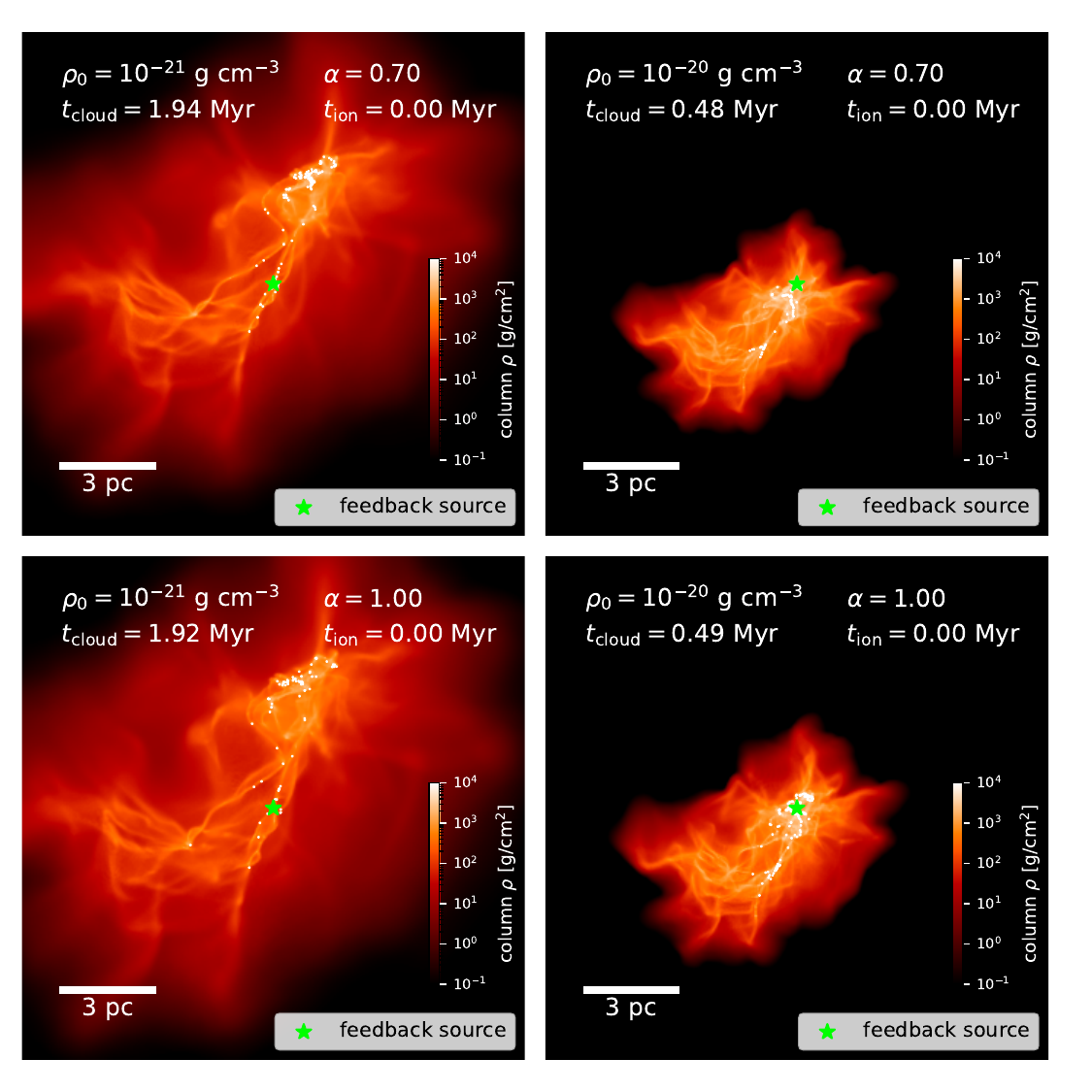}
    \vspace{-0.3cm}
    \caption{Column density of four GMC models with initial densities $10^{-21}\ \mathrm{g\ cm^{-3}}$ (\textit{left}) and $10^{-20}\ \mathrm{g\ cm^{-3}}$ (\textit{right}), and virial ratios $\alpha = 0.7$ (\textit{top}) and $\alpha = 1.0$ (\textit{bottom}). The timestamp $t_\mathrm{cloud}$ indicates the time elapsed since the collapse from uniform ellipsoids, and $t_\mathrm{ion}$ is the time since ionizing feedback is injected. Sink particles are indicated by white dots. The most massive sink particle formed in each model is indicated by a green star. }
    \label{fig:clouds_rho_init}
    \vspace{0.3cm}
    \includegraphics[width=\linewidth]{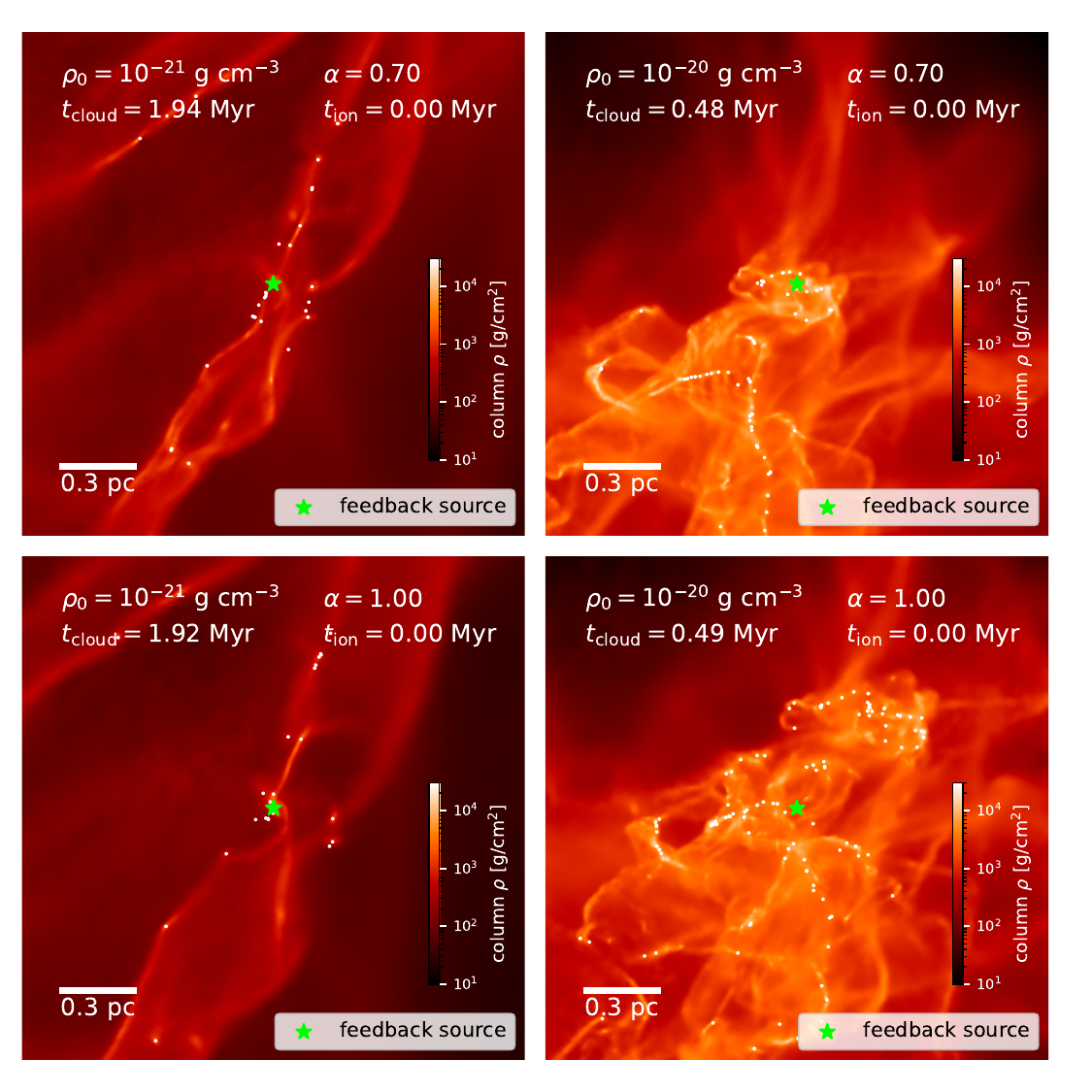}
    \vspace{-0.3cm}
    \caption{Same as Fig.~\ref{fig:clouds_rho_init} but zoomed-in on the most massive sink particle. Note the change in column density colour scale. }
    \label{fig:clouds_rho_init_zoomed}
    \vspace{0.3cm}
\end{figure}

The clouds are evolved for approximately $0.7-0.9\ t_\mathrm{ff}$, beyond which the timesteps were found to be significantly constrained\footnote{Turbulent clouds can normally evolve significantly beyond $1\ t_\mathrm{ff}$, however the free-fall time here is only approximated by assuming the cloud to be a sphere and neglecting the envelope mass. The `true' $t_\mathrm{ff}$ may differ.}. Fig.~\ref{fig:clouds_rho_init} shows the column density of each GMC immediately before ionizing feedback injection. The sinks are indicated by small white dots. The most massive sink in each cloud is marked by a green star, with masses ranging from $10\ \mathrm{M_\odot}$ to $81\ \mathrm{M_\odot}$. Some clouds appear to have discrete boundaries. This is largely due to the surrounding non-turbulent warm envelope that exerts a slight compression onto the cloud edges, despite not being seen in the column density render. 

Fig.~\ref{fig:clouds_rho_init_zoomed} zooms-in on their local environment. In the $10^{-21}\ \mathrm{g\ cm^{-3}}$ runs, their most massive stars are both located at an intersection between two interwound filaments that span North-South across the GMC, parallel to the direction of the cloud's longer extents \citep[e.g.][]{andre14}. This may be a hub-filament system \citep[e.g.][]{wang20}. The regions surrounding these filaments are rather diffuse. The $10^{-20}\ \mathrm{g\ cm^{-3}}$ runs, on the other hand, have their most massive stars located within dense stellar clusters in the Northern part of the GMCs. These runs are more in agreement with the competitive accretion scenario \citep[e.g.][]{bonnell01}.

\subsection{Photoionization feedback} \label{sec:injecting_rad}

The green stars marked in Fig.~\ref{fig:clouds_rho_init} are set as ionizing sources. The radiation follows a blackbody spectrum, with stellar surface temperature estimated using the sink mass and its radius approximated as $10\ \mathrm{R_\odot}$ for a typical O star. To compensate for the disregard of other nearby massive stars, which are meant to be radiating too, we assume an ionizing photon flux of $Q = 10^{51}\ \mathrm{s^{-1}}$. This $Q$ corresponds to the Lyman-continuum photon rate from Giant H {\scshape ii} regions made by the most massive OB clusters in our Galaxy \citep[e.g.][]{debuizer23}. For the present study, we focus on one single feedback-driven cavity that is carved by the SN progenitor itself, though setting multiple sources is possible.

\subsubsection{Evolution and morphologies of the H {\scshape ii} regions} \label{sec:evol_morph_hii_regions}

The radiation initially creates compact H {\scshape ii} regions around the embedded ionizing sources. After several thousands of years, the ionized gas expands and breaks out of its parent filament or clump. This process could have been sped up if stellar winds or radiation pressure are included. Once the ionization front leaves the clump, it rapidly (perhaps supersonically) expands towards regions with low optical depth \citep[e.g.][]{tenoriotagle79}. This is usually along the planes perpendicular to the filaments, or along the gaps in between the cluster sub-structures \citep[][]{bonnell03}. 

\begin{figure*}
    \centering
    \begin{minipage}{.48\textwidth}
        \centering
        \includegraphics[width=\linewidth]{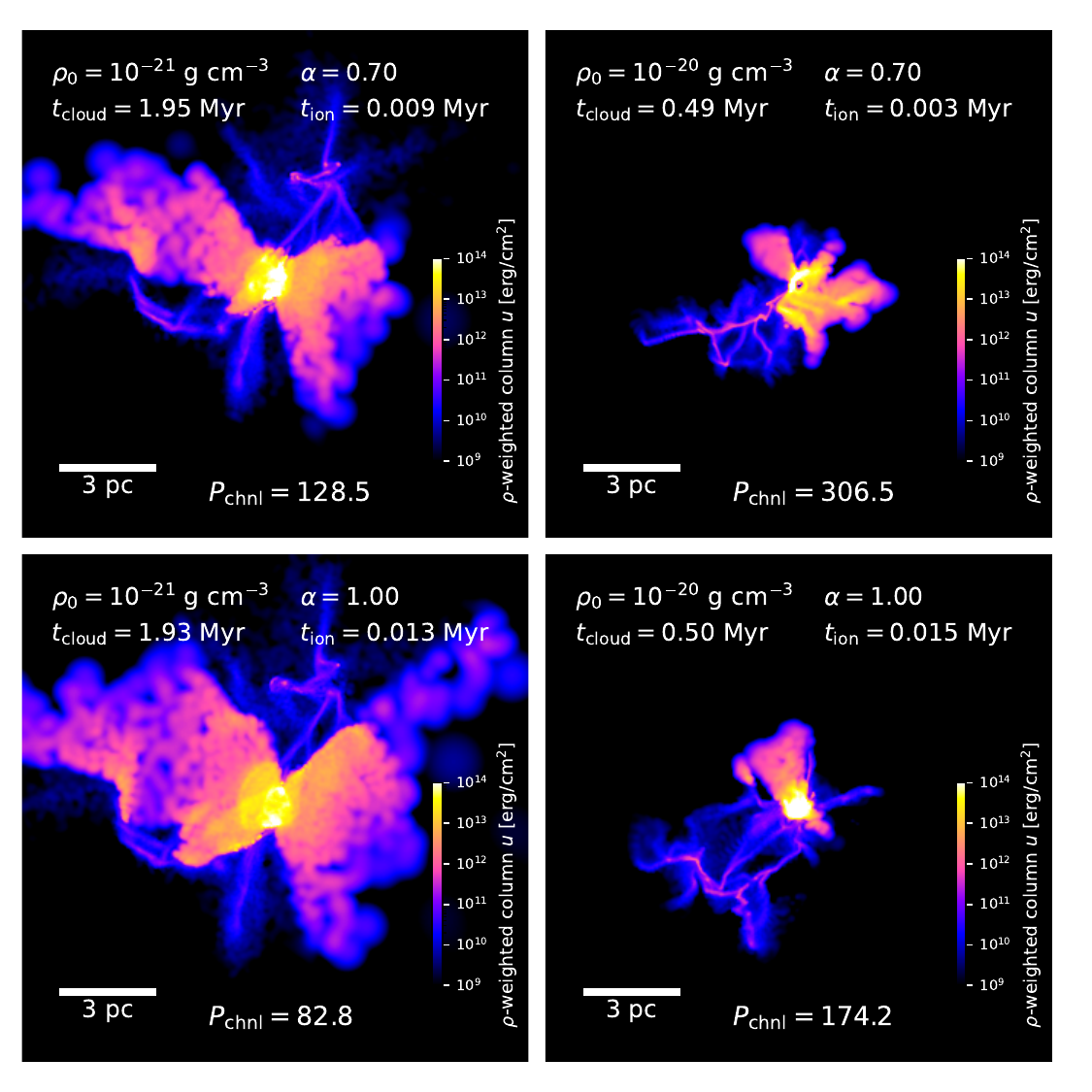}
        \subcaption{Compact environments}
        \label{fig:cloud_slice_u_intermed}
    \end{minipage}
    \begin{minipage}{.48\textwidth}
        \centering
        \includegraphics[width=\linewidth]{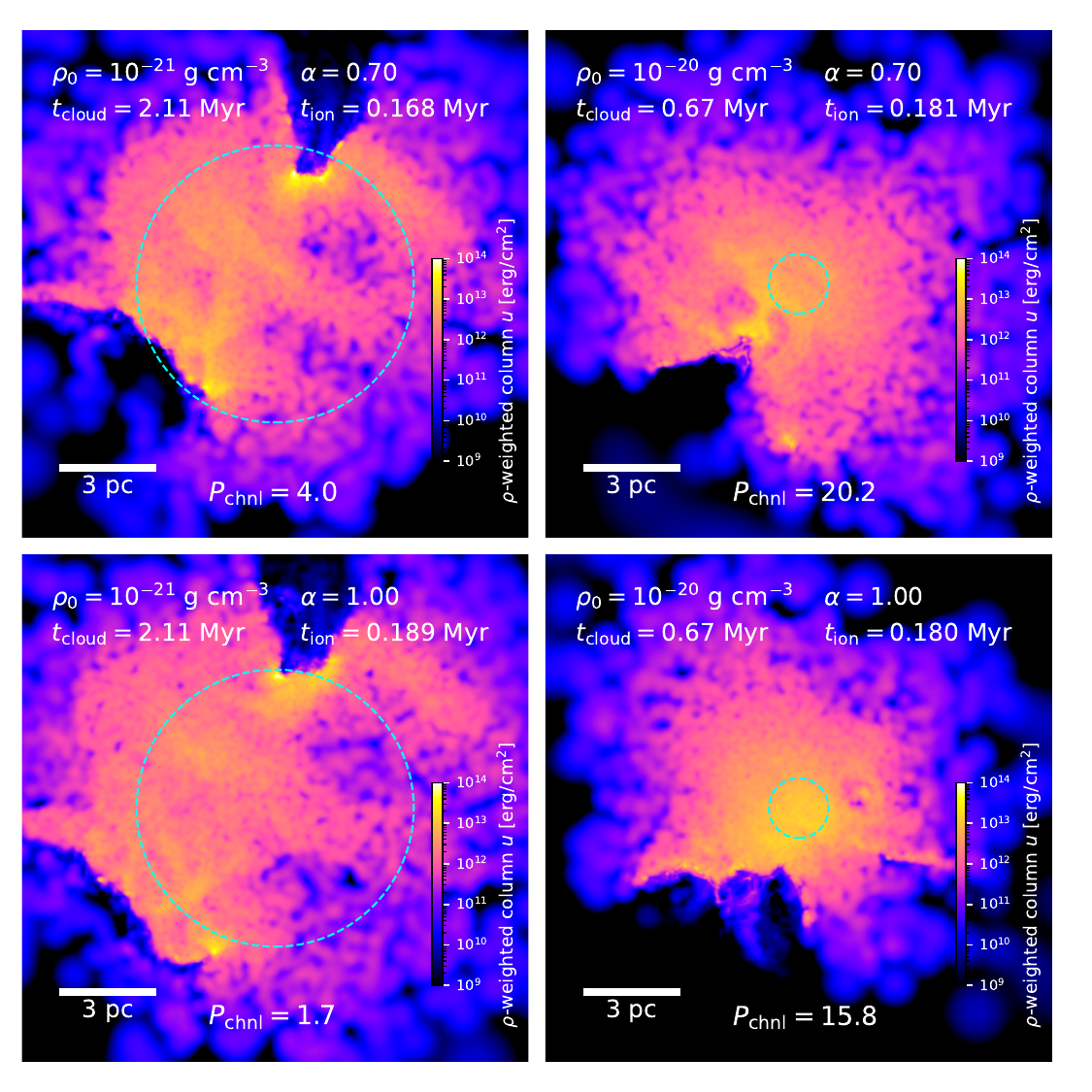}
        \subcaption{Dispersed environments}
        \label{fig:cloud_slice_u_final}
    \end{minipage}
    \caption{Column internal energy of particles within a thin slice of thickness 0.1 pc around the ionizing source, created by adding photoionization into the GMC models in Fig.~\ref{fig:clouds_rho_init} and evolved until (a) the early times, when the ionized regions show peculiar structures (\textit{left}: compact environments), and until (b) the later times, when the cloud is largely dispersed and the H {\scshape ii} regions have expanded (\textit{right}: dispersed environments). $t_\mathrm{ion}$ is the time elapsed since ionizing feedback is injected. $P_\mathrm{chnl}$ provides the channelling parameter (see Section~\ref{sec:channelling_param}) of the cloud in the snapshot. The estimated Str\"{o}mgren radius is shown as a circle in cyan.}
\end{figure*}

Fig.~\ref{fig:cloud_slice_u_intermed} demonstrates how the ionized regions appear during the early times. We cut out a slice with thickness 0.1 pc around the progenitor in each simulation and plot their column internal energies. The ionization-heated regions precisely trace the deeply embedded low-density cavity and channel structures around the sink. At this stage, these regions have highly irregular morphologies. In runs where the ionizing source lies along a filament (or in a hub-filament), large butterfly-shaped channels are formed; whereas in runs where the source is embedded within a cluster, the channels exhibit more complex and peculiar structures. Bearing in mind that we have used an exceptionally large value for $Q$, clouds with less OB stars may require more time for these structures to develop. 

\begin{figure*}
    \centering
    \begin{minipage}{.48\textwidth}
        \centering
        \includegraphics[width=\linewidth]{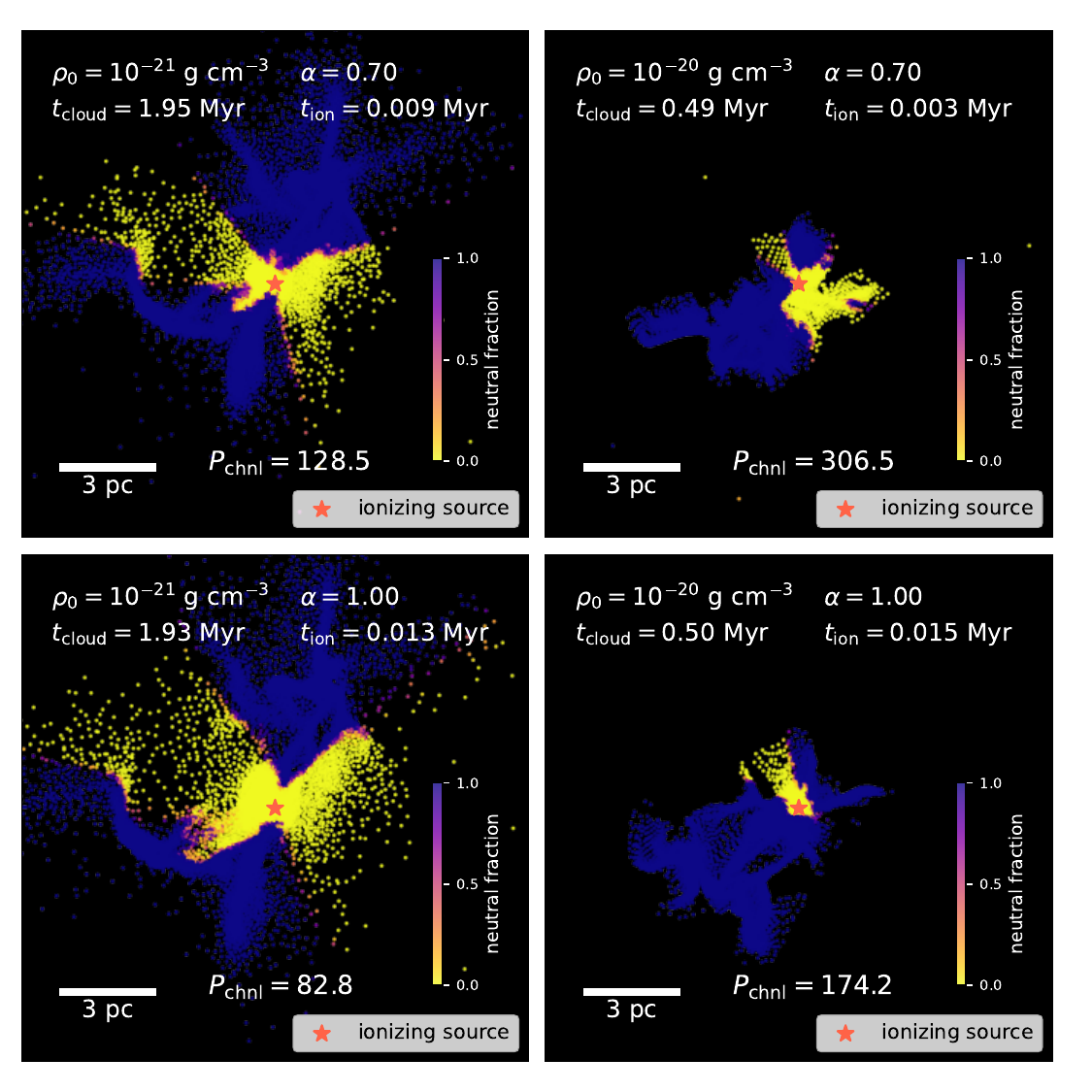}
        \subcaption{Compact environments}
        \label{fig:cloud_slice_ionicfrac_intermed}
    \end{minipage}
    \begin{minipage}{.48\textwidth}
        \centering
        \includegraphics[width=\linewidth]{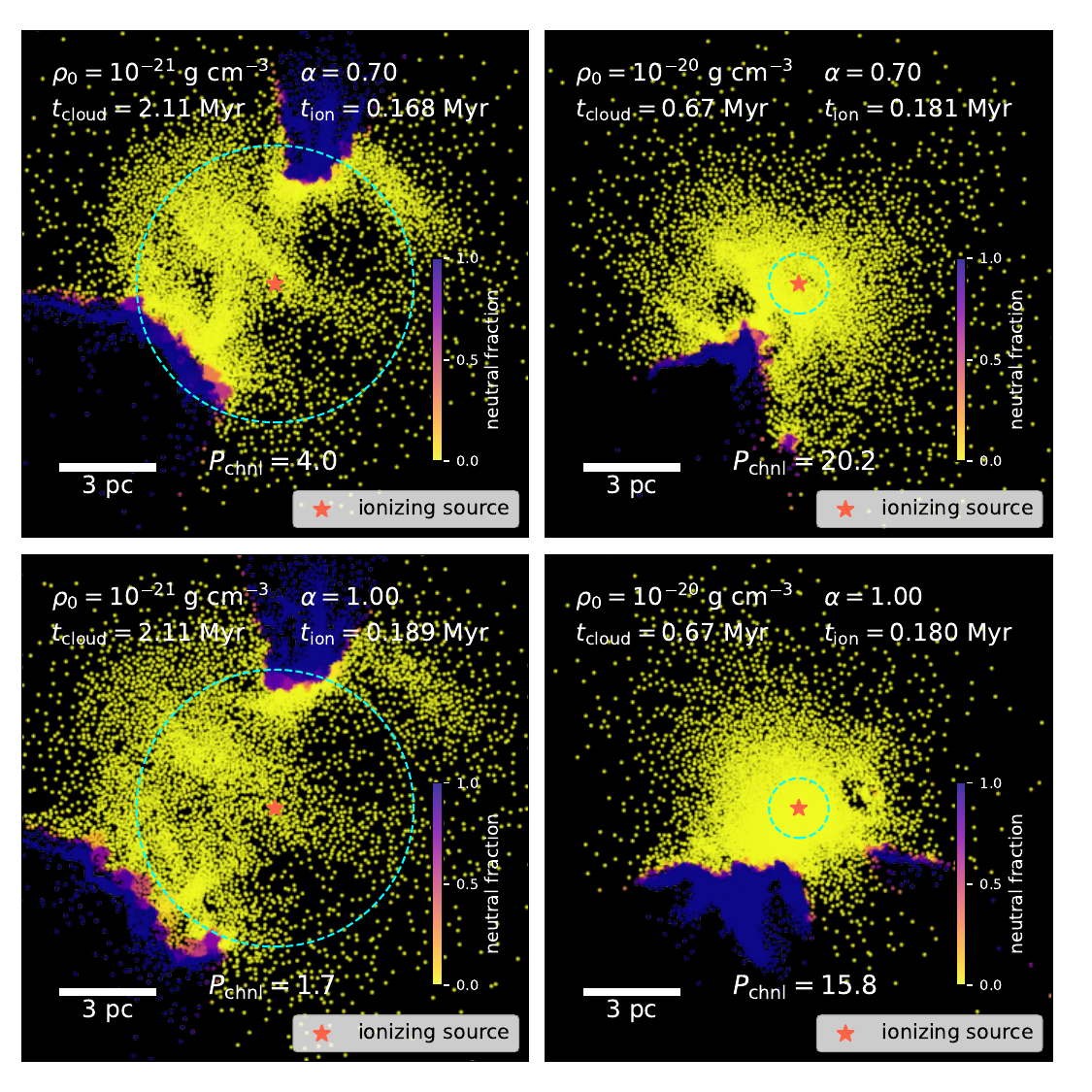}
        \subcaption{Dispersed environments}
        \label{fig:cloud_slice_ionicfrac_final}
    \end{minipage}
    \caption{Neutral fraction of the particles shown in Fig.~\ref{fig:cloud_slice_u_intermed} and \ref{fig:cloud_slice_u_final} (without SPH interpolation). Yellow dots indicate fully-ionized particles, and blue dots indicate fully-neutral. The transition zones are seen as thin borders in pink. The estimated Str\"{o}mgren radius is shown in cyan.}
\end{figure*}

We also plot in Fig.~\ref{fig:cloud_slice_ionicfrac_intermed} the neutral fractions of the particles taken from the same slices as those in Fig.~\ref{fig:cloud_slice_u_intermed}. The yellow particles are fully ionized and the blue ones are fully neutral. The thin transition zones are seen in pink; their widths correspond to the photons' mean free paths \citep{osterbrock74}. They trace the boundaries\footnote{The ionized regions in Fig.~\ref{fig:cloud_slice_ionicfrac_intermed} appear to have open boundaries towards the outside of the GMCs, but this is only due to the lack of resolution in the warm envelope. All H {\scshape ii} regions in steady-state have finite boundaries. } of the cavities and channels, providing a clear view of their geometries. It also demonstrates their resemblance with the idealised semi-confined SN model presented in Paper~I. We hereafter label this set of ionized GMCs at young ages as the `compact environments'.

The four GMCs are then further evolved to reach approximately 1 $t_\mathrm{ff}$. By this time, the H {\scshape ii} regions have expanded beyond the clouds' initial domains in all runs. The ionized gas has dispersed a significant portion of the gas near the ionizing sources, leaving only the densest filaments in the GMCs. Filamentary structures that immediately surround the source are smoothed out by the expanding shells. The internal energies and neutral fractions of these snapshots are shown in Fig.~\ref{fig:cloud_slice_u_final} and Fig.~\ref{fig:cloud_slice_ionicfrac_final} respectively. The ionized regions are now approximately spherical bubbles, with only a small portion of their surfaces obstructed by the remains of the dense GMCs, appearing similar to that in \citet{kimostriker15}. 

Fig.~\ref{fig:cloud_slice_u_final} also demonstrates how expanding H {\scshape ii} regions create compressed gas layers at the rim of the existing dense structures. This is particularly obvious in the $\rho_0 = 10^{-21}\ \mathrm{g\ cm^{-3}}$ runs, where illuminated borders are seen around the remaining parts of the GMCs. They closely resemble the bright-rimmed clouds \citep[e.g.][]{ortega13} which are believed to be produced by the molecular (or partially ionized) outflows driven by strong ionizing sources \citep[e.g.][]{sugitani91,sugitani94}. Such outflows may trigger star formation via radiation-driven implosion \citep[e.g.][]{kesseldeynetburkert03,saha22a,saha22b}. Whilst the definition of a `triggered star' remains debatable \citep[][]{dale15b}, the effect of ionization on the ongoing star formation activity in our simulations shall be examined in a future paper. 

Furthermore, Fig.~\ref{fig:cloud_slice_u_final} and ~\ref{fig:cloud_slice_ionicfrac_final} reveal that the overall size and compactness of the H {\scshape ii} regions are approximately correlated with the GMC's initial density. This result agrees with the predictions of \citet{stromgren39}, who derived the radius of an H {\scshape ii} region in a uniform medium upon reaching ionization equilibrium. The Str\"{o}mgren radius is given by 
\begin{equation} 
    R_\mathrm{St} = \left( \frac{3Q}{4 \pi \alpha_B n_{\rm H}^2} \right) ^{1/3},
    \label{eq:stromgren_radius}
\end{equation}
where $n_\mathrm{H} \approxeq \rho/m_\mathrm{p}$ is the number density of the medium. For GMC densities $10^{-21}\ \mathrm{g\ cm^{-3}}$ and $10^{-20}\ \mathrm{g\ cm^{-3}}$, their Str\"{o}mgren radii are approximately 4.375 pc and 0.943 pc. We plot them in Fig.~\ref{fig:cloud_slice_u_final} and Fig.~\ref{fig:cloud_slice_ionicfrac_final} as dashed circles. Obviously, the ionization fronts in the simulations far exceed their predicted Str\"{o}mgren radii. This is because the GMCs are porous - the cold gas gradually mixes with the low-density warm envelope as ionizing feedback unbinds the cloud. We hereafter label these four photoionized GMCs at late-stages as the `dispersed environments'\footnote{Note however that, despite the name `dispersed environments', the clouds are not completed dispersed. We shall see in Fig.~\ref{fig:chnl_density_final} that the cavities and channels are still present, but bigger and less distinctive. }.

\subsubsection{The channelling parameter} \label{sec:channelling_param}

Fig.~\ref{fig:cloud_slice_u_intermed} and ~\ref{fig:cloud_slice_u_final} illustrate the huge variety of H {\scshape ii} region morphologies, each tracing a different density field. To quantitatively interpret their geometries, we propose defining a `channelling parameter' $P_\mathrm{chnl}$. In broad terms, it measures the extent to which the pre-SN feedback bubble resembles the `semi-confined SN' model in Paper~I, wherein the small ionized cavity is connected to narrow ionized tubes that thread through the cold neutral GMC gas. More specifically, 
\begin{enumerate}[wide=0pt, labelwidth=!,itemindent=!,labelindent=!, leftmargin=!, label=(\roman*), parsep=0pt]
    \item $P_\mathrm{chnl}$ should increase with the ratio of ionized mass in the channels to those in the cavity; 
    \item $P_\mathrm{chnl}$ should increase with the ratio of neutral mass around the channels to the ionized mass within the channels; and
    \item $P_\mathrm{chnl}$ should decrease when the cavity is so large that it almost erases all structures in its parent cloud. 
\end{enumerate}
A visual representation of the above is included in Appendix~\ref{appen:pchnl_visual}. We therefore choose to define it as a dimensionless parameter, 
\begin{equation}
    P_\mathrm{chnl} = \frac{m_\mathrm{neu,chnl}}{m_\mathrm{ion,cav}}, 
    \label{eq:p_chnl}
\end{equation}
where $m_\mathrm{ion,cav}$ is the total mass of ionized particles within a cavity radius $r_\mathrm{cav}$, and $m_\mathrm{neu,chnl}$ is the neutral mass that lie between $r_\mathrm{cav}$ and the channels' maximum extent $r_\mathrm{chnl}$. This parameter reflects how well the channels penetrate in between the filamentary structures and the star-forming clumps, and is hereafter used in this paper to compare the GMCs. Their values do not represent any physical quantities and it is merely dimensionless index that quantifies the compactness of the ionized gas. However, we shall demonstrate later (especially in Fig.~\ref{fig:cumulative_mass_distri}) that $P_\mathrm{chnl}$ correlates with the SN outflow properties. 

\begin{figure*}
    \centering
    \begin{minipage}{.47\textwidth}
        \centering
        \includegraphics[width=\linewidth]{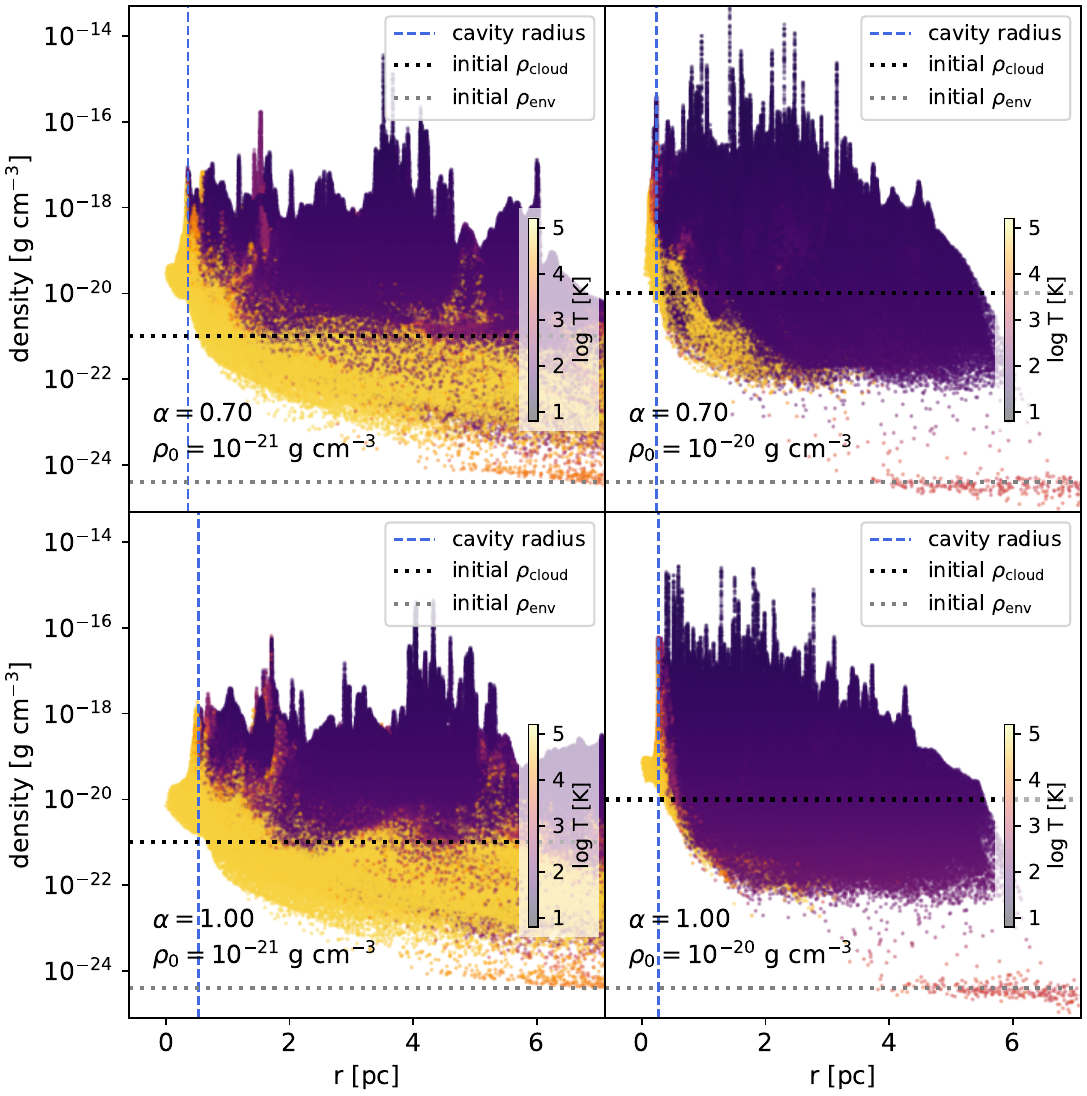}
        \subcaption{Compact environments}
        \label{fig:chnl_density_intermed}
    \end{minipage}
    \begin{minipage}{.47\textwidth}
        \centering
        \includegraphics[width=\linewidth]{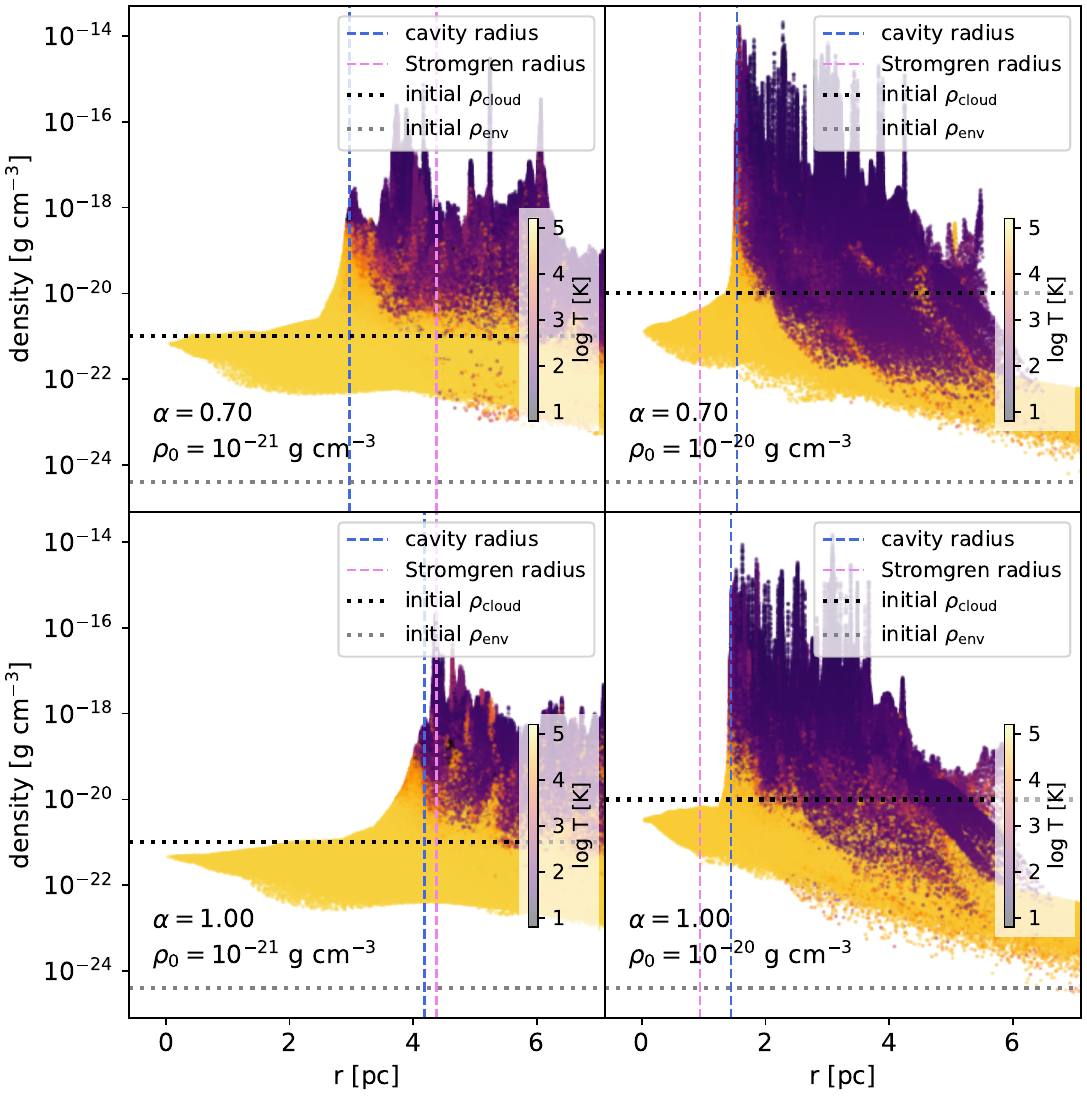}
        \subcaption{Dispersed environments}
        \label{fig:chnl_density_final}
    \end{minipage}
    \caption{Particle density plotted against their distance to the ionizing source for the GMCs in Fig.~\ref{fig:clouds_rho_init} at (a) the early times, when the cavity is small and some ionized gas escaped via narrow low-density channels (\textit{left}: compact environments), and at (b) a later time, when cavities and channels have significantly expanded (\textit{right}: dispersed environments). Colour indicates particle temperature. Vertical dashed line in blue marks the cavity radius $r_\mathrm{cav}$, defined as the position of the leftmost density-peak. The Str\"{o}mgren radius (equation~\ref{eq:stromgren_radius}) is shown as a purple vertical dashed line. Initial density of the cloud and the envelope are indicated by horizontal dotted lines in black and grey respectively. }

    \begin{minipage}{.48\textwidth}
        \centering
        \includegraphics[width=\linewidth]{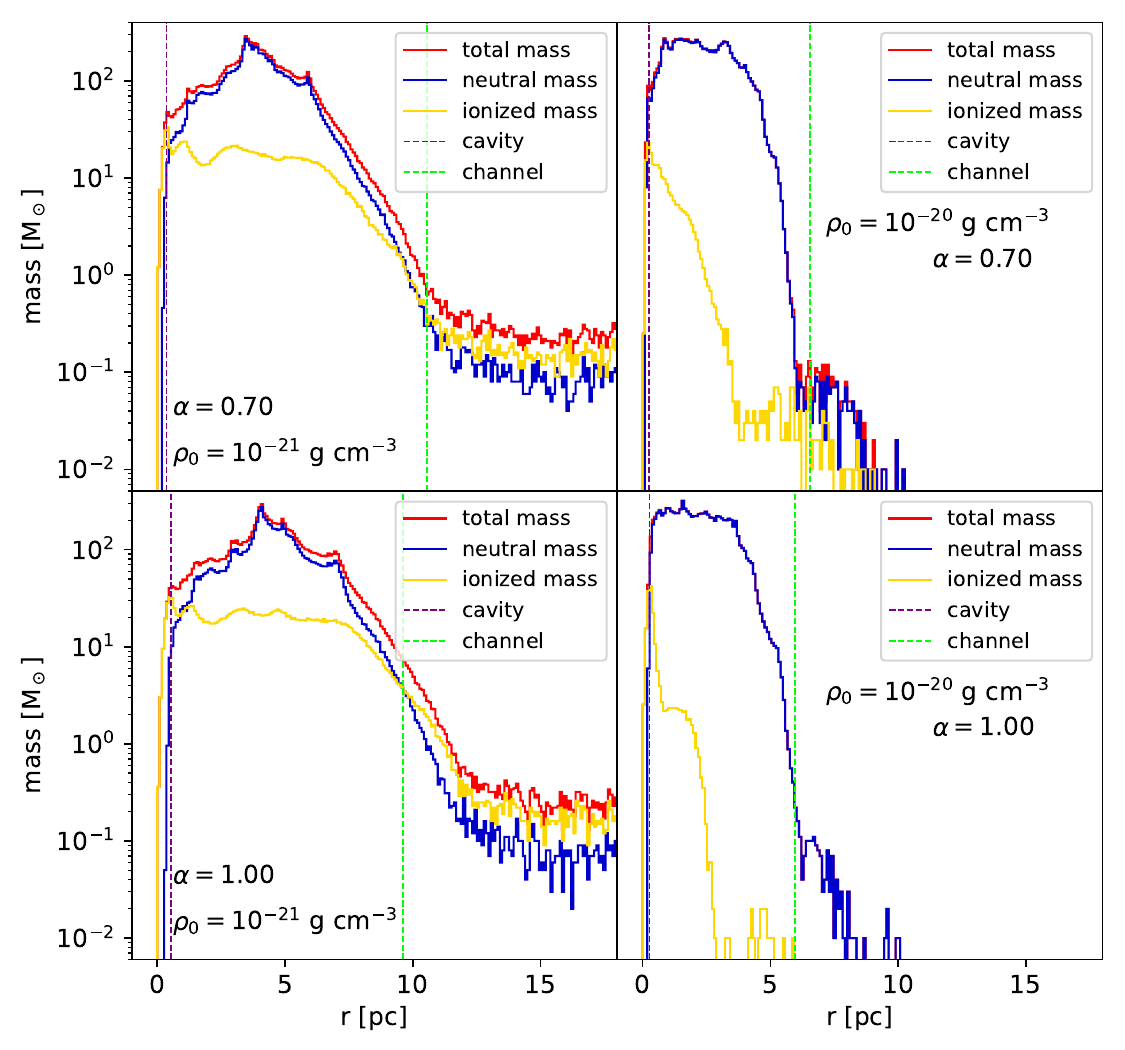}
        \subcaption{Compact environments}
        \label{fig:chnl_ionization_intermed}
    \end{minipage}
    \begin{minipage}{.48\textwidth}
        \centering
        \includegraphics[width=\linewidth]{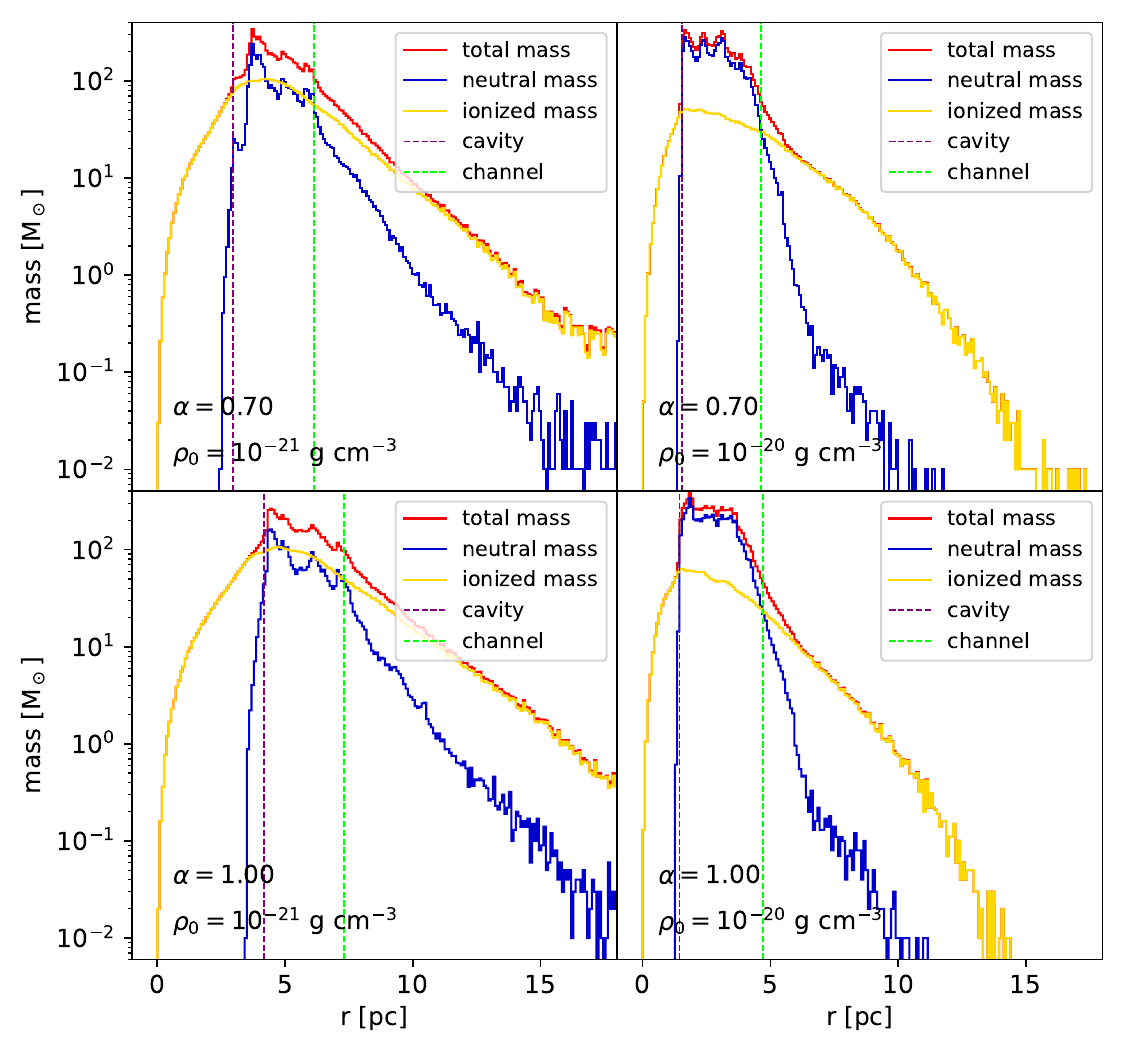}
        \subcaption{Dispersed environments}
        \label{fig:chnl_ionization_final}
    \end{minipage}
    \caption{Radial distribution of total mass (red), ionized mass (yellow) and neutral mass (blue) relative to the ionizing source in the same snapshots as in Fig.~\ref{fig:chnl_density_intermed} and ~\ref{fig:chnl_density_final}. Note the change in distance scale. Purple vertical dashed line indicates the cavity radius $r_\mathrm{cav}$ found in Fig.~\ref{fig:chnl_density_intermed} and ~\ref{fig:chnl_density_final}. Green vertical dashed line marks the channel radius $r_\mathrm{chnl}$, defined as the intersection between the ionized (yellow) curve and the neutral (blue) curve. }
\end{figure*}

Cavity radius $r_\mathrm{cav}$ may be trivially determined by plotting the density of the particles against their distances to the ionizing source, as shown in Fig.~\ref{fig:chnl_density_intermed} and \ref{fig:chnl_density_final}. Colours indicate the temperatures, allowing those exposed to the UV radiation (yellow) to be identified. An immediate observation is that the majority of the ionized gas `tunnels' through the `overhead' GMC gas, providing strong evidence for the presence of feedback-driven low-density channels. The `hill slopes' on the left correspond to the sculpted cavities. We define the radius of the leftmost density-peak to be $r_\mathrm{cav}$, which marks the position of the closest shell swept by the expanding H {\scshape ii} region. The $r_\mathrm{cav}$'s are denoted by blue vertical dashed lines. 

Another aspect to note from Fig.~\ref{fig:chnl_density_final} is that the Str\"{o}mgren radii (indicated by purple vertical dashed lines) indeed provide good estimates of $r_\mathrm{cav}$ for most of the time, but there are exceptions. Those with $r_\mathrm{cav} > R_\mathrm{St}$ indicate that the H {\scshape ii} regions have expanded due to overpressure \citep[][]{hosokawainutsuka06}, but those with $r_\mathrm{cav} << R_\mathrm{St}$ highlight the influence of inhomogeneous density fields. The insight from here is that the use of Str\"{o}mgren radius, or any other analytically-derived feedback bubble radius, may become ineffective with the presence of large channels. 

To estimate the channels' maximum extent $r_\mathrm{chnl}$, we bin the ionized and neutral particles (i.e. gas mass) by their distance to the ionizing source. The results are presented in Fig.~\ref{fig:chnl_ionization_intermed} and ~\ref{fig:chnl_ionization_final}, with the total mass distribution curves included as reference. We see that the cavity radii $r_\mathrm{cav}$ obtained from Fig.~\ref{fig:chnl_density_intermed} and ~\ref{fig:chnl_density_final} precisely match the peak of the ionized mass distributions. This peak is also where the ionized curve intersects with that of the total mass, indicating that gas below $r_\mathrm{cav}$ are almost fully ionized. 

Whilst cavities and bubbles may be well-defined boundaries, clouds do not \citep[e.g.][]{chevance23}. In an ideal picture, the channels `end' at where the cold molecular gas (in which they embed) transitions to warm phase. Yet with their irregular morphologies and the turbulent mixing that occurs at the interfaces, the transition radius becomes ambiguous, let alone observationally defining them. As such, for the purpose here, we arbitrarily define $r_\mathrm{chnl}$ to be where the ionized curve intersects with that of the neutral. That is, where the escaped ionized gas has outweighed the existing cold clumps and merged with the external ISM. In situations where they do not cross, we take the maximum reach of the ionized curve. $r_\mathrm{chnl}$'s are indicated by the green vertical dashed lines in Fig.~\ref{fig:chnl_ionization_intermed} and ~\ref{fig:chnl_ionization_final}.

The resulting $P_\mathrm{chnl}$ values computed with the above-described methods accurately reflect the cavity compactness, boundness, and the overall porosity of the GMCs. We demonstrate in Appendix~\ref{appen:pchnl_vs_covarea} that it also correlates with the `sky cover fraction' in the perspective of the progenitor. During the early stages, the morphology of the ionized gas (hence $P_\mathrm{chnl}$) is predominantly determined by the cloud's turbulent density structure. However, Fig.~\ref{fig:chnl_ionization_final} shows that the radial distribution of ionized gas towards later evolutionary stages approach a similar shape, with only minor density-dependent variations. This could imply that the expansion of H {\scshape ii} regions will become rather insensitive to the clumpy structures in its natal cloud as soon as the ionization front reaches the external ISM.

\subsection{Supernova feedback} \label{sec:injecting_sn}

Finally, the ionizing sources in the eight snapshots with a range of $P_\mathrm{chnl}$ values (see Fig.~\ref{fig:cloud_slice_u_intermed} and ~\ref{fig:cloud_slice_u_final}) are detonated as SN progenitors. We classify our SN simulations into `compact environment runs' and `dispersed environment runs' according to the snapshot used as the starting condition. We stress again that whilst the dispersed environment runs place SNe in more evolved H {\scshape ii} regions, the medium is not uniform nor homogeneous, only less compact and more porous. All SN simulations are run for approximately $0.15\ \mathrm{Myr}$, beyond which the shocks almost completely dissipate. Table~\ref{tab:sims_summary} summarises the SN simulations presented in this paper. The feedback injection timings are included, but they do not have much significance to the objective of our current study.

\begin{table*}
	\centering
	\caption{Summary of the eight SN simulations, with varied initial cloud density $\rho$, initial virial ratio $\alpha$, and compactness of the progenitor's environment (as reflected in $P_\mathrm{chnl}$ of the ionized gas structures). Other derived parameters include total mass of the cloud $M$, length of semi-major axis $R_\mathrm{max}$, initial root-mean-square velocity $v_\mathrm{rms}$ (determined by $M_\mathrm{rms}$) and the approximated free-fall time $t_\mathrm{ff}$. $\Delta t_\mathrm{cloud}$ denotes the time over which the cloud is evolved without feedback, $\Delta t_\mathrm{ion}$ is the time duration of photoionization, and $\Delta t_\mathrm{SN}$ is the time over which the SN remnant is evolved in the simulation. }
    \vspace{2mm}
	\label{tab:sims_summary}
	\begin{tabular}{ccccccccccc} 
		\hline
		$\rho$ & $M$ & $R_\mathrm{max}$ & $\alpha$ & $v_\mathrm{rms}$ & $t_\mathrm{ff}$ & $\Delta t_\mathrm{cloud}$ & $\Delta t_\mathrm{ion}$ & SN run label & $P_\mathrm{chnl}$ & $\Delta t_\mathrm{SN}$   \\
        $\mathrm{[g\ cm^{-3}]}$ & $\mathrm{[M_\odot]}$ & [pc] &  & $\mathrm{[km\ s^{-1}]}$ & [Myr] & [Myr] & [Myr] & (progenitor environment) &  & [Myr] \\
		\hline

		$10^{-21}$ & $10^4$ & 8.65 & 0.7 & 2.44 & 2.11 & 1.94 & 0.009 & compact & 128.5 & 0.15 \\
        .. & .. & .. & .. & .. & .. & .. & 0.168 & dispersed & 4.0 & 0.15 \\

        $10^{-21}$ & $10^4$ & 8.65 & 1.0 & 2.90 & 2.11 & 1.92 & 0.013 & compact & 82.8 & 0.15 \\ 
        .. & .. & .. & .. & .. & .. & .. & 0.189 & dispersed & 1.7 & 0.15 \\

        $10^{-20}$ & $10^4$ & 4.01 & 0.7 & 3.63 & 0.67 & 0.48 & 0.003 & compact & 306.5 & 0.15 \\ 
        .. & .. & .. & .. & .. & .. & .. & 0.181 & dispersed & 20.2 & 0.15 \\

        $10^{-20}$ & $10^4$ & 4.01 & 1.0 & 4.38 & 0.67 & 0.49 & 0.015 & compact & 174.2 & 0.15 \\ 
        .. & .. & .. & .. & .. & .. & .. & 0.180 & dispersed & 15.8 & 0.15 \\
		\hline
	\end{tabular}
    \vspace{2mm}
\end{table*}

\subsubsection{Evolution and morphologies of SN remnants} \label{sec:evol_morph_snr}

\begin{figure*}
    \centering
    \begin{minipage}{.48\textwidth}
        \centering
        \includegraphics[width=\linewidth]{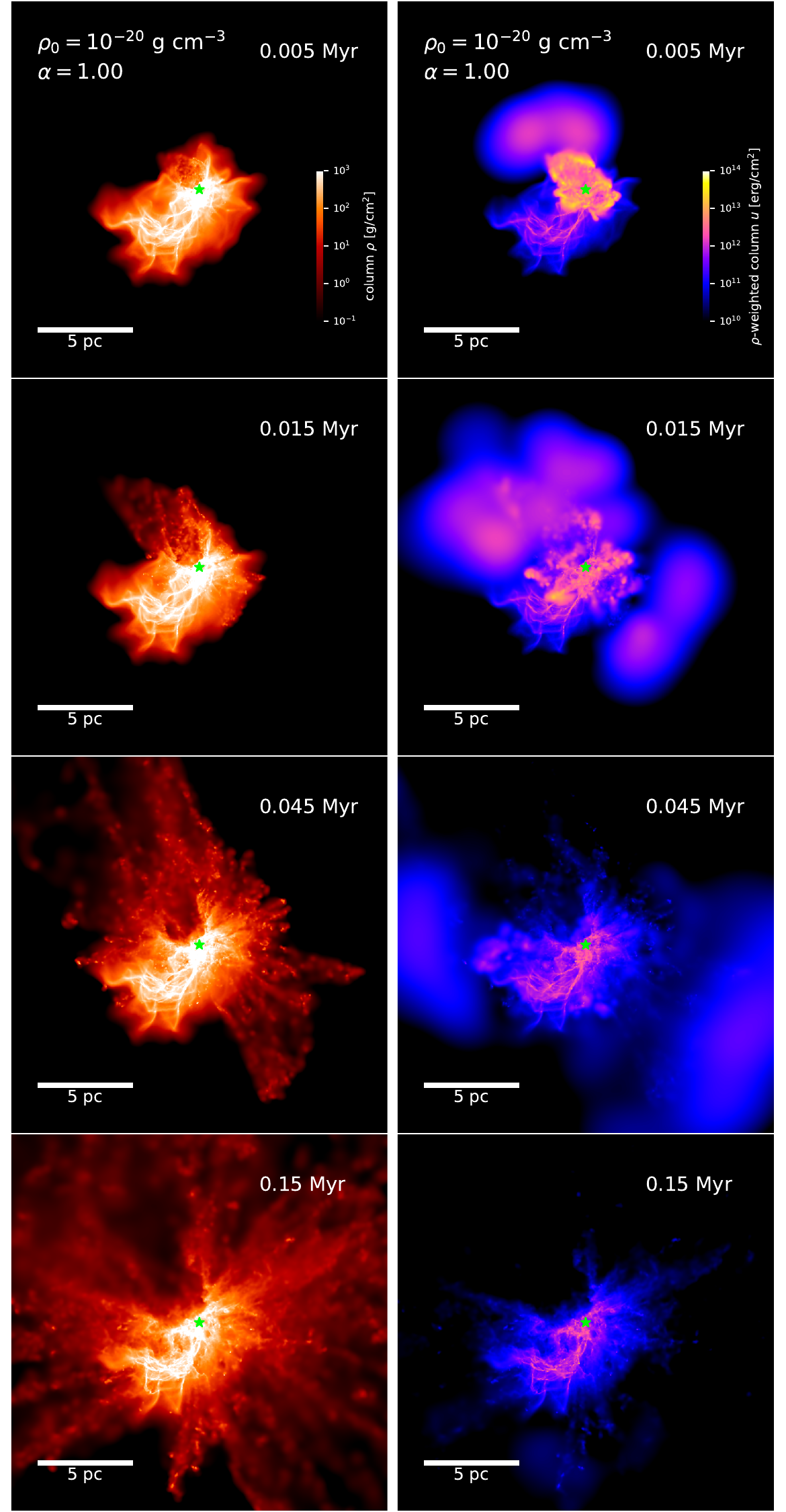}
        \subcaption{Compact environment run}
        \label{fig:radsn_20_10_evol_intermed}
    \end{minipage}
    \begin{minipage}{.48\textwidth}
        \centering
        \includegraphics[width=\linewidth]{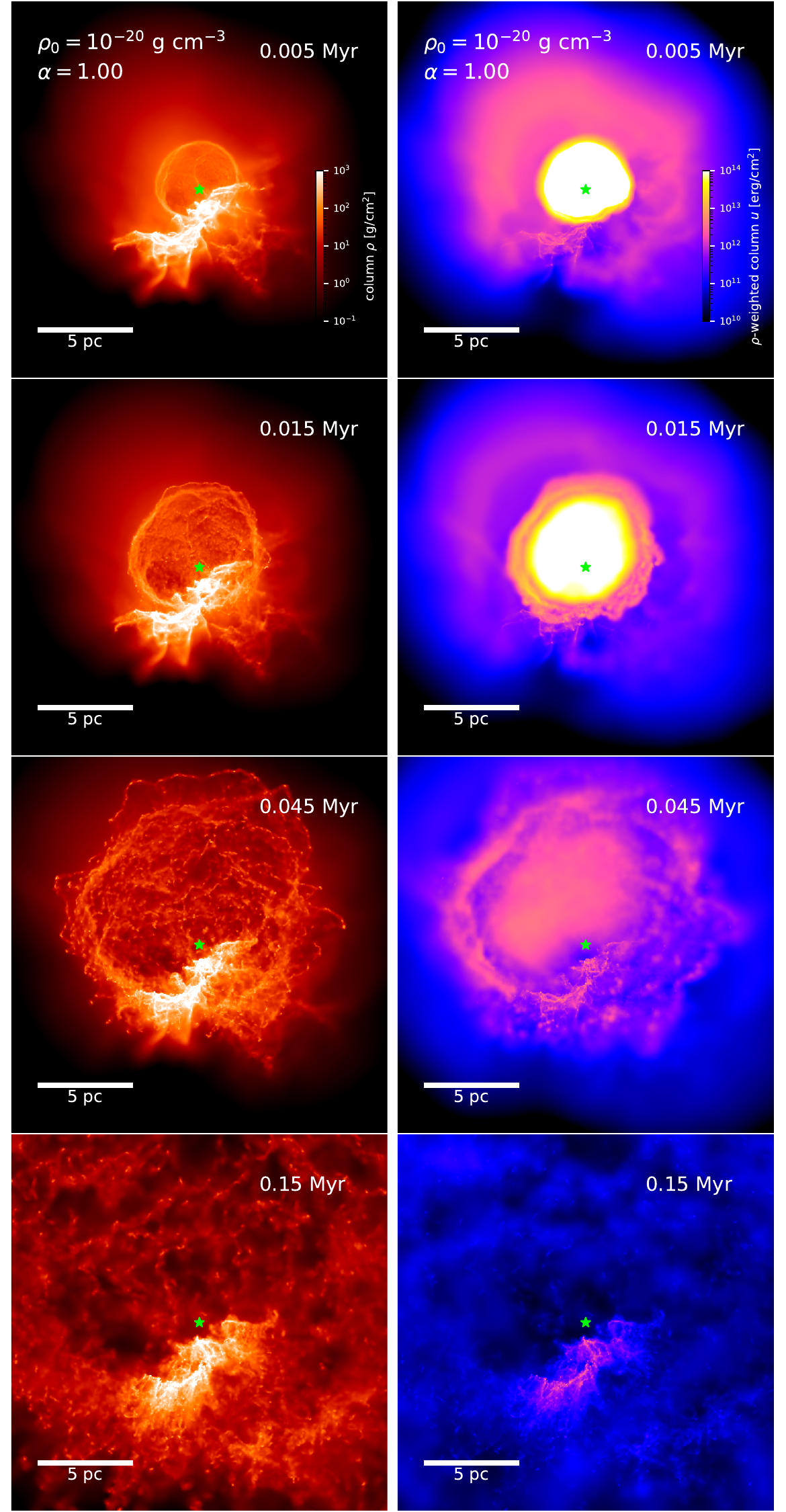}
        \subcaption{Dispersed environment run}
        \label{fig:radsn_20_10_evol_final}
    \end{minipage}
    \caption{\textit{Top to bottom}: Evolution of SNR in the $10^{-20}\ \mathrm{g\ cm^{-3}}|\alpha=1.0$ GMC, where the progenitor is detonated within (a) a small cavity surrounded by clumpy density structures (\textit{left}: the compact environment run), and (b) a large cavity created by pre-SN ionizing feedback (\textit{right}: the dispersed environment run). Red colours show the column density, and blue colours show the column internal energy. }
\end{figure*}

Fig.~\ref{fig:radsn_20_10_evol_intermed} presents snapshots of column density and column internal energy of a remnant (hereafter SNR), taken from the $10^{-20}\ \mathrm{g\ cm^{-3}}|\alpha=1.0$ compact environment run. By 5000 years, the shock-heated gas escapes from one side of the GMC, creating an outflow of energy that resembles a champagne flow \citep[e.g.][]{bodenheimer79}. It also resembles the idealised `semi-confined SN' from Paper~I. Soon after, another blowout emerges from the opposite side of the cloud, forming two lobes of outflowing gas by 45000 years. These outflows clear depart from the behaviour of a spherical blast. 

Additionally, there appears to be a systematic time discrepancy between the escape of the shock front (as seen in the internal energy render) and the unbinding of GMC gas (as seen in the density render). The gas do not seem to be driven by any sweeping SNR shells; instead, they appear to be hauled by the outgoing postshock flows. An alternative explanation is that the cold gas is being `kicked out' of the GMC only upon complete transfer of momentum from the SN ejecta, hence causing the time delay. 

Contrary to the above, the SNR in the dispersed environment run (Fig.~\ref{fig:radsn_20_10_evol_final}) retains an overall spherical symmetry, and has a similar appearance to SNRs evolved in a homogeneously-clumpy medium \citep[e.g.][]{guo25}. This is likely owing to the large spherical cavity carved by pre-SN feedback. The low-density environment within the cavity also helped preserve the thermal energy in the SNR. It demonstrates how the small-scale density structures immediately around the progenitors could determine the SNR morphologies, in particular, with how the outflows escape.

The evolution of the six other SNR runs (the compact and dispersed runs for GMCs with $10^{-21}\ \mathrm{g\ cm^{-3}}|\alpha=0.7$, $10^{-21}\ \mathrm{g\ cm^{-3}}|\alpha=1.0$ and $10^{-20}\ \mathrm{g\ cm^{-3}}|\alpha=0.7$) are presented in Appendix~\ref{appen:other_sn_runs}. The $10^{-21}\ \mathrm{g\ cm^{-3}}$ compact runs tend to have less significant differences to their dispersed counterparts. No SNR in the compact runs produced coherent explosions, as outflows preferentially travel down the low-density channels in the GMC.

\section{Results} \label{sec:results}

Recall from Section~\ref{sec:introduction} that the purpose of this paper is to (i) verify the analytical model proposed in Paper~I and (ii) examine how the outflows vary with the cloud density structures. To first validate the premise of our study, we demonstrate in Sections~\ref{sec:sky_plots} and \ref{sec:impacted_columnden} that the SN energies are indeed preferentially deposited in the low-density ISM. We then examine the kinematic properties of the cloud and the outflows in Sections~\ref{sec:outflow_vel}, \ref{sec:turbulence} and \ref{sec:energy_momentum} to compare against the findings from Paper~I. Emphasis is placed on examining the properties of the outflows that escaped the cloud in Sections~\ref{sec:mass_removal} and \ref{sec:escape_mass_energy_momen}, which provides crucial information for SN sub-grid modelling.

\subsection{Preferential removal of gas in low-density channels} \label{sec:sky_plots}

\begin{figure*}
    \centering
    \begin{minipage}{.4\textwidth}
        \centering
        \includegraphics[width=\linewidth]{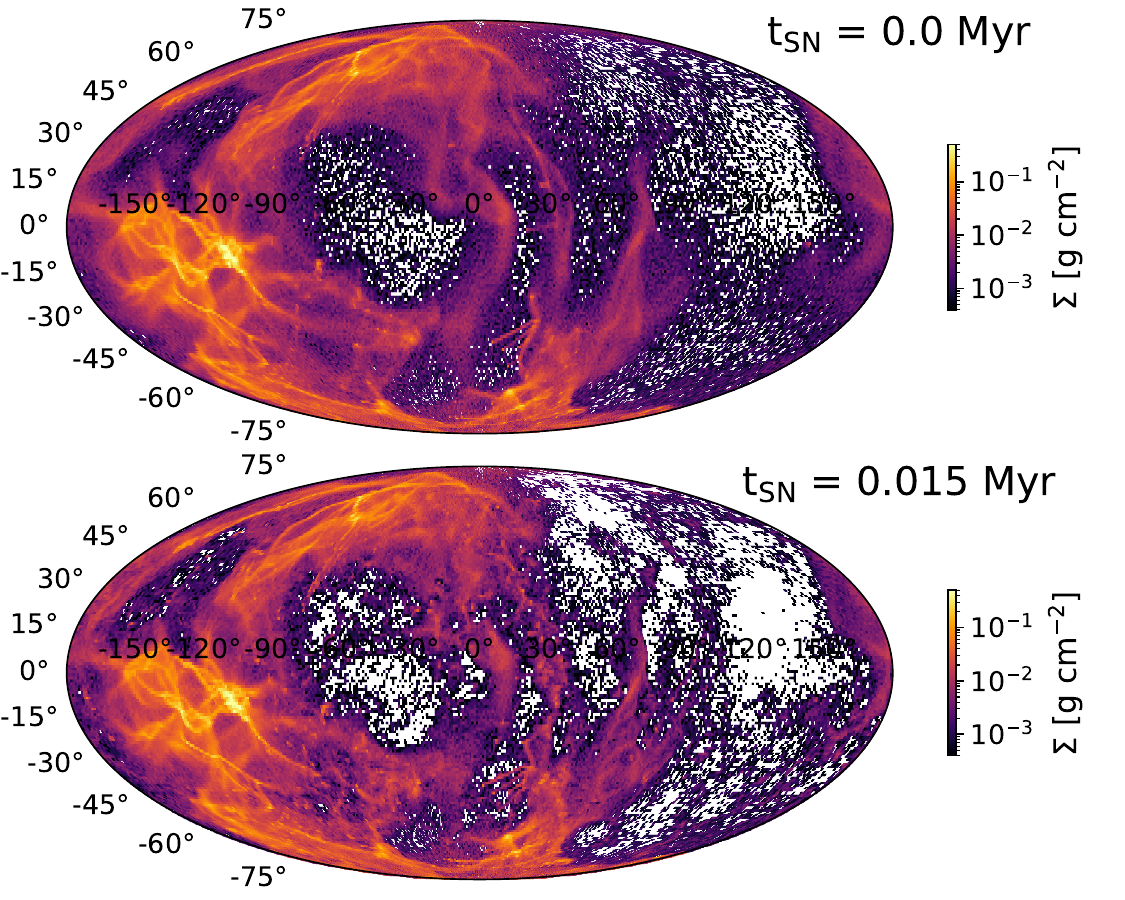}
        \subcaption{Compact environment run ($R_\mathrm{max}=4\ \mathrm{pc}$)}
        \label{fig:radsn_20_10_rho_sky_intermed}
    \end{minipage}
    \begin{minipage}{.4\textwidth}
        \centering
        \includegraphics[width=\linewidth]{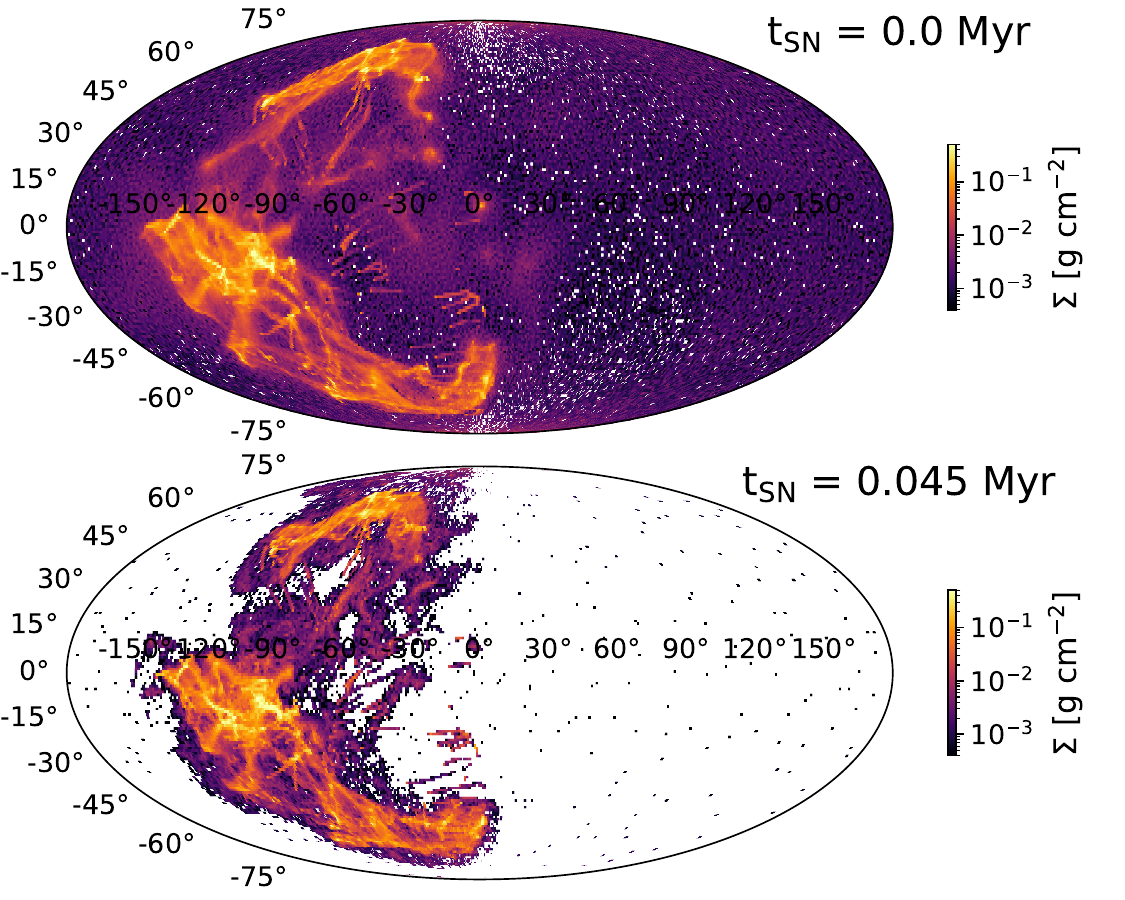}
        \subcaption{Dispersed environment run ($R_\mathrm{max}=4\ \mathrm{pc}$)}
        \label{fig:radsn_20_10_rho_sky_final}
    \end{minipage}
    \caption{Angular distribution of column density $\Sigma$ across the `sky' at 4 pc distance from the progenitor in the $10^{-20}\ \mathrm{g\ cm^{-3}}|\alpha=1.0$ GMC, for (a) the compact environment run (\textit{left}) and (b) the dispersed environment run (\textit{right}), before SN (\textit{top}) and after SN (\textit{bottom}). Spherical maps are displayed with the Mollweide projection method. Empty cells indicate the lack of SPH interpolation points but not necessarily the lack of gas. }
\end{figure*}

\begin{figure*}
    \centering
    \begin{minipage}{.4\textwidth}
        \centering
        \includegraphics[width=\linewidth]{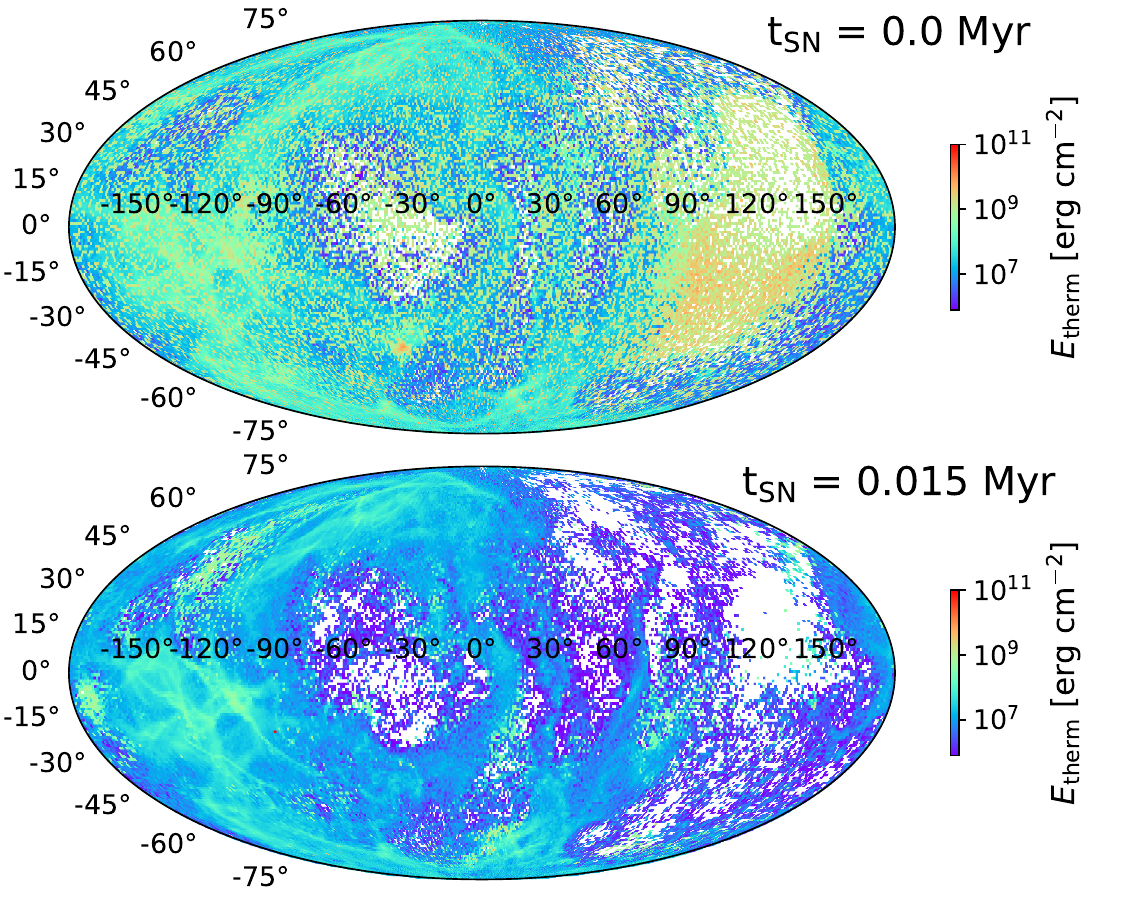}
        \subcaption{Compact environment run ($R_\mathrm{max}=4\ \mathrm{pc}$)}
        \label{fig:radsn_20_10_etherm_sky_intermed}
    \end{minipage}
    \begin{minipage}{.4\textwidth}
        \centering
        \includegraphics[width=\linewidth]{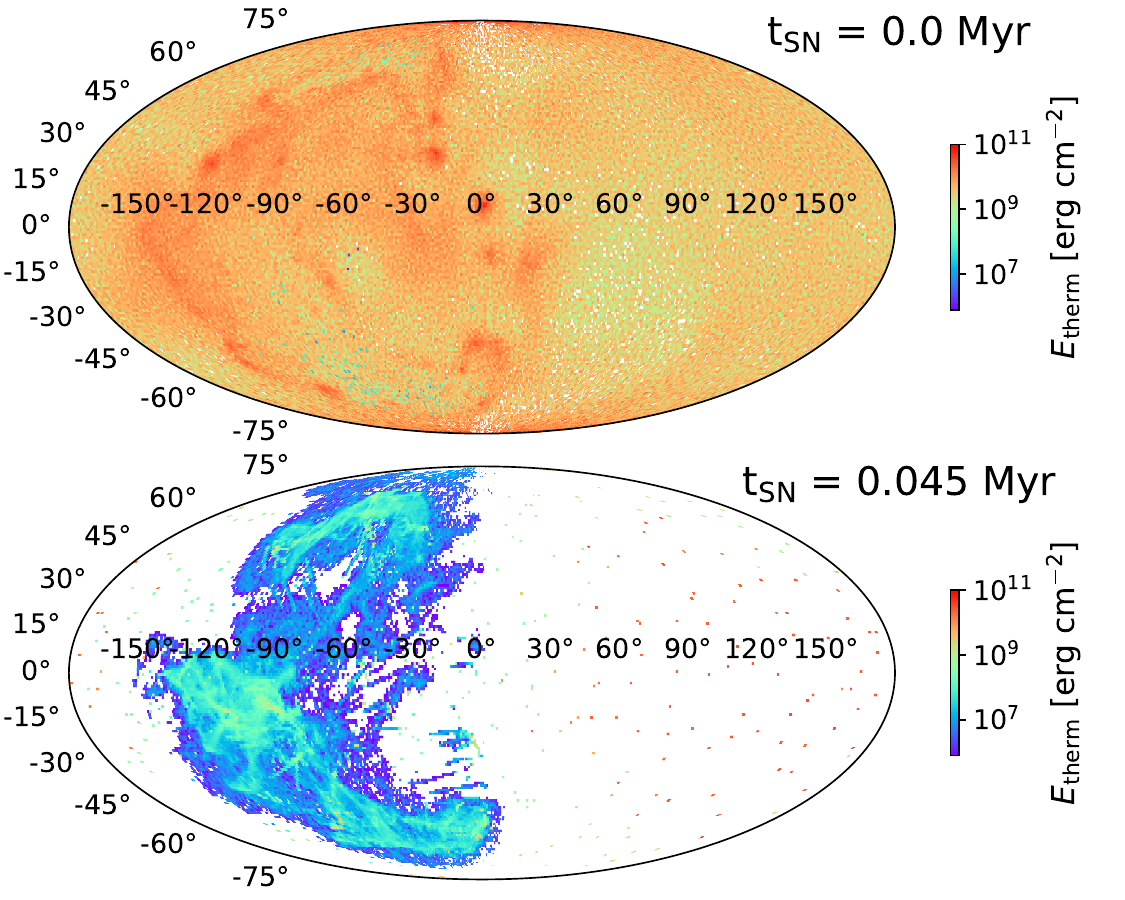}
        \subcaption{Dispersed environment run ($R_\mathrm{max}=4\ \mathrm{pc}$)}
        \label{fig:radsn_20_10_etherm_sky_final}
    \end{minipage}
    \caption{Angular distribution of column thermal energy $E_\mathrm{therm}$ in the $10^{-20}\ \mathrm{g\ cm^{-3}}|\alpha=1.0$ GMC before (\textit{top}) and after (\textit{bottom}) SN. }
\end{figure*}

\begin{figure*}
    \centering
    \begin{minipage}{.4\textwidth}
        \centering
        \includegraphics[width=\linewidth]{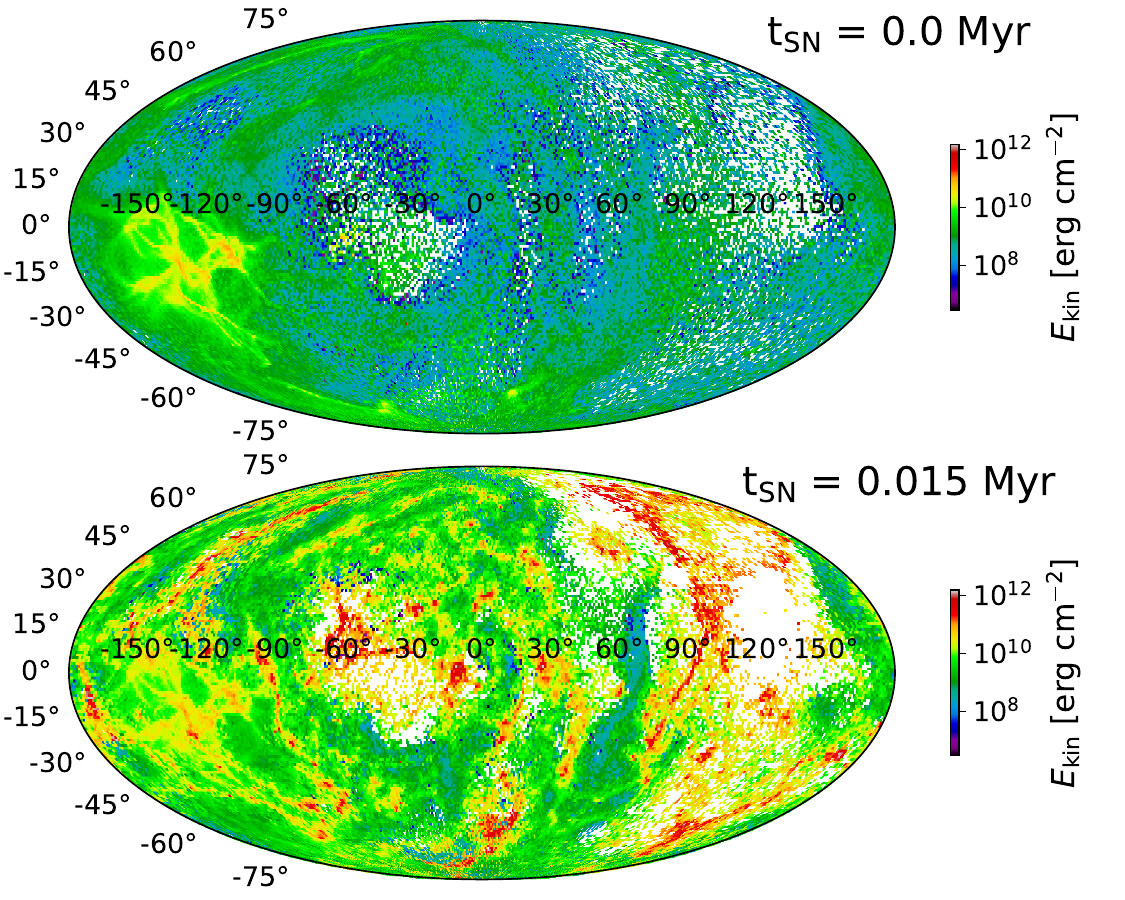}
        \subcaption{Compact environment run ($R_\mathrm{max}=4\ \mathrm{pc}$)}
        \label{fig:radsn_20_10_ekin_sky_intermed}
    \end{minipage}
    \begin{minipage}{.4\textwidth}
        \centering
        \includegraphics[width=\linewidth]{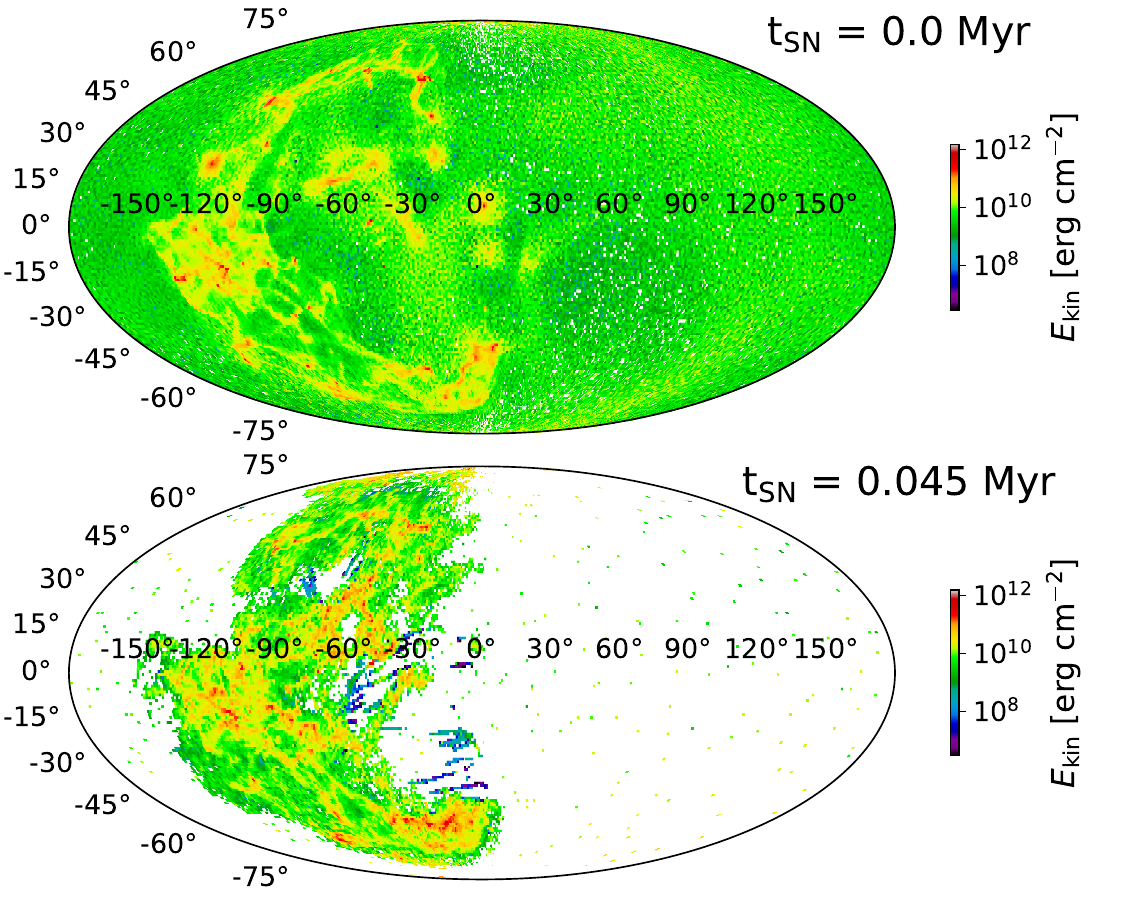}
        \subcaption{Dispersed environment run ($R_\mathrm{max}=4\ \mathrm{pc}$)}
        \label{fig:radsn_20_10_ekin_sky_final}
    \end{minipage}
    \caption{Angular distribution of column kinetic energy $E_\mathrm{kin}$ in the $10^{-20}\ \mathrm{g\ cm^{-3}}|\alpha=1.0$ GMC before (\textit{top}) and after (\textit{bottom}) SN. }
\end{figure*}

We present the angular distribution of mass, kinetic energy and thermal energy from the position of the progenitor before and after SN injection. We extract particles up to 4 pc radius from the progenitor and bin their positions by spherical angular coordinates\footnote{A caveat with this tessellation method is that the unequal volumes of each angular bin may introduce minor biases in the results, particularly if SPH smoothing is neglected. However the uncertainties are significant only when the number of particles in the bin is very small.} $(\theta, \phi)$. In each bin, we sum the particles' mass and energies, weighted by the surface area bound by the bin widths $\Delta \theta$ and $\Delta \phi$. This procedure is repeated for each of the eight SN simulations to produce 2-D sky grids. Fig.~\ref{fig:radsn_20_10_rho_sky_intermed} to Fig.~\ref{fig:radsn_20_10_ekin_sky_final} show the results for the $10^{-20}\ \mathrm{g\ cm^{-3}}|\alpha=1.0$ compact and dispersed runs, plotted with the Mollweide projection method. Note that some cells are empty because we neglected SPH interpolations; it does not necessarily imply the absence of gas.

The column density maps in Fig.~\ref{fig:radsn_20_10_rho_sky_intermed} clearly demonstrate the preferential removal of gas from regions that are already diffuse before the explosion takes place. At $t_\mathrm{SN}=0.015\ \mathrm{Myr}$, a large hole is seen centred at around $(120^\circ,20^\circ)$, with another slightly smaller hole at $(-40^\circ, 0^\circ)$. The borders of these `emptied' patches precisely trace the outer boundaries of the dense areas seen in at $t_\mathrm{SN}=0.0\ \mathrm{Myr}$. This phenomenon is observed not only around the large holes, but also for the very narrow gaps that lie in between the filamentary structures. Dense filaments with $\Sigma > 0.1\ \mathrm{g\ cm^{-2}}$ appear almost intact, and these regions are likely where star formation may continue. It illustrates the asymmetry of SN `vents' in compact GMC environments. 

Meanwhile, those with intermediate densities ($\Sigma \sim 0.01\ \mathrm{g\ cm^{-2}}$) develop small fractal structures, especially if they lie at the edge of the holes through which the SN outflows escaped. We suggest that this is due to the turbulence driven by fluid instabilities, such as the Kelvin--Helmholtz instability \citep[e.g.][]{fielding20}, which arises from hot gas streaming pass the cold channel walls. Additionally, these fractal structures appear even more clumpy when both sides of the filament (in projection) are subjected to SN shock flows. 

The internal energy map at $t_\mathrm{SN}=0.0\ \mathrm{Myr}$ in Fig.~\ref{fig:radsn_20_10_etherm_sky_intermed} shows that the large holes are exactly where the ionized gas gathered before the SN explodes. The fact that the regions exposed to photoionization coincide with those impacted by SNe suggests that both, if not all, kinds of feedback pass predominantly through the same channels seeded by the inhomogeneous density structure. Once photoionization is switched off and the SN shock front exits the cloud, the GMC rapidly cools to its initial temperatures, as seen in the $t_\mathrm{SN}=0.015\ \mathrm{Myr}$ plot. No hot remnants are kept due to the high GMC densities in this run. 

Finally, we inspect the distribution of kinetic energy in Fig.~\ref{fig:radsn_20_10_ekin_sky_intermed}. At $t_\mathrm{SN} = 0.0\ \mathrm{Myr}$, the map captures the turbulent motions within the GMC, seen as a bright patch on the left in the upper plot. Of more importance is that after SN injection, the kinetic energy from the shock is preferentially deposited at around the two largest holes identified in Fig.~\ref{fig:radsn_20_10_rho_sky_intermed}. These high-energy areas are likely tracing the postshock flows that follow the SN blowouts. In contrast, regions with high column densities are well-shielded and poorly coupled to the SN kinetic energy, which is consistent with the findings of \citet{lucas20}. 

On the other hand, the progenitor in the dispersed environment run is much more exposed to the outer ISM prior to detonation. The dense filaments in the GMC occupy only a very small area on its sky, and are thus difficult to interfere with the SNR bubble expansion in any significant way. Like in the compact runs, all dense regions survived the SN explosion, as shown in Fig.~\ref{fig:radsn_20_10_rho_sky_final}. Note that the snapshots presented here are taken at 0.045 Myr instead of 0.015 Myr, since the SN blowouts in the dispersed runs occur at a relatively later time. Perhaps the cooling at the shock-cloud interface rapidly condensed the hot gas, hence prevented the GMC from being completely destroyed. This run resembles the scenario described in e.g. \citet{kim17} as opposed to the `semi-confined' situation described in Paper~I.

\subsection{Impacted column densities vs cloud porosity} \label{sec:impacted_columnden}

\begin{figure}
    \centering
    \includegraphics[width=0.9\linewidth]{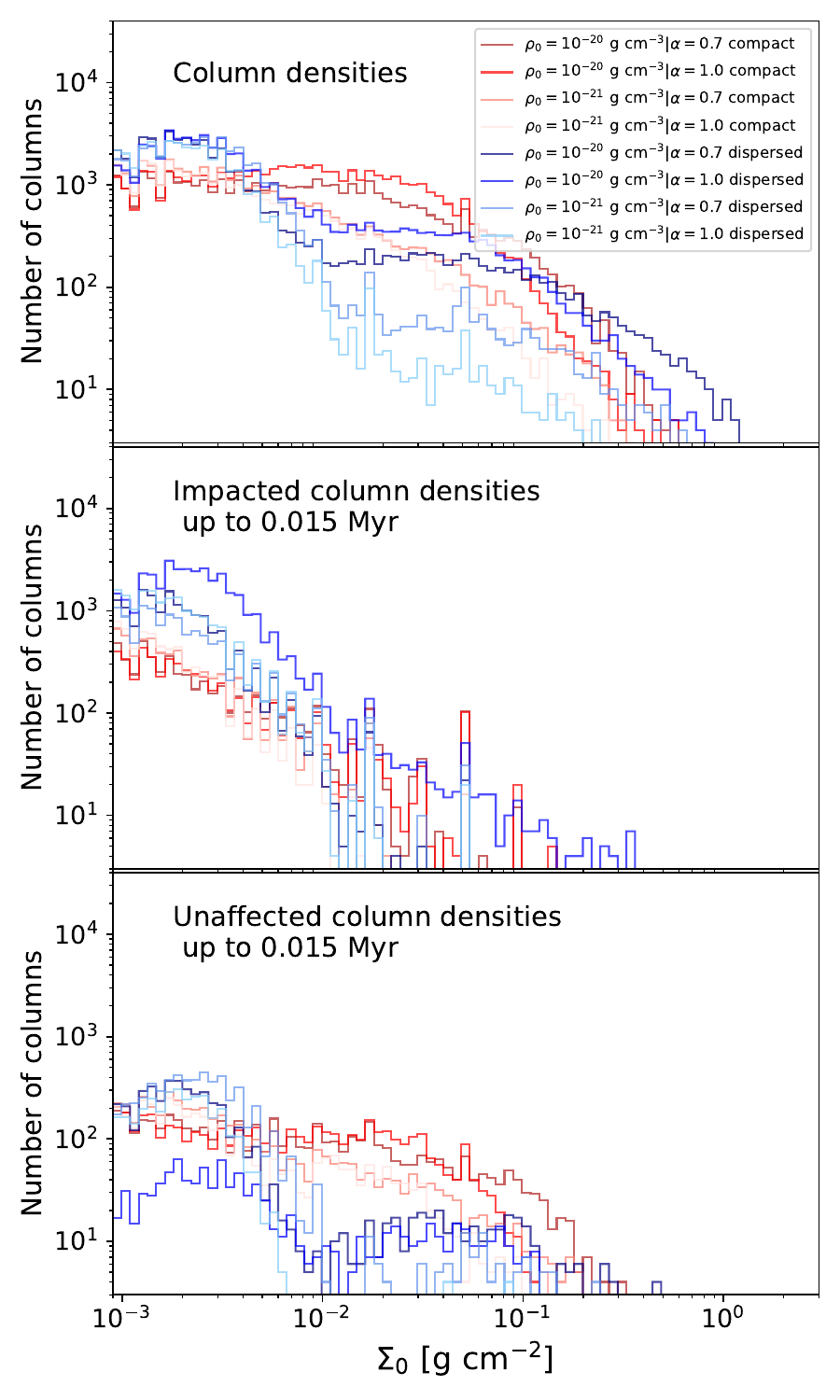}
    \caption{\textit{Top}: Distribution of column densities within 4 pc radius relative to the progenitor in the compact environment runs (red colours) and the dispersed environment runs (blue colours) before SN injection. \textit{Middle}: Distribution of column densities that were significantly impacted by SN explosion. \textit{Bottom}: Distribution of column densities that remained unaffected by $t_\mathrm{SN}=0.015\ \mathrm{Myr}$. }
    \label{fig:columnden_0015Myr}
\end{figure}

Quick observations on Fig.~\ref{fig:radsn_20_10_rho_sky_intermed} and ~\ref{fig:radsn_20_10_rho_sky_final} appear to suggest that there exist a common threshold in column density below which the gas will be dispersed by SN blowouts, independent of their size or angular coverage around the progenitor. To examine this conjecture, we extract, from our eight simulations, the column densities of angular bins that were significantly emptied (up to 4 pc) after SN, and those which had not been affected. Let $\Sigma_{0}$ denote the column density before SN and $\Sigma_{0.015}$ for that at $t_\mathrm{SN}=0.015\ \mathrm{Myr}$. We define `impacted column densities' as those with $\Sigma_{0.015} < 0.01 \Sigma_{0}$, and `unaffected column densities' as those with $0.95<\Sigma_{0.015} / \Sigma_{0}<1.05$. The results are presented in Fig.~\ref{fig:columnden_0015Myr}.

The third panel reveals that the unaffected column densities span fully across their initial range before SN injection, as shown in the first panel. The overall shape of the curves also appear similar, only with amplitudes reduced by approximately one order of magnitude, meaning both high-density gas and low-density gas have around 10\% of its initial amount almost completely unaffected by the SN. Those with high column densities may be self-shielded, whereas those with lower column densities could be (i) shielded by small clumps on the side facing the progenitor or (ii) is already diffuse enough for SN outflows to pass without causing much disruption. 

To the contrary, the distributions of impacted column densities peak strongly towards the smallest values, demonstrating the preferential removal of low-density gas as previously discussed. The maximum impacted column density is different in each run. It does not appear to be correlated with the maximum values in their original distributions, nor it correlates with the porosity of the GMC. This result suggests that there is no universal density threshold below which the gas must be dispersed, or over which the gas must be able to withstand the SN feedback. 

\begin{figure}
    \centering
    \includegraphics[width=\linewidth]{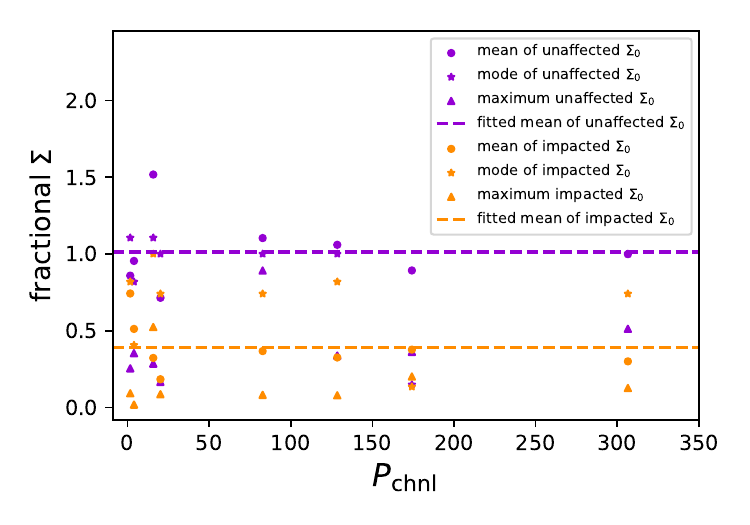}
    \caption{Fractional mean (circle), mode (star) and maximum value (triangle) of the impacted column densities (orange) and unaffected column densities (purple) relative to their original values before SN, plotted against $P_\mathrm{chnl}$ of the GMC. Horizontal dashed lines are fitted to the mean values. }
    \label{fig:pchnl_columnden}
\end{figure}

Knowing that there is no universal threshold, we proceed to investigate the relation between the impacted column densities and their initial states. For simplicity, we consider only the mean, the mode and the maximum value of the distributions in Fig.~\ref{fig:columnden_0015Myr}. We then compare these statistics of the impacted and unaffected column densities to that of the original by calculating the fractional change. Fig.~\ref{fig:pchnl_columnden} presents the results, plotted against $P_\mathrm{chnl}$ of the GMC found previously in Section~\ref{sec:channelling_param}.

Surprisingly, the fractional statistics of the impacted and unaffected column densities appear independent of the clouds' porosity. The mean of those impacted lie almost consistently at 40\% of their initial values. What it could imply is that whenever an embedded progenitor `sees' a range of column densities across its line-of-sights, its feedback would preferentially `attack' those in the lower half of the column density distribution. If so, the density threshold is relative, and varies with the local environment of the SNR. The other implication is that as long as the cloud has an inhomogeneous density structure, feedback inevitably removes a portion of the gas in regions with relatively lower densities, regardless of the GMC's overall gravitational boundness.

\subsection{Outflow velocities} \label{sec:outflow_vel}

Having demonstrated that the SNe in the compact runs indeed behave like the semi-confined SN model described in Paper~I, we now examine their outflow velocities to compare against the predictions from the analytical model. Key components considered in the model are (i) the movement of the cavity shell upon collision with the shock, and (ii) the venting of energy during the quasi-steady pressure-relief phase. The SNR was assumed to be completely thermalised within the cavity, and that the outflows are driven only by the pressure gradient across the channels. 

Here, no coherent shells are formed in the compact runs; the equations of motion considered in component (i) might not apply. From Fig.~\ref{fig:energy_momentum_insphere} it also appears that the outflows are largely driven by kinetic energy and momentum, rather than by thermal pressure alone as previously assumed. Nevertheless, we shall demonstrate below that the simulation results agree with the semi-confined model despite its idealised simplifications. 

\begin{figure*}
    \centering
    \begin{minipage}{.48\textwidth}
        \centering
        \includegraphics[width=\linewidth]{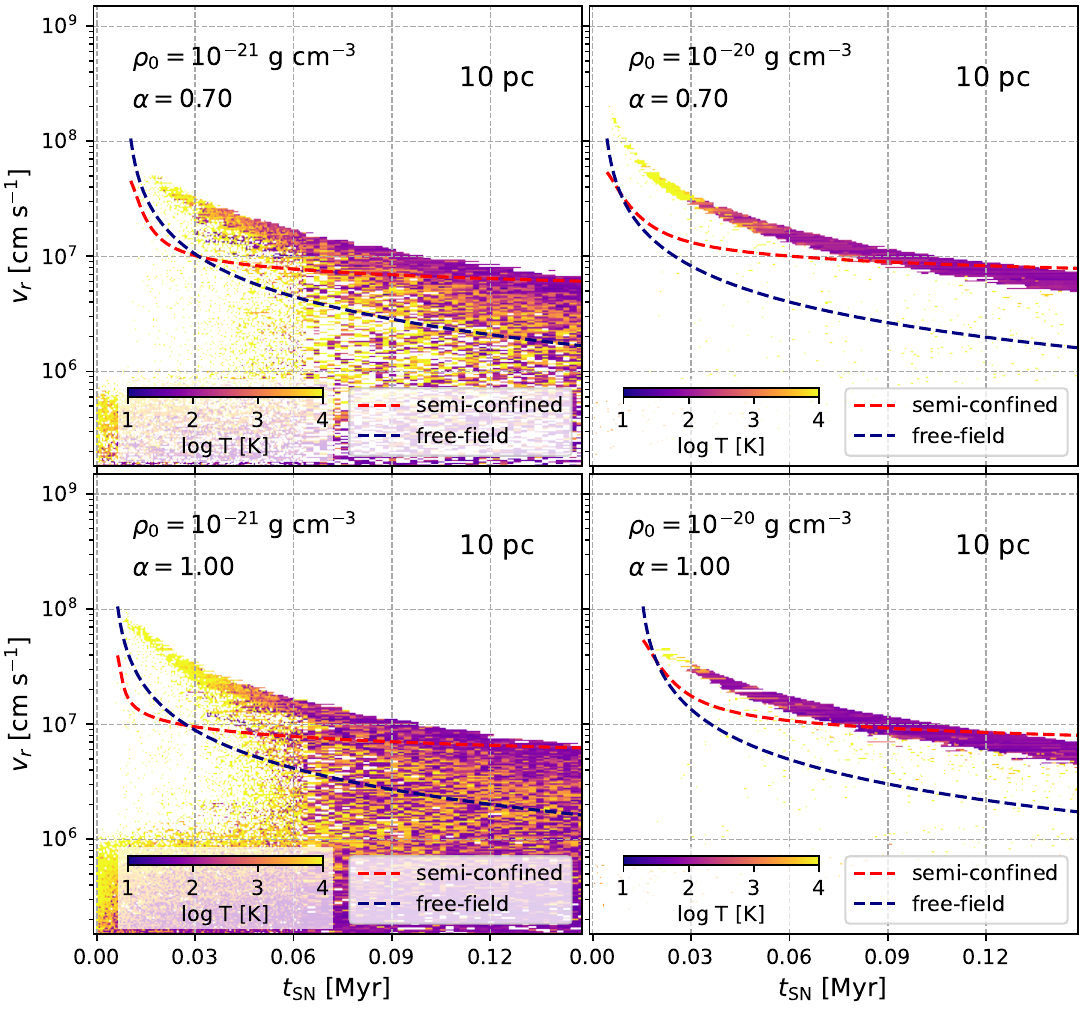}
        \subcaption{Compact environment runs ($r_\mathrm{detect}=10\ \mathrm{pc}$)}
        \label{fig:outflow_vel_10pc_intermed}
    \end{minipage}
    \begin{minipage}{.48\textwidth}
        \centering
        \includegraphics[width=\linewidth]{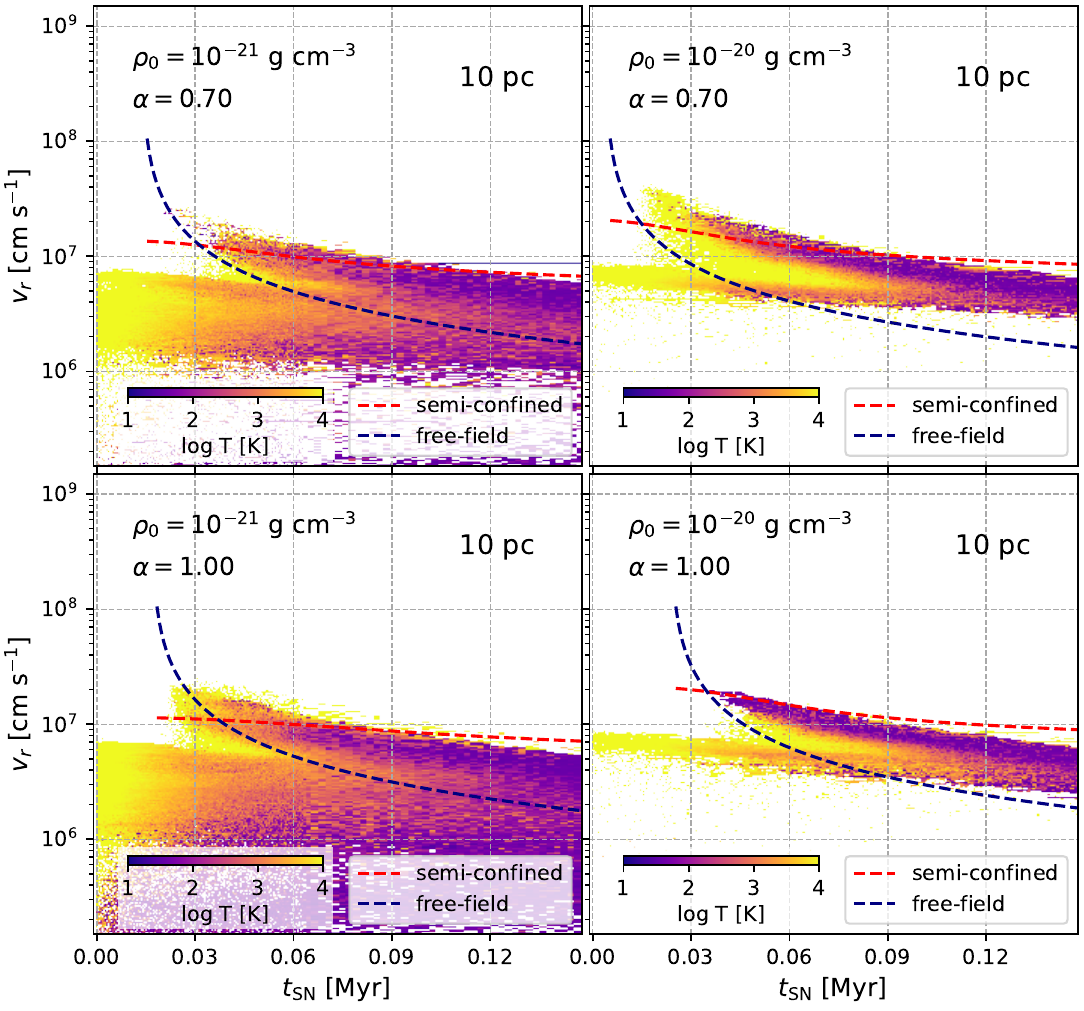}
        \subcaption{Dispersed environment runs ($r_\mathrm{detect}=10\ \mathrm{pc}$)}
        \label{fig:outflow_vel_10pc_final}
    \end{minipage}
    \caption{Radial velocity of outflows measured at 10 pc from the progenitor after injecting SN into (a) the highly bound GMCs in Fig.~\ref{fig:chnl_ionization_intermed} with narrow ionized channels (\textit{left}: compact environment runs), and into (b) the more evolved GMCs in Fig.~\ref{fig:chnl_ionization_final} with larger cavities (\textit{right}: dispersed environment runs). Colours indicate the average temperature of particles within the velocity bin. Analytically-derived velocities computed with the semi-confined SN and the free-field SN models from Paper~I are shown in red and blue dashed lines. }
\end{figure*}

Fig.~\ref{fig:outflow_vel_10pc_intermed} and \ref{fig:outflow_vel_10pc_final} plot the outflow velocities measured at 10 pc radius in the compact runs and the dispersed runs, with colours indicating their temperatures. These plots are produced by, first, extracting all particles within a distance of $10 \pm 1\ \mathrm{pc}$ relative to the progenitor. We then bin them by radial velocity and compute the average temperature of particles in each velocity bin. This step is repeated for each snapshot throughout the SNR evolution. 

Velocities computed with the semi-confined and free-field analytical models are plotted on top for comparison. Physical parameters involved in the calculation, including the radius and density of the cloud, cavity and shell, are estimated using Fig.~\ref{fig:chnl_density_intermed} to Fig.~\ref{fig:chnl_ionization_final}. Channel size is kept constant in all models, but the influence should be negligible according to the findings from Paper~I. We assume a uniform density of $4 \times 10^{-25}\ \mathrm{g\ cm^{-3}}$ in the free-field model, though the choice of density has a minimal effect on the curve. Since the models define time-zero to be the point where blowouts begin, we manually shift the curves horizontally to match where the high-velocity outflows first appear on the plot.

Initially, the outflow velocities are often bimodally distributed, particularly in the $10^{-21}\ \mathrm{g\ cm^{-3}}$ runs. The distinctive high-velocity component corresponds to the SN blowouts, whereas the broad low-velocity component likely originates from the gas near the cloud edge, which diffused to larger distances after receiving the thermal energies from the SNe. 

Soon after, the distributions merge due to the mixing of the plumes, but we can still distinguish the individual components from their temperatures. The high-velocity SN blowouts from channels appear as cold narrow stripes; their rapid cooling could be a result of the Joule-Thomson effect. Warm gas with slightly lower velocities likely corresponds to the bulk outflows from outer regions in the GMC where cooling (or shielding, in reality) is less efficient. Finally, the slow cold gas is probably the molecular gas that just got unbound from the GMC's potential. 

Since the outflows from channels are of sole interest in this study, we consider only the evolution of the maximum velocity, which reflects the maximal perturbation imposed by the SNe. Quick comparisons between Fig.~\ref{fig:outflow_vel_10pc_intermed} and \ref{fig:outflow_vel_10pc_final} reveal that the evolution of the maximum velocity, after approximately 0.06 Myr, are very similar to each other. This is true for all runs in spite of their differences in the initial blowout velocity, shock arrival times, and the breadth of the velocity distribution underneath the curve. Not only that these results support the semi-confined model, they also suggest that the late-time behaviour of SN outflows are rather independent of the porosity of its parent GMC. 

\begin{figure*}
    \centering
    \begin{minipage}{.48\textwidth}
        \centering
        \includegraphics[width=\linewidth]{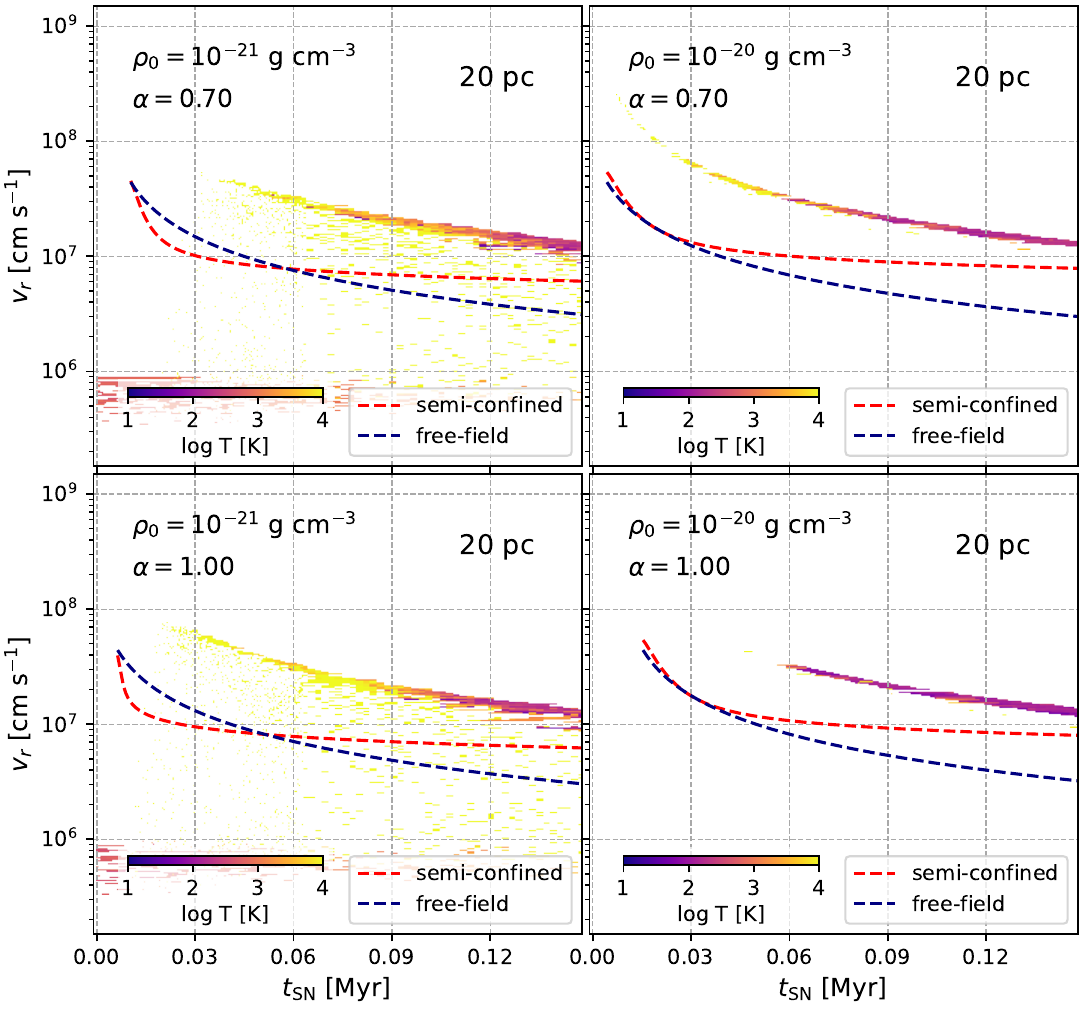}
        \subcaption{Compact environment runs ($r_\mathrm{detect}=20\ \mathrm{pc}$)}
        \label{fig:outflow_vel_20pc_intermed}
    \end{minipage}
    \begin{minipage}{.48\textwidth}
        \centering
        \includegraphics[width=\linewidth]{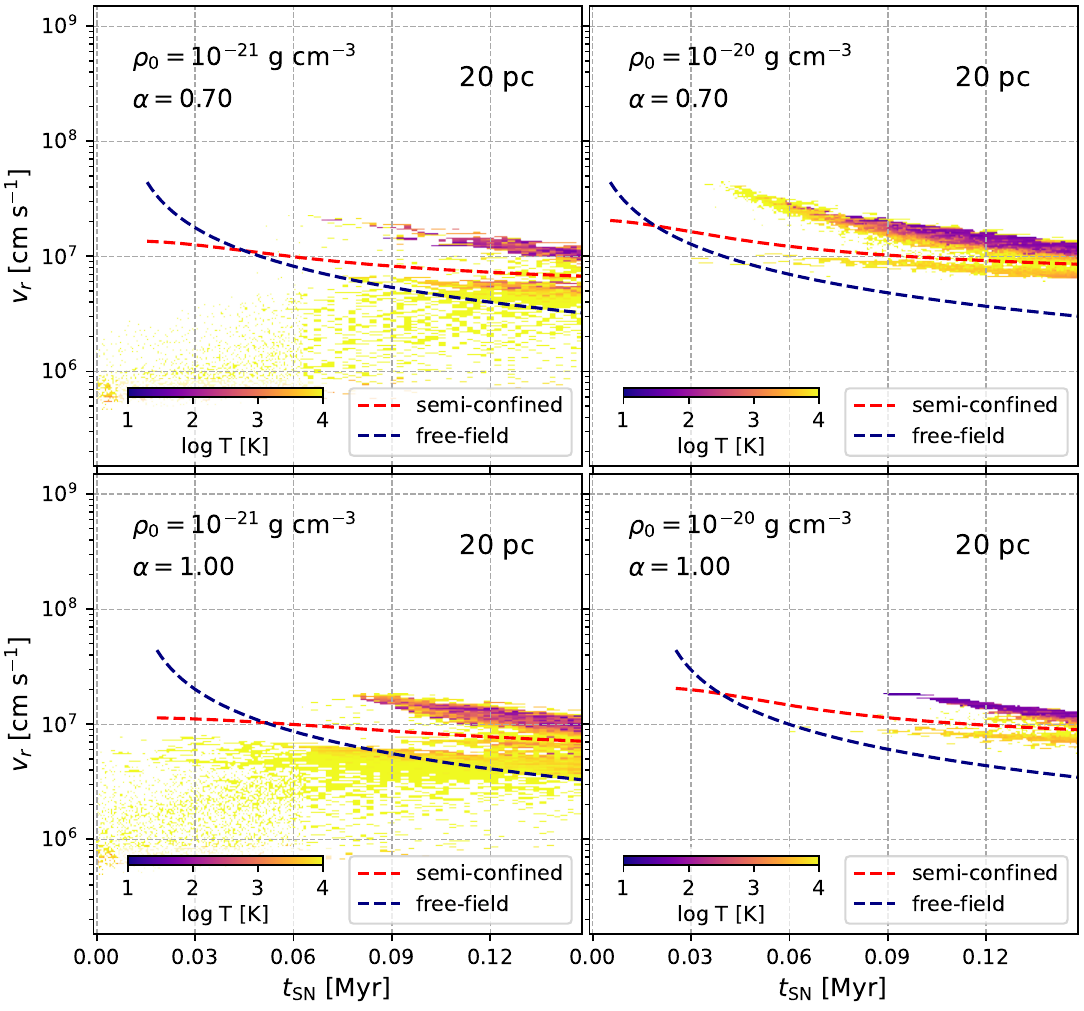}
        \subcaption{Dispersed environment runs ($r_\mathrm{detect}=20\ \mathrm{pc}$)}
        \label{fig:outflow_vel_20pc_final}
    \end{minipage}
    \caption{Same as Fig.~\ref{fig:outflow_vel_10pc_intermed} and Fig.~\ref{fig:outflow_vel_10pc_final} but measured at 20 pc away from the progenitor.  }
\end{figure*}

Better agreement with the model is found when we move the `detector' to 20 pc away from the progenitor in the simulation. The results are presented in Fig.~\ref{fig:outflow_vel_20pc_intermed} and \ref{fig:outflow_vel_20pc_final}. 20 pc is near the boundaries of the initially-placed envelope, but the gas expands beyond this distance once they are heated by feedback. The low-velocity component (cloud gas) is no longer observed. We also see that the high-velocity outflows tend to arrive earlier in the compact GMCs, revealing that SN energy escaping as plumes from narrow channels can stream rapidly to larger distances. Nevertheless, the more important result here is that the maximum velocities at later stages are alike in all runs. Significant differences are seen only during the initial blowout, or at least before 0.1 Myr.

\begin{figure}
    \centering
    \includegraphics[width=0.9\linewidth]{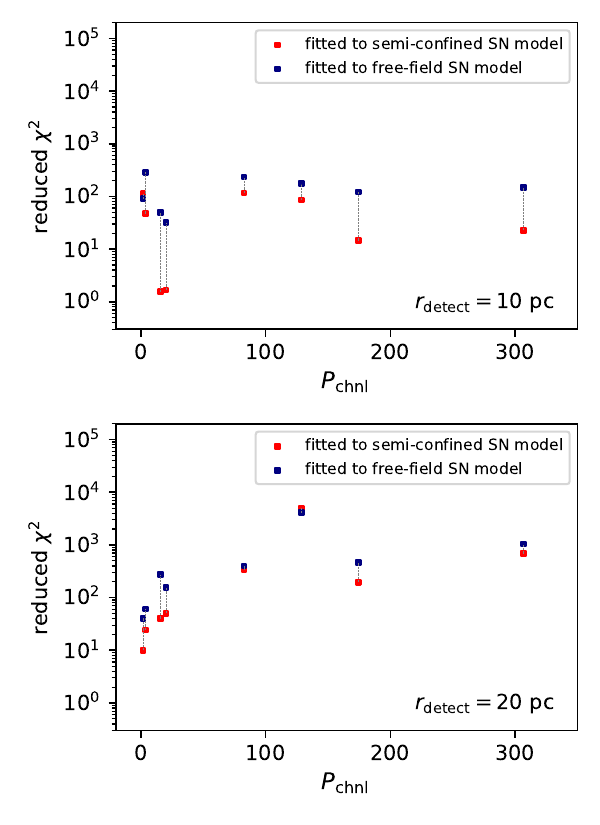}
    \caption{\textit{Top}: Reduced $\chi^2$ values from fitting the maximum outflow velocities, at 10 pc distance, to the semi-confined SN model (red) and to the free-field SN model (blue) from Paper~I, plotted against $P_\mathrm{chnl}$ of the GMC. \textit{Bottom}: Same but measured at 20 pc from the progenitor. }
    \label{fig:chi2_pchnl_outflow}
\end{figure}

Fig.~\ref{fig:chi2_pchnl_outflow} explicitly evaluates whether the semi-confined model or the free-field model provides a better fit to the maximum velocities. We compute reduced $\chi^2$ values for each run to compare the simulation results against their analytic calculations. Uncertainties are estimated as the standard deviation in velocity of particles with $T < 10^3\ \mathrm{K}$, with the temperature filter being imposed to isolate the channel outflows. 

What we find is that the $\chi^2$ values fitted to the semi-confined models are always lower than those fitted to the free-field models, suggesting that SNe in inhomogeneous environments may be better described as partially-contained explosions, as opposed to being spherical blasts. The fact that the $\chi^2$ values remain approximately constant across different $P_\mathrm{chnl}$ also suggests that the kinematic properties of SNe outflows are, in fact, fairly insensitive to the local environment of the progenitor.

\subsection{Turbulence} \label{sec:turbulence}

If the outflow velocities are similar in all runs, what about the induced turbulence? Previously in Paper~I, we found, with decomposed kinetic energy power spectra, that semi-confined SNe can sustain stronger solenoidal turbulence in the post-shock regions compared to spherical blasts in a uniform medium. This is particularly the case at the smaller length-scales, where vortices develop when the shocks collide with inhomogeneous structures \citep[e.g.][]{dobbsbonnell08,padoan16}. It could have also been produced by the Rayleigh--Taylor fingers that protrude into the cold cavity walls upon collision with the shocks \citep[e.g.][]{blondinellison01}.

\begin{figure}
    \vspace{0.1in}
    \centering
    \includegraphics[width=0.85\linewidth]{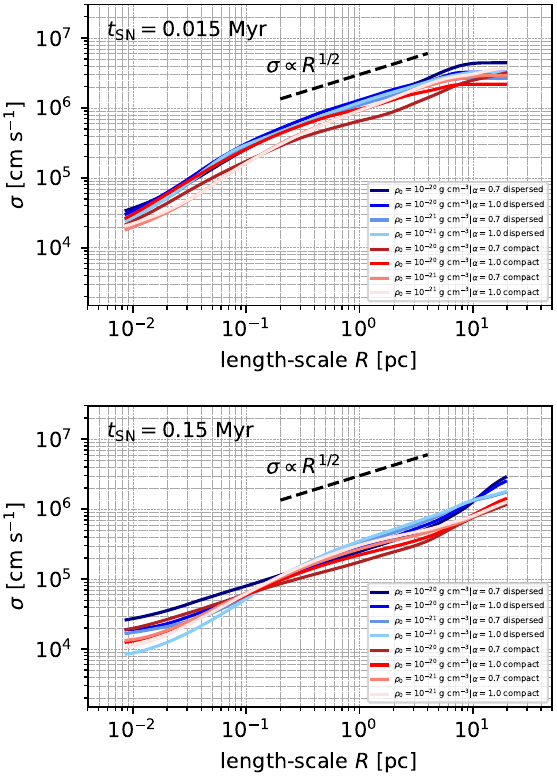}
    \caption{\textit{Top}: Velocity dispersion against length-scale $R$ at $t_\mathrm{SN} = 0.015\ \mathrm{Myr}$ in the compact environment runs (red colours) and the dispersed environment runs (blue colours). \textit{Bottom}: Velocities dispersions at $t_\mathrm{SN} = 0.15\ \mathrm{Myr}$. Black dashed lines show the Larson relation $\sigma \propto R^{1/2}$.}
    \label{fig:sne_vel_dispersion}
\end{figure}

In this work, all GMCs are turbulent, and all SNe collide with a clumpy medium. Perhaps the only differentiating factor which would influence the turbulent properties is the size of the channels, which likely influences the strength of the Kelvin--Helmholtz instabilities as the hot outflows stream pass them. To investigate whether or not GMC porosity has an effect on the feedback-induced eddies, we plot the velocity dispersions in the eight simulations, at $t_\mathrm{SN} = 0.015\ \mathrm{Myr}$ and at $t_\mathrm{SN} = 0.15\ \mathrm{Myr}$. For each particle, we pre-define a set of radii ranging from 0.01 pc to 30 pc, and measure the 3-D velocity (magnitude) dispersion within them. Fig.~\ref{fig:sne_vel_dispersion} plots the dispersion at each radius averaged over all particles. The Larson relation, $\sigma \propto R^{1/2}$ \citep{larson81}, is shown for reference.

It is apparent that their overall gradients at size-scales of $0.1 - 10\ \mathrm{pc}$ remain in agreement with the Larson relation, owing to the power spectrum that underpins the turbulent velocities imposed during setup. The linear correlation is slightly weakened beneath 0.1 pc and above 20 pc. The latter is due to finite GMC sizes, whereas the former likely indicates that dissipation becomes non-negligible at scales below 0.1 pc in our simulation. At 0.15 Myr, their amplitudes systematically shifted downwards as the remnants cool. 

An important fact revealed in Fig.~\ref{fig:sne_vel_dispersion} is that the velocity dispersions in the compact runs and in the dispersed runs remain almost indistinguishable from each other for most of the time in the SNR's evolution. In general, the dispersed runs tend to produce slightly higher dispersions on scales beyond 1 pc, however this is likely only because the dispersed GMCs have shallower potentials, which in turned allowed the gas to migrate to larger distances. On scales below 0.5 pc, the overlap is even stronger. The SNe in the compact environments have not given rise to any unique small-scale turbulent features that are not observed in the dispersed runs. 

One possible implication from the results above, along with those in Section~\ref{sec:outflow_vel}, is that the kinematic perturbations induced by a single SN explosion is rather independent of the host cloud's density structure as long as the medium is not homogeneous. This is perhaps analogous to the results reported in \citet{walch12}, who found that the global parameters of the outflows driven by ionizing radiation do not strongly depend on the cloud's fractal dimension, even with different H {\scshape ii} region shell structures.

\subsection{Energy and momentum} \label{sec:energy_momentum}

\begin{figure*}
    \centering
    \includegraphics[width=0.9\linewidth]{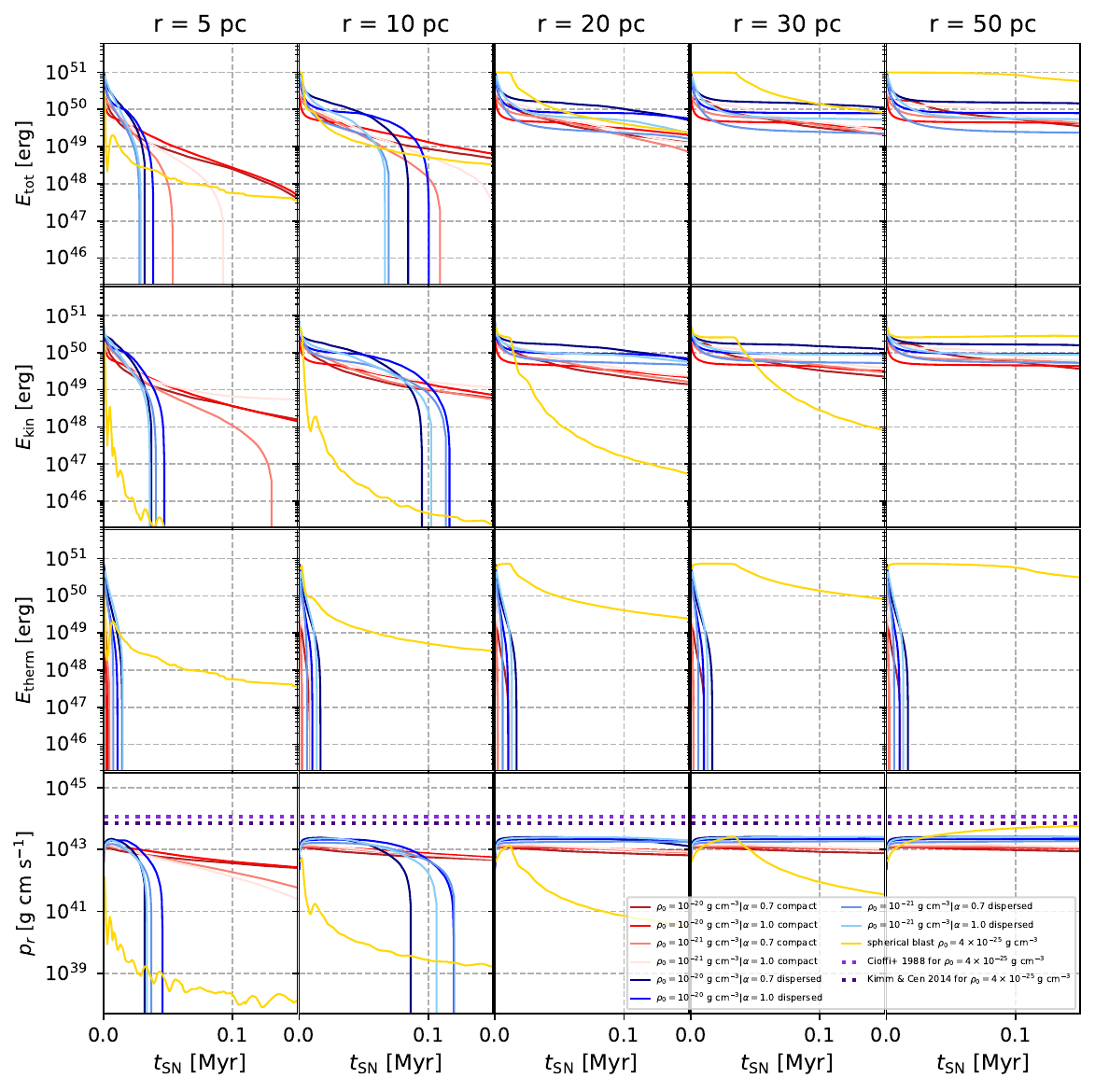}
    \caption{Evolution of total energy (\textit{first row}), kinetic energy (\textit{second row}), thermal energy (\textit{third row}) and radial momentum (\textit{fourth row}) of SNR within 5 pc radius to within 50 pc radius (\textit{left to right}) in the compact environment runs (red colours) and in the dispersed environment runs (blue colours). Results from a spherical SN in uniform medium are shown in yellow. Analytical solutions for terminal momentum from \citet{cioffi88} and \citet{kimmcen14} are plotted in grey and black dotted lines respectively.}
    \label{fig:energy_momentum_insphere}
    \vspace{0.3cm}

    \includegraphics[width=0.9\linewidth]{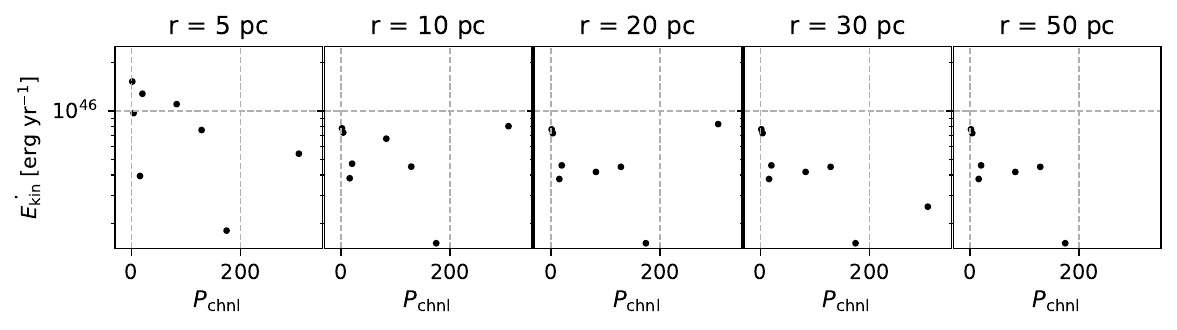}
    \caption{Instantaneous rate of kinetic energy loss at $t_\mathrm{SN} = 0.01\ \mathrm{Myr}$ measured from the results in Fig.~\ref{fig:energy_momentum_insphere}, plotted against the starting snapshot's $P_\mathrm{chnl}$.  }
    \label{fig:dEdt_pchnl}
\end{figure*}

We turn to investigate whether or not the energy released from the SNe in different environments are also hardly distinguishable like the velocities. We plot in Fig.~\ref{fig:energy_momentum_insphere} the kinetic energy, thermal energy (or internal energy) and the radial momentum enclosed within a range of threshold radii. It covers gas particles from both the GMC and the envelope, but excludes the energy and momentum carried in the sinks. We also show the results from a spherical blast simulation, with ambient medium density $\rho_0 = 4 \times 10^{-25}\ \mathrm{g\ cm^{-3}}$, which corresponds to that of the warm envelope. 

The threshold radii correspond to the typical peak resolutions achieved in current cosmological simulations. Here, 5 pc covers the vast majority of the dense filamentary parts in the GMC in both the $10^{-20}\ \mathrm{g\ cm^{-3}}$ runs and the $10^{-21}\ \mathrm{g\ cm^{-3}}$ runs (see Fig.~\ref{fig:chnl_ionization_intermed}, \ref{fig:chnl_ionization_final}). 10 pc covers the majority of the cloud gas prior to SN explosion (see Fig.~\ref{fig:chnl_ionization_intermed}, \ref{fig:chnl_ionization_final}). 20 pc encloses the maximum reach of the SN outflows. 30 pc and 50 pc are beyond the initial radius of the warm envelope, but some gas may be driven to such large distances by the SN blowouts. We plot results up to 50 pc mostly for the spherical blast case, wherein the shock front stalls at around 60 pc after undergoing rapid expansion. 

The measured energy and momentum in each radius are then subtracted by their initial amounts to isolate those injected by the SN. Total energy, as shown in the first row of Fig.~\ref{fig:energy_momentum_insphere}, is the sum of kinetic energy and thermal energy; their initial values are $10^{51}\ \mathrm{erg}$ as required. The SNe were injected in pure kinetic form, but the shocks immediately dissipate half of their energy at the beginning of the simulations, causing the initial kinetic energies to appear lower.

\begin{figure*}
    \centering
    \includegraphics[width=0.9\linewidth]{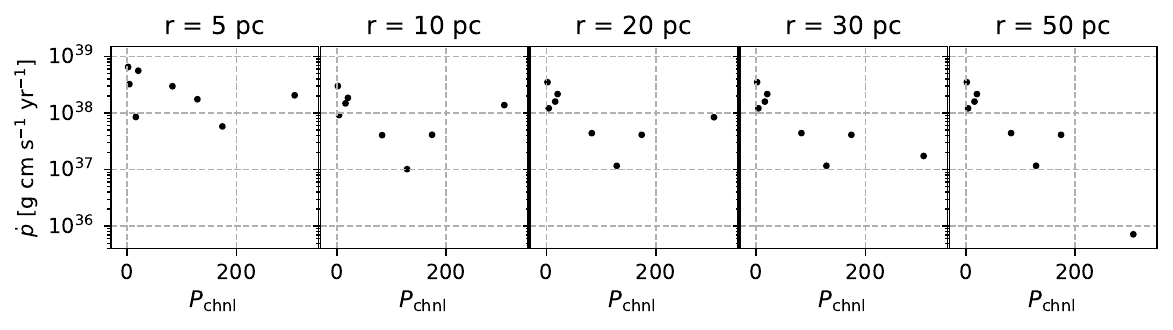}
    \caption{Instantaneous rate of change in momentum at $t_\mathrm{SN} = 0.01\ \mathrm{Myr}$ measured from the results in Fig.~\ref{fig:energy_momentum_insphere}, plotted against the starting snapshot's $P_\mathrm{chnl}$.  }
    \label{fig:dpdt_pchnl}
\end{figure*}

\subsubsection{Comparing the GMC runs against the spherical blast} \label{sec:GMCsSN_vs_sphericalSN}

In comparison to the results from Paper~I, the SNR in this work have undergone rapid radiative cooling, owing to the fact that the cavities here are much denser (see Fig.~\ref{fig:chnl_ionization_intermed},\ref{fig:chnl_ionization_final}) than the $10^{-23} \ \mathrm{g\ cm^{-3}}$ bubble adopted in our previous idealised models\footnote{Cavities in reality would have their densities further reduced by stellar winds, jets and radiation pressure, all of which were neglected in our study.}. In addition, the shock mixes with the turbulent cavity walls upon collision, which further removes heat from the SNR interior. Hence the thermal energy curves sharply decline within 0.01 Myr after injection, and most gas have cooled off within 5 pc before the outflows escape the parent cloud.  

Nevertheless, the rapid cooling of the SNR have not completely hampered its coupling with the external ISM. We found that the SNe that detonated within GMCs, in both the compact runs and the dispersed runs, retain significantly more kinetic energy at smaller radii compared to the spherical blast. The majority of their momentum is also deposited at regions closer to the progenitor, within around 10 pc, as opposed to the spherical blast where momentum is primarily deposited at 50 pc away. 

This result agrees with the findings from Paper~I, that the radial distributions of energy and momentum delivered by partially-confined SNR in GMCs largely differ to those in a uniform medium. But what we further showed here is that this is also true for the more dispersed GMCs that have been substantially disrupted by pre-SN feedback. In any case, using spherical blast models may significantly overestimate the true impact radii of SNe. 

Another aspect to note is that, on larger scales, the overall amount of kinetic energy and momentum injected by spherical blasts are in fact slightly higher than those in the GMCs. At 50 pc, the radial momentum carried in the spherical SN correctly reaches the terminal momentum predicted by \citet{kimmcen14}, with the ambient medium density term set to $4 \times 10^{-25}\ \mathrm{g\ cm^{-3}}$. Contrary to the results from Paper~I, none of the GMC runs in here reached this terminal momentum\footnote{However, the ambient medium in the GMC runs here are much more blended with the cloud gas which has likely increased the overall density, in which case the terminal momentum is expected to become smaller.}, let alone exceeding it. Similar trends are seen in the kinetic energies.

\subsubsection{Comparing the compact runs against the dispersed runs} \label{sec:compSN_vs_dispSN}

We see in the previous section how SNe in GMCs differ to those in a uniform medium, but in what ways do a SN in a compact environment behave differently to those in a dispersed environment? From Fig.~\ref{fig:energy_momentum_insphere} it is apparent that their differences are distinctive only at scales below 20 pc. The compact runs deposit more energy and momentum within 5 pc, whereas the dispersed runs deposit more at around $10-20$ pc. It may be interpreted as that the majority of the energy and momentum is transferred when the swept-up materials encounter the cavity walls. 

Bearing in mind that many cosmological simulations resolve only a few tens of parsecs at best, the 50-pc measurements here are perhaps more informative to sub-grid modelling. An important result from Fig.~\ref{fig:energy_momentum_insphere} is that the energies and momentum of the compact runs (red) and the dispersed runs (blue) gradually become indistinguishable beyond 30 pc. At 50 pc, their variations are less than an order of magnitude, consistent with the findings of \citet{iffrighennebelle15}. It implies that the overall kinetic energies and momentum input from a SN are not highly sensitive to the cloud-scale density structures. We can conclude that the analytical SN terminal momentum, which has been heavily relied on in past studies, remains a good approximate. 

Nonetheless, minor differences are still observed. The dispersed runs retain an overall higher amount of kinetic energy than those in compact environments, especially towards the later stages. Such differences are more apparently seen in the radial momentum plots. We may attribute this to the cavity density, that momentum coupling has a weak but noticeable dependence on the progenitors' local ambient medium density. Indeed, such dependency is also reflected in the terminal momentum equations \citep[e.g.][]{cioffi88}. 

Knowing that there exist slight variations across the GMC runs, we turn to inspect whether or not the speed of energy transfer to the ISM relates to the cloud's porosity. Fig.~\ref{fig:dEdt_pchnl} plots the rate of kinetic energy loss at $t_\mathrm{SN} = 0.01\ \mathrm{Myr}$ (approximately where blowout velocities peak) against $P_\mathrm{chnl}$ for each radius considered in Fig.~\ref{fig:energy_momentum_insphere}. We found an overall negative correlation with $P_\mathrm{chnl}$ at the 5-pc scale, but the rest show very weak to no correlations. The same analysis was repeated for radial momentum in Fig.~\ref{fig:dpdt_pchnl}. This time we found that the rate of momentum deposition exhibits much stronger negative trends with $P_\mathrm{chnl}$ at all scales. 

Such discrepancies may originate from the differences in mass of the ejected cloud gas. In Section~\ref{sec:outflow_vel} we demonstrated that SN outflow velocities have low sensitivity to the host cloud's porosity. The fact that kinetic energy is more dominated by the velocity term explains why they show weaker correlations with $P_\mathrm{chnl}$ than in momentum. What differentiates them must be the mass term.

\subsection{Mass removal} \label{sec:mass_removal}

\begin{figure*}
    \centering
    \includegraphics[width=0.9\linewidth]{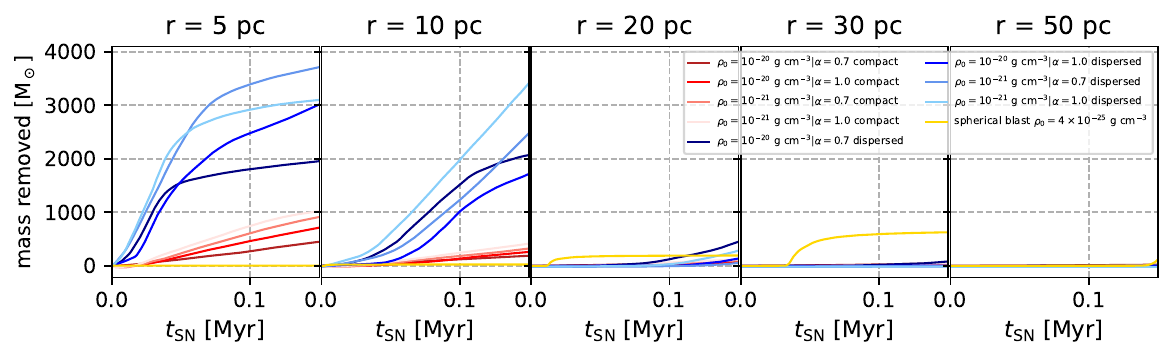}
    \caption{Cumulative mass removed from within 5 pc radius to within 50 pc radius (\textit{left to right}) in the compact environment runs (red colours) and the dispersed environment runs (blue colours). Results from the spherical blast run are shown in yellow. }
    \label{fig:mass_insphere}
    \vspace{0.3cm}
    
    \includegraphics[width=0.9\linewidth]{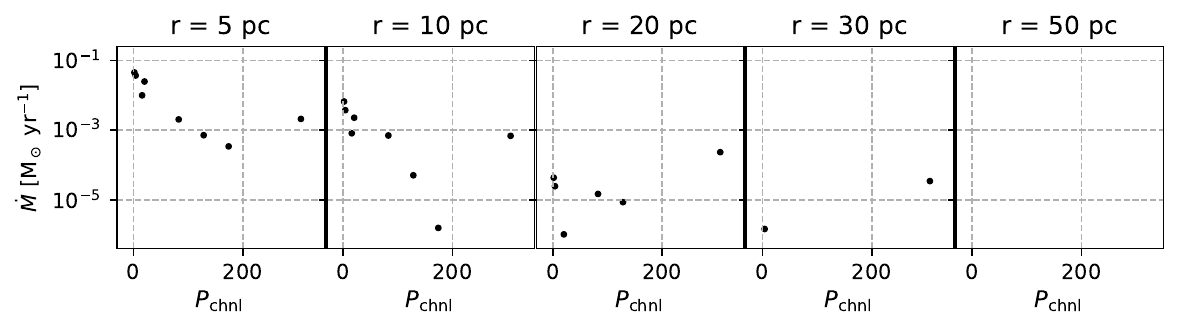}
    \caption{Instantaneous rate of mass loss at $t_\mathrm{SN} = 0.01\ \mathrm{Myr}$ measured from the results in Fig.~\ref{fig:mass_insphere}, plotted against the starting snapshot's $P_\mathrm{chnl}$.  }
    \label{fig:dmdt_pchnl}
\end{figure*}

As such, we investigate how the removal of materials by SN feedback associates with the GMCs' properties. Fig.~\ref{fig:mass_insphere} shows the cumulative amount of mass removed over time from within each radius in Fig.~\ref{fig:energy_momentum_insphere}. The differences between the compact runs and the dispersed runs now indeed become much more prominent. The compact runs retain a significant portion of their mass within $5-10$ pc. Mass removal from the GMCs occur at a slow but steady rate, with less than $10^3\ \mathrm{M_\odot}$ lost by 0.015 Myr. Almost no particles reach 20 pc, but this is likewise the case for the dispersed runs. However, the dispersed runs show much higher mass removal rates at $5-10$ pc, indicating that their materials are being driven relatively further away.

This result is consistent with the findings from Fig.~\ref{fig:energy_momentum_insphere}, that the maximum reach of the materials carried in the SN blowouts correlate with their cavity sizes. As before, we inspect the relation between mass loss rates and the GMCs' $P_\mathrm{chnl}$ in Fig.~\ref{fig:dmdt_pchnl}. Strong negative correlations are seen on scales below 10 pc, evidencing that local mass loss is suppressed when $P_\mathrm{chnl}$ is large, that is, when distinctive channels are present in the density structure. Similar results have been reported in \citet{lucas20}, that the amount of mass removed from the GMCs can be reduced when the channels are well-formed (e.g. sculpted by both ionizing radiation and winds). They attributed it to the preferential momentum transfer onto low-density gas.

But in fact, there is a simpler explanation. The way we choose to define $P_\mathrm{chnl}$, with ratios of ionized to neutral mass inside and outside the cavity, happened to describe the GMC's gravitational boundness. This correlation is no incident. H {\scshape ii} regions in more compact environments naturally give rise to distinctive tube-like channels along the narrow gaps between the shielded dense clumps. A highly bound cloud also implies a deeper potential, which makes it harder for its gas to `climb out' and escape to the wider galactic environment. Thus, that $P_\mathrm{chnl}$ and mass loss rates appear correlated may be explained by their common association with the clouds' compactness, rather than being a direct consequence of the channel properties. 

In either case, the amount of mass driven to the external medium is most sensitive to the cloud porosity, meaning that GMC boundness may influence the spatial (and perhaps temporal) fluctuation of metal distribution in the ISM. Though our simulations do not follow chemical evolution, Type II SNe ejecta in reality carry a significant amount of $\alpha$ element isotopes such as $^{16}\mathrm{O}$, $^{24}\mathrm{Mg}$ and $^{32}\mathrm{S}$ \citep[e.g.][]{nakamura99,zhang25}, which are distributed to the galaxy via SNe. If only a few hundred solar masses of materials leave the cloud, and are transported to distances less than 10 pc from their injection sites, an immediate consequence is the production of poorly-mixed ISM patches, noting that turbulent mixing is slower on parsec scales \citep[e.g.][]{deavillezmaclow02}. 

Whether or not this situation is common depends on the mass of its host galaxy. It occurs only when the SN progenitor is embedded in the GMC, or confined by thermal pressures exerted by the intercloud medium. But regardless of its frequency of occurrence, if the spread of the metals depends sensitively on the local cloud density (which is often unresolved in cosmological simulations), it modifies the metallicity distribution. It implies a metal injection radius that (i) is spatially variable and (ii) can be much smaller than that predicted by the spherical blast wave equations\footnote{In e.g. \citet{krumholzting18}, metal injection radius is set to be where the SN blast wave expansion velocity becomes equal to the velocity dispersion at ISM scale height.}. Reduced injection radii would skew the metal field correlation function towards smaller length scales \citep{krumholzting18}.

\subsection{Escaped mass, energy and momentum} \label{sec:escape_mass_energy_momen}

The measurements performed in Sections~\ref{sec:turbulence} to \ref{sec:mass_removal} cover both the GMC and the external warm diffuse medium. We intend those results to help inform the large-scale simulations whose fluid resolution elements blend all structures underneath $5-50$ pc. But to understand the greater impact of SNe on galaxy evolution, such as the correlation statistics of metals \citep[e.g.][]{zhang25} or the driving of (inverse) turbulent cascades \citep[e.g.][]{beattie25}, examining the perturbed gas beyond the cloud would be helpful.

\begin{figure}
    \centering
    \includegraphics[width=0.8\linewidth]{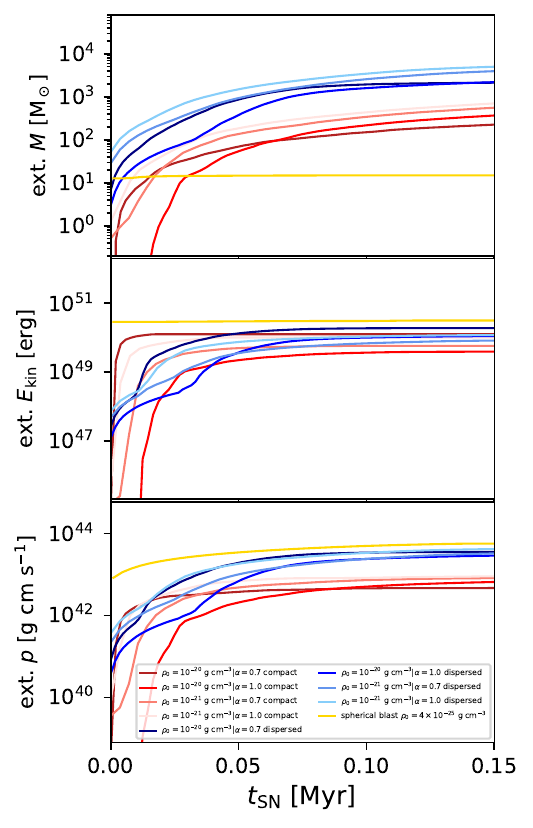}
    \caption{Evolution of total mass (\textit{top}), kinetic energy (\textit{middle}) and momentum (\textit{bottom}) enclosed between 8 pc to 50 pc from the SN progenitor in the compact runs (red colours), the dispersed runs (blue colours) and the spherical blast run (yellow). }
    \label{fig:external_outflow_properties}
\end{figure}

Fig.~\ref{fig:external_outflow_properties} shows the evolution of mass, kinetic energy and momentum of particles that lie in the $8-50$ pc distance range to the progenitor. $8\ \mathrm{pc}$ approximately covers the GMC dense gas in all runs, so we extract only the gas in the outer medium. We also subtract them by their initial values at $t_\mathrm{SN}=0$ to consider only those driven outwards after SNe. Once again we compare the compact runs against the dispersed runs. Results from the spherical blast run are included for reference, but note that their total mass is significantly less than the runs with GMCs.  

Here, the previously-identified trends become even more evident: (i) kinetic energy of the outflows are similar across all runs; (ii) mass of the outflows clearly differentiates between the compact and the dispersed runs, which subsequently lead to (iii) their distinctive differences in momentum. One unique observation from Fig.~\ref{fig:external_outflow_properties} is that, towards $t_\mathrm{SN}=0.15\ \mathrm{Myr}$, the mass and momentum carried in the escaped gas alone are neatly sorted in descending order of GMC compactness (as reflected in the curves' colour tones). By omitting the cloud itself, a strong correlation between the escaped mass and the channels' morphology is revealed.

\begin{figure}
    \centering
    \vspace{0.2in}
    \includegraphics[width=\linewidth]{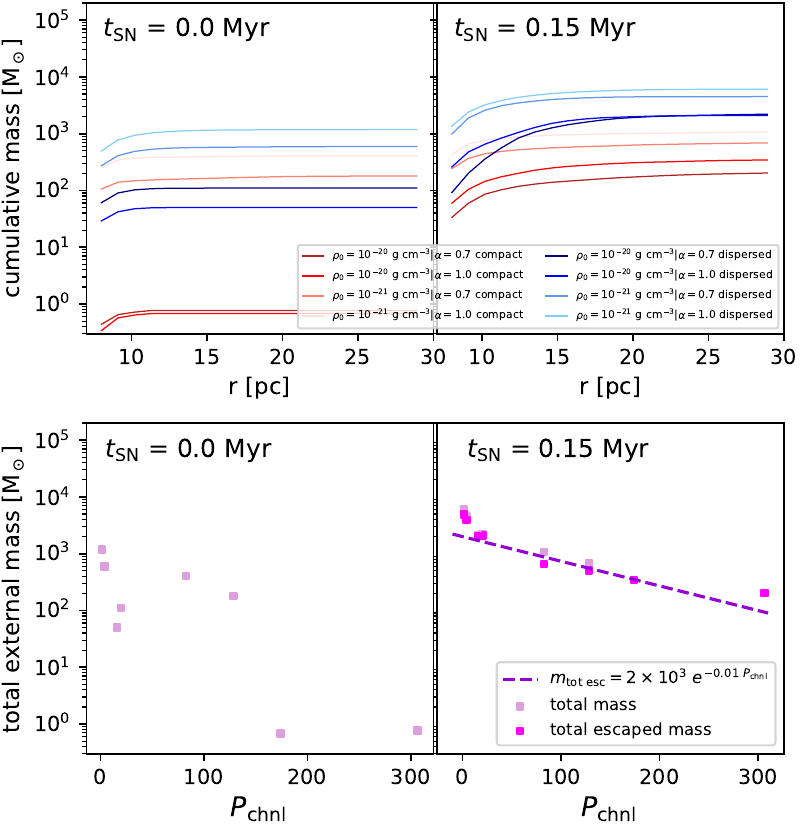}
    \caption{\textit{Top}: Radial cumulative mass distribution beyond 8 pc relative to the SN progenitor in the compact runs (red colours) and the dispersed runs (blue colours), immediately before SN injection (\textit{left}), and at 0.15 Myr after SN injection (\textit{right}). \textit{Bottom}: Total mass beyond 8 pc (pink) and those amongst which that entered this domain only after SN explosion (magenta), plotted against $P_\mathrm{chnl}$. Purple dashed line provides an approximate fit to the escaped mass (equation~\ref{eq:m_esc}).  }
    \label{fig:cumulative_mass_distri}
\end{figure}

We explicitly verify this by examining the spatial distribution of mass in the diffuse medium. The upper panels in Fig.~\ref{fig:cumulative_mass_distri} plot the radial cumulative mass distribution in the $8-30$ pc range, at $t_\mathrm{SN}=0$ and $0.15\ \mathrm{Myr}$. At both times, we observe negligible rise in the cumulative mass beyond 25 pc across all runs. The value at $30\ \mathrm{pc}$ thus gives the total external mass at the time indicated. These are then plotted against $P_\mathrm{chnl}$ of the cloud in the lower panels. We further subtract the results at $t_\mathrm{SN}=0.15\ \mathrm{Myr}$ by their initial values to isolate those which escaped due to SN feedback.

A log-linear relation between escaped mass and $P_\mathrm{chnl}$ is found. The correlation is weak before SN injection, yet strengthens significantly after the SNR has evolved. We impose a best fit line to empirically derive a formula for the mass driven out of the GMC by one SN, 
\begin{equation}
    m_\mathrm{esc} \ [\mathrm{M_\odot}] = 2 \times 10^3 \exp(-0.01 \ P_\mathrm{chnl}). 
    \label{eq:m_esc}
\end{equation}
The reader is reminded that $P_\mathrm{chnl}$ is only a dimensionless parameter that measures how prominent the ionized channels appear. Whilst it is not a physical property, it nonetheless accurately reflects the GMC's boundness. Equation~(\ref{eq:m_esc}) thereby provides an alternative method to estimate the amount of feedback-driven materials released from individual clouds using the 3-D morphology of the ionized regions. Whether or not this can be extended to 2-D projections for comparison to observations warrants further investigation.

\section{Conclusions} \label{sec:conclusion}

In this study, we present four turbulent GMCs with varied density and virial ratio. In each evolved cloud, we select the most massive dynamically-formed sink to be the ionizing source, whose ionizing photon flux is equivalent to that from a massive OB cluster. During the early times, these strong UV radiation carves cavities and channels along the paths of least resistance in the GMCs. We select a snapshot from each cloud at this stage to be `the compact environments'. 

The photoionization simulations are continued until the H {\scshape ii} regions expand beyond the GMCs and partially blend into the external diffuse ISM. Snapshots are again selected at this stage to become `the dispersed environments', giving a total of eight ionized clouds with different density structures. We quantify the compactness of the ionized regions by defining a channelling parameter $P_\mathrm{chnl}$. 

A SN carrying $10^{51}\ \mathrm{erg}$ of kinetic energy is subsequently detonated at the position of the ionizing source in each of the eight clouds. We then compare the properties and impact of the outflows across different GMC environments. As a first observation, our results show that SN shocks indeed preferentially remove gas from regions with relatively lower line-of-sight column density. That the energy is weakly coupled with the shielded dense filaments implies star formation may continue even after the first SN explosion. 

Several important findings have been derived. In comparison to the simple spherical blasts in uniform medium, we find that SNe in realistic turbulent GMC environments can
\begin{enumerate}[wide=0pt, labelwidth=!,itemindent=!,labelindent=!, leftmargin=!, label=(\roman*), parsep=0pt]
    \item deposit the majority of their ejecta mass, energy and momentum at much shorter distances to the progenitor; and
    \item expel gas at higher velocities, which is better described by the semi-confined SN model from Paper~I than as a spherical shock wave propagation. 
\end{enumerate}
Thus, sub-grid calculations that rely on 1-D blast equations may overestimate the impact radii but underestimate the amount of dynamical perturbation locally around the SNR. This is particularly the case for the massive galaxies, where GMCs are likely more gravitationally bound and resistant to disruptions by early feedback. 

Comparisons amongst the compact and dispersed runs reveal even more interesting relations between the properties of the SNe outflows and their host clouds. Key findings are summarised below: 
\begin{enumerate}[wide=0pt, labelwidth=!,itemindent=!,labelindent=!, leftmargin=!, label=(\roman*), parsep=0pt]
    \item The kinematic properties of the SN plumes, particularly on length scales beyond $10\ \mathrm{pc}$, are not highly sensitive to the GMC's porosity. This includes both the bulk velocity and the turbulent velocity dispersions. Bipolar blowouts from compact clouds do not leave any prominently unique signatures on the maximum radial velocities or the kinetic energy deposition. It also explains why in Paper~I the outflow velocities are rather independent of the number of channels, nor the channel sizes. 
    \item Contrary to the above, the mass of materials escaped to the outer environment is most sensitive to the GMC's porosity. It decreases log-linearly with the compactness of the cloud, which can be characterised by the presence of distinctive ionized channel structures (larger $P_\mathrm{chnl}$). The reason could be to do with the deeper potential in highly bound clouds, or the well-formed channels that help suppress the removal of denser gas. 
    \item The fact that mass-loss varies with $P_\mathrm{chnl}$ means radial momentum also has a weak dependence on the host cloud's porosity. However, their variations appear almost negligible beyond $20\ \mathrm{pc}$, meaning that the terminal momentum equation \citep[e.g.][]{kimmcen14} remains effective. Indeed, the momentum injection method is often found to be most accurate for reproducing the observed stellar mass-metallicity relation on cosmological scales \citep[e.g.][]{ibrahimkobayashi23}. 
\end{enumerate}

In essence, materials carried in the SN outflows from compact clouds contain less mass and migrate shorter distances, but at higher velocities. Whilst the compactness of individual clouds may not be resolved in large-scale simulations, it can still be gauged using the local gas densities. In an inhomogeneous galactic disc with density fluctuations at (kilo-)parsec scales, this finding translates to a variable metal injection radius and a variable turbulent driving scale. 

Shorter feedback impact radii naturally result in a clumpier ISM \citep[e.g.][]{krumholzting18}, though faster outflows could possibly allow for more efficient spreading and in turn shorten the ISM mixing timescale. Nevertheless, reduction in amount of materials returned to the ISM per SN explosion could slow down the chemical enrichment process, which may perturb how we deduce the galaxies' formation environments \citep[e.g.][]{scannapieco05}. We therefore argue that the structure and boundness of GMCs should also be taken into account in feedback sub-grid modelling.

One possible extension to this work would be to formulate outflow mass as a function of $P_\mathrm{chnl}$ with increased cloud mass (e.g. $10^5-10^6\ \mathrm{M_\odot}$), increased pre-SN feedback (e.g. including stellar winds), increased blast energy (e.g. $10^{52}\ \mathrm{erg}$ hypernovae), or with multiple SNe, since massive stars likely have high multiplicity \citep[e.g.][]{zinneckeryorke07,grellmann13}. Establishing clear relations between these parameters and the SN impact radii will be highly practical for the purpose of constructing sub-grid models.

\section*{Acknowledgements}

This work was supported by the National Science and Technology Council (NSTC grant no. 115-2124-M-002-014), the Ministry of Education (Higher Education Sprout Project NTU-115L104022-1), and the National Center for Theoretical Sciences of Taiwan. CSCL would like to thank the comments provided by Matthew Bate and Moira Jardine on the PhD thesis of which this paper was based on. The simulations were performed using the HPC Hypatia, operated by University of St Andrews. The SPH figures in this paper were created using the python package {\scshape sarracen} \citep[][]{sarracen23}.

\section*{Data Availability}

The simulation outputs underpinning this publication can be accessed at: [DOI]. 
Newest versions of the simulations codes are publicly available at \url{https://github.com/Cheryl-Lau/phantom} and \url{https://github.com/Cheryl-Lau/CMacIonize}.



\bibliographystyle{mnras}
\bibliography{phd_ref_list}



\appendix

\section{Pictorial definition of the channelling parameter} \label{appen:pchnl_visual}

\begin{figure}
    \centering
    \includegraphics[width=0.85\linewidth]{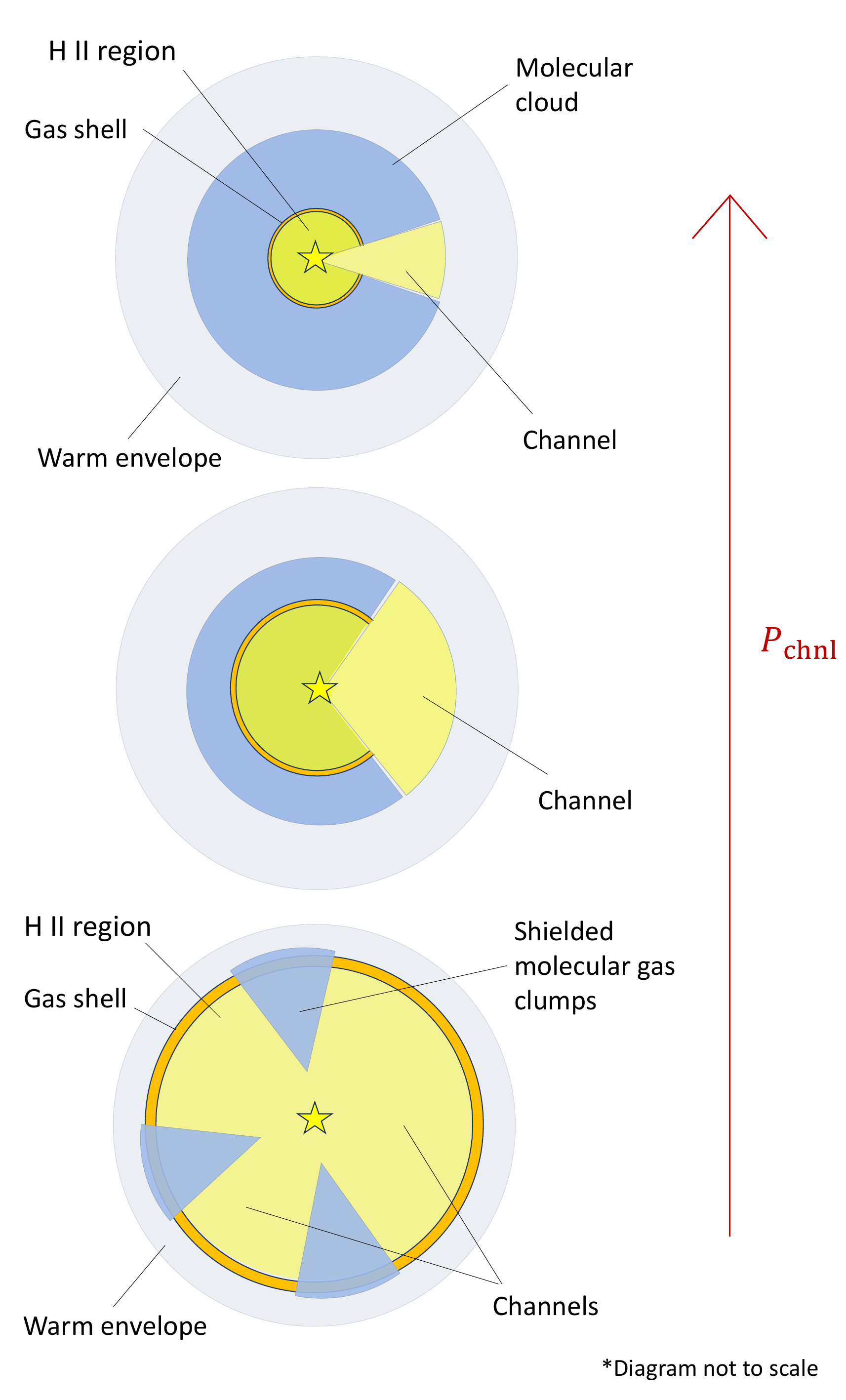}
    \caption{\textit{Top}: Illustration of a high $P_\mathrm{chnl}$ scenario, where both the ionized cavity and the channels (yellow) are embedded within a dense cloud (see Fig.~1 in \citet{laubonnell25} for details). \textit{Bottom}: Illustration of a low $P_\mathrm{chnl}$ scenario, where the cloud has been mostly dispersed by feedback and the channels are no longer seen as narrow paths that thread through the neutral gas. \textit{Middle}: An intermediate situation, where the cavity and channels have expanded but are yet to destroy the parent cloud. }
    \label{fig:pchnl_diagram}
\end{figure}

Fig.~\ref{fig:pchnl_diagram} illustrates how the channelling parameter $P_\mathrm{chnl}$ is defined (see Section~\ref{sec:channelling_param}). Ionized regions are coloured in yellow. A high $P_\mathrm{chnl}$ indicates strong resemblance with the semi-confined SN model in \citet{laubonnell25}, where the cavity is embedded within a dense cloud. Narrow channels that thread through the neutral gas help connect the cavity to the external ISM, allowing feedback outflows to travel through them. To the contrary, a low $P_\mathrm{chnl}$ means that the cavity has significantly expanded, and that the feedback source has become exposed to the external environment due to cloud dispersal.

\section{Relation between channelling parameter and cloud porosity} \label{appen:pchnl_vs_covarea}

We demonstrate that the channelling parameter $P_\mathrm{chnl}$ can accurately reflect the porosity of the GMCs shown in Fig.~\ref{fig:cloud_slice_u_intermed} and \ref{fig:cloud_slice_u_final}. 

\begin{figure}
    \centering
    \includegraphics[width=\linewidth]{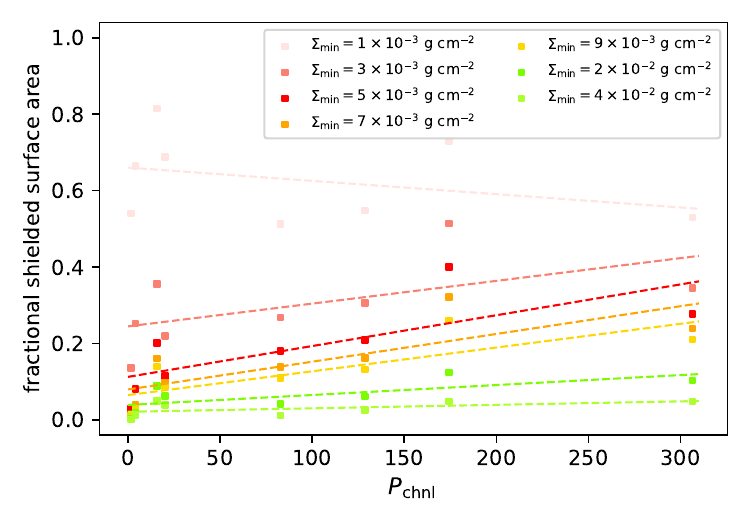}
    \caption{Fractional surface area on a sphere of radius $4\ \mathrm{pc}$ with high column density ($\Sigma > \Sigma_\mathrm{min}$) relative to the ionizing source, plotted against $P_\mathrm{chnl}$. Colours indicate the adopted threshold column density $\Sigma_\mathrm{min}$ above which the volume element was considered sufficiently shielded. Dashed lines provide linear best fits for each set of results with a different threshold. }
    \label{fig:pchnl_covarea}
\end{figure}

Here, porosity is gauged by considering a sphere of radius $4\ \mathrm{pc}$ from the ionizing source. The sphere is tessellated by azimuthal and elevation angles, and we calculate the column density in each volume element (see Fig.~\ref{fig:radsn_20_10_rho_sky_intermed} and \ref{fig:radsn_20_10_rho_sky_final}). Fig.~\ref{fig:pchnl_covarea} plots the summed surface area of all elements whose column densities are above a certain threshold $\Sigma_\mathrm{min}$ (i.e. more shielded from the source), taken as a fraction of the full spherical surface area. These are then plotted against $P_\mathrm{chnl}$. We also investigate how the results vary with the chosen $\Sigma_\mathrm{min}$; a linear best fit line is provided for each test value.

Overall, we observe strong upward trends only when $\Sigma_\mathrm{min}$ is around $5-9 \times 10^{-3}\ \mathrm{g\ cm^{-2}}$. This small window pinpoints the typical column density thresholds above which gas becomes well-shielded from strong photoionization feedback. Crucially, that $P_\mathrm{chnl}$ rises with the fractional shielded surface area signifies that prominent channel-like structures are probable indicators of low GMC porosity.

\section{Evolution of supernova remnants in Giant Molecular Clouds} \label{appen:other_sn_runs}

\begin{figure*}
    \centering
    \begin{minipage}{.48\textwidth}
        \centering
        \includegraphics[width=\linewidth]{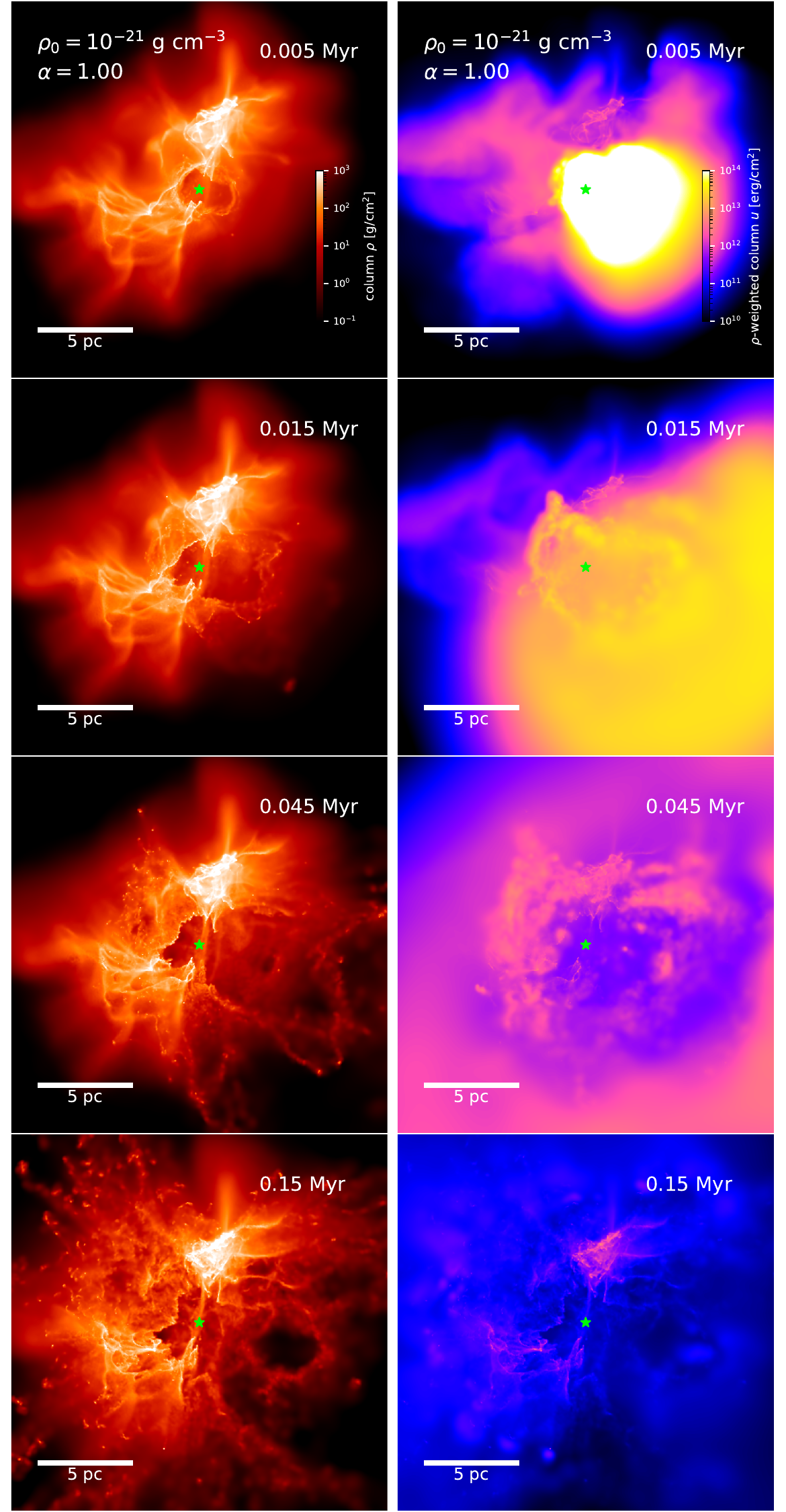}
        \subcaption{Compact environment run}
        \label{fig:radsn_21_10_evol_intermed}
    \end{minipage}
    \begin{minipage}{.48\textwidth}
        \centering
        \includegraphics[width=\linewidth]{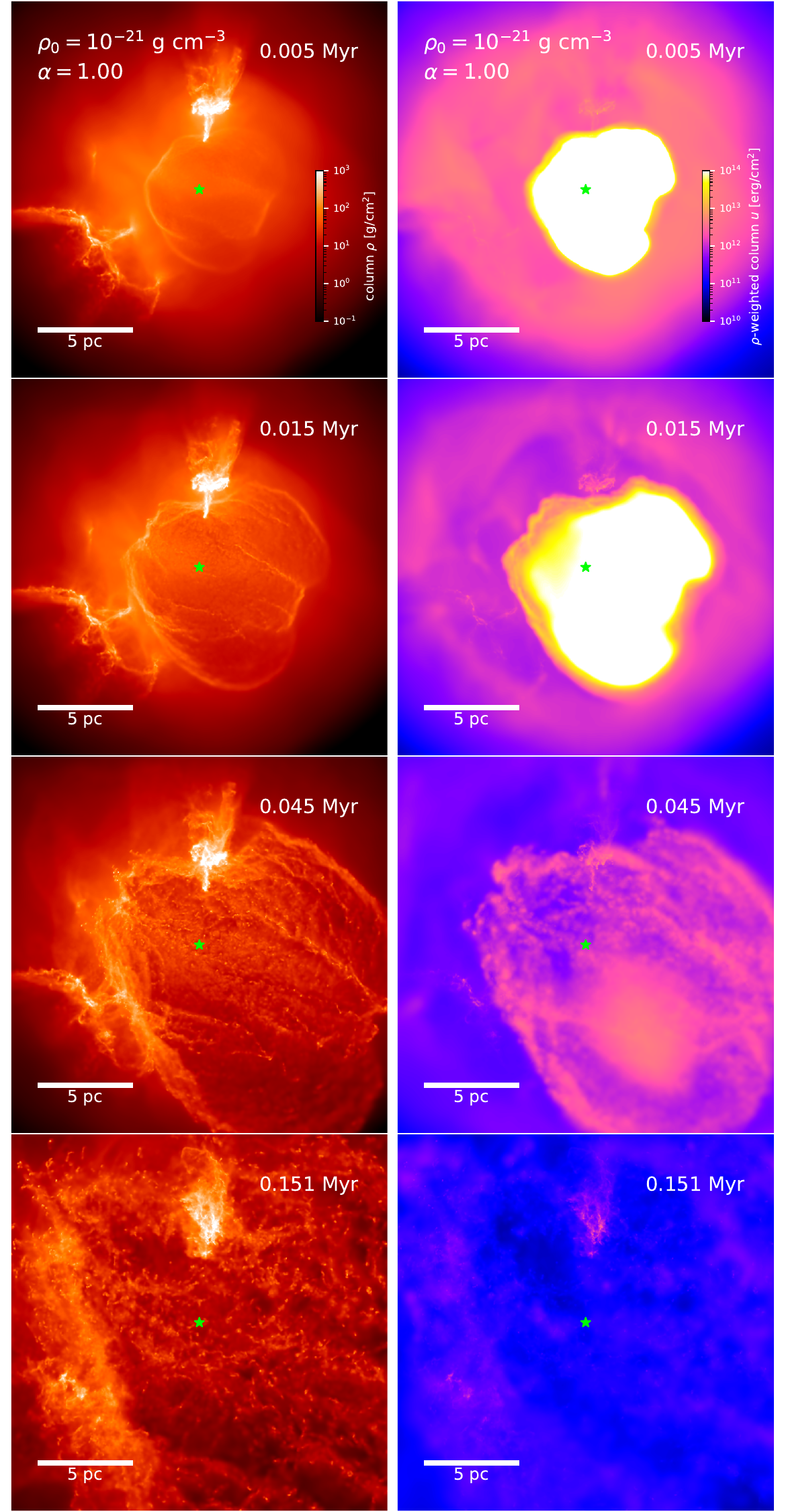}
        \subcaption{Dispersed environment run}
        \label{fig:radsn_21_10_evol_final}
    \end{minipage}
    \caption{\textit{Top to bottom}: Evolution of SNR in the $10^{-21}\ \mathrm{g\ cm^{-3}}|\alpha=1.0$ GMC, where the progenitor is detonated within (a) a small cavity surrounded by clumpy density structures (\textit{left}: the compact environment run), and (b) a large cavity created by pre-SN ionizing feedback (\textit{right}: the dispersed environment run). Red colours show the column density, and blue colours show the column internal energy. }
\end{figure*}

\begin{figure*}
    \centering
    \begin{minipage}{.48\textwidth}
        \centering
        \includegraphics[width=\linewidth]{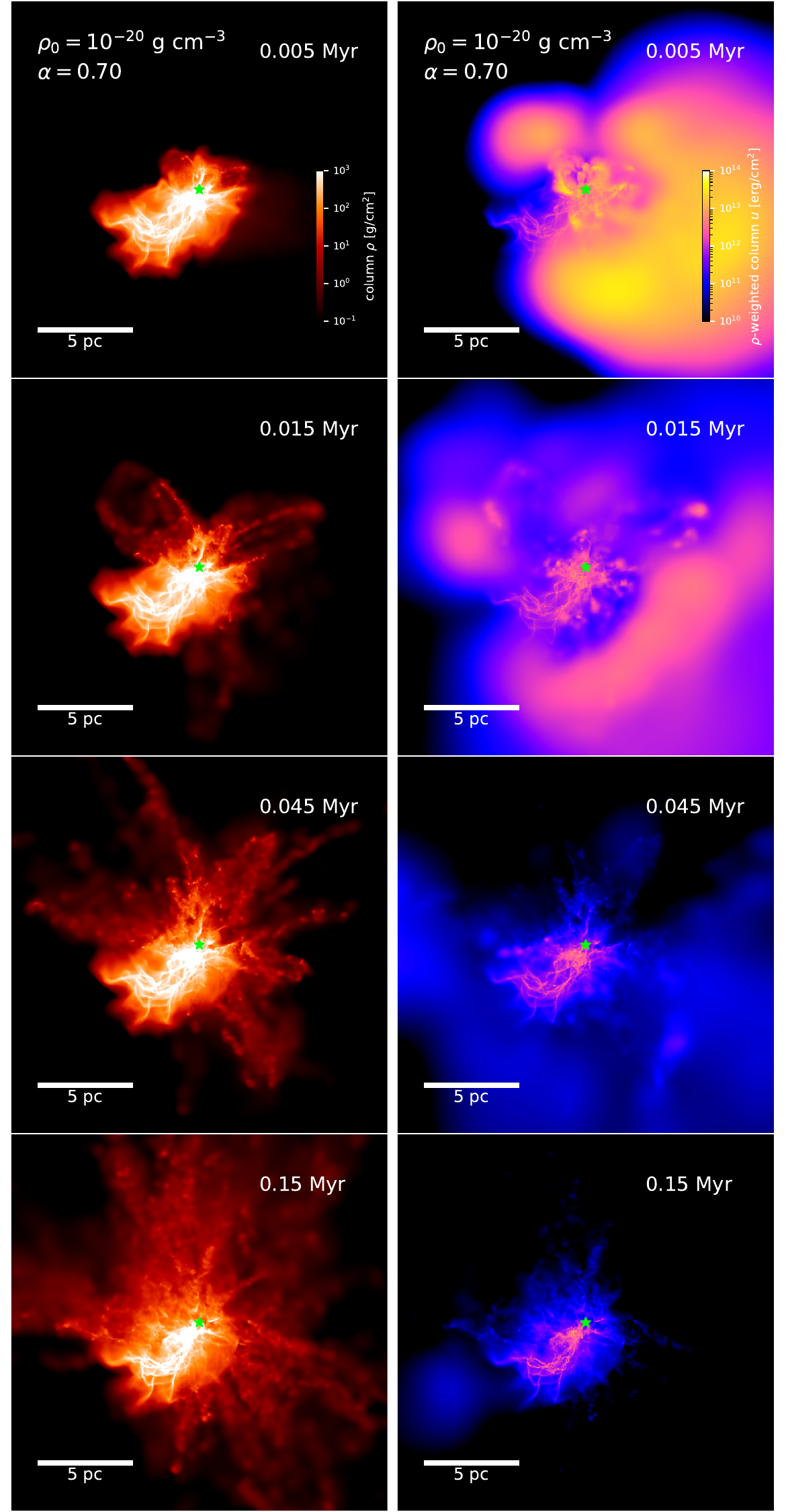}
        \subcaption{Compact environment run}
        \label{fig:radsn_20_07_evol_intermed}
    \end{minipage}
    \begin{minipage}{.48\textwidth}
        \centering
        \includegraphics[width=\linewidth]{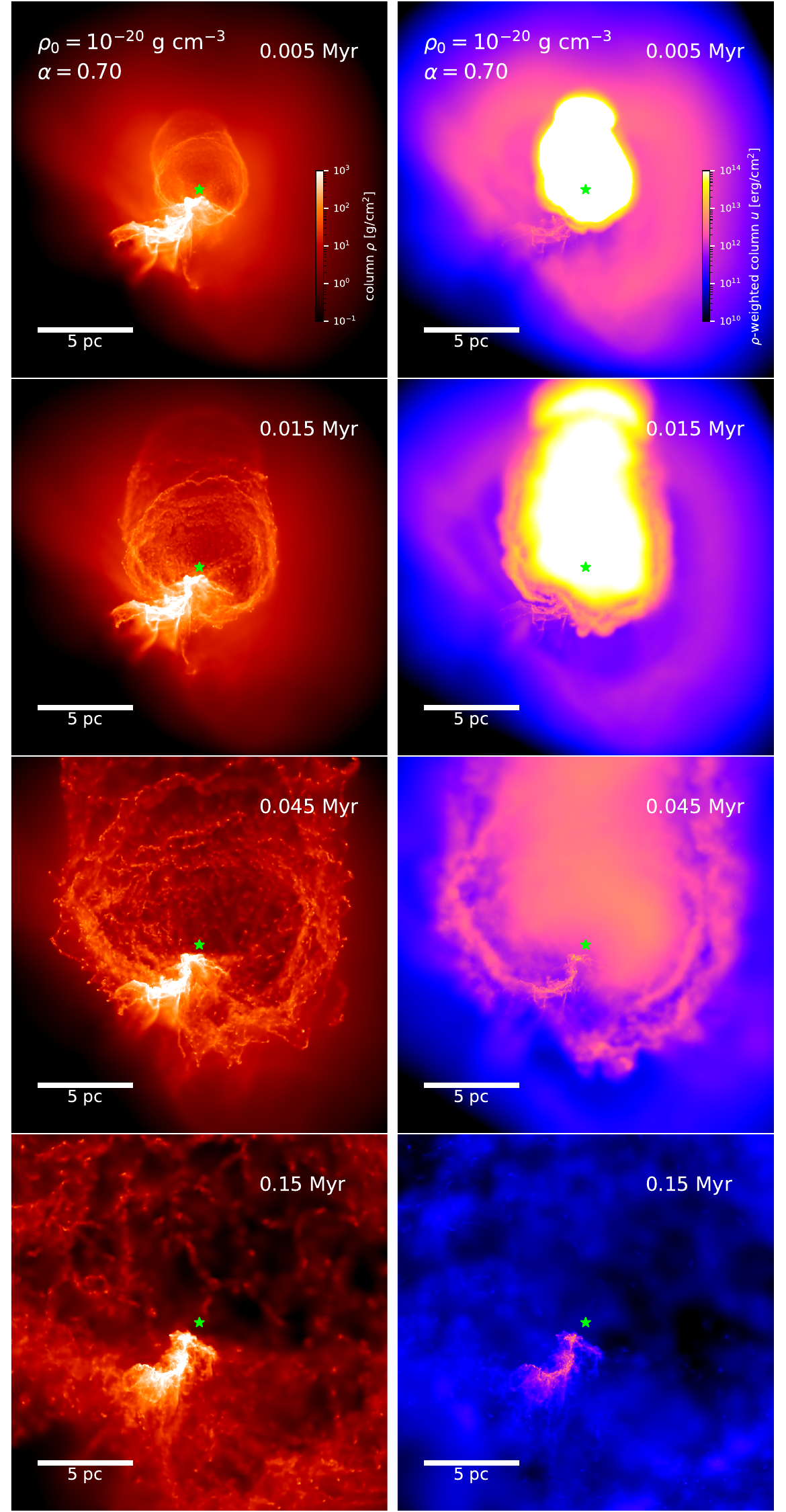}
        \subcaption{Dispersed environment run}
        \label{fig:radsn_20_07_evol_final}
    \end{minipage}
    \caption{Column density and column internal energy evolution of SN in the $10^{-20}\ \mathrm{g\ cm^{-3}}|\alpha=0.7$ GMC, within (a) a compact environment and (b) a dispersed environment. }
\end{figure*}

\begin{figure*}
    \centering
    \begin{minipage}{.48\textwidth}
        \centering
        \includegraphics[width=\linewidth]{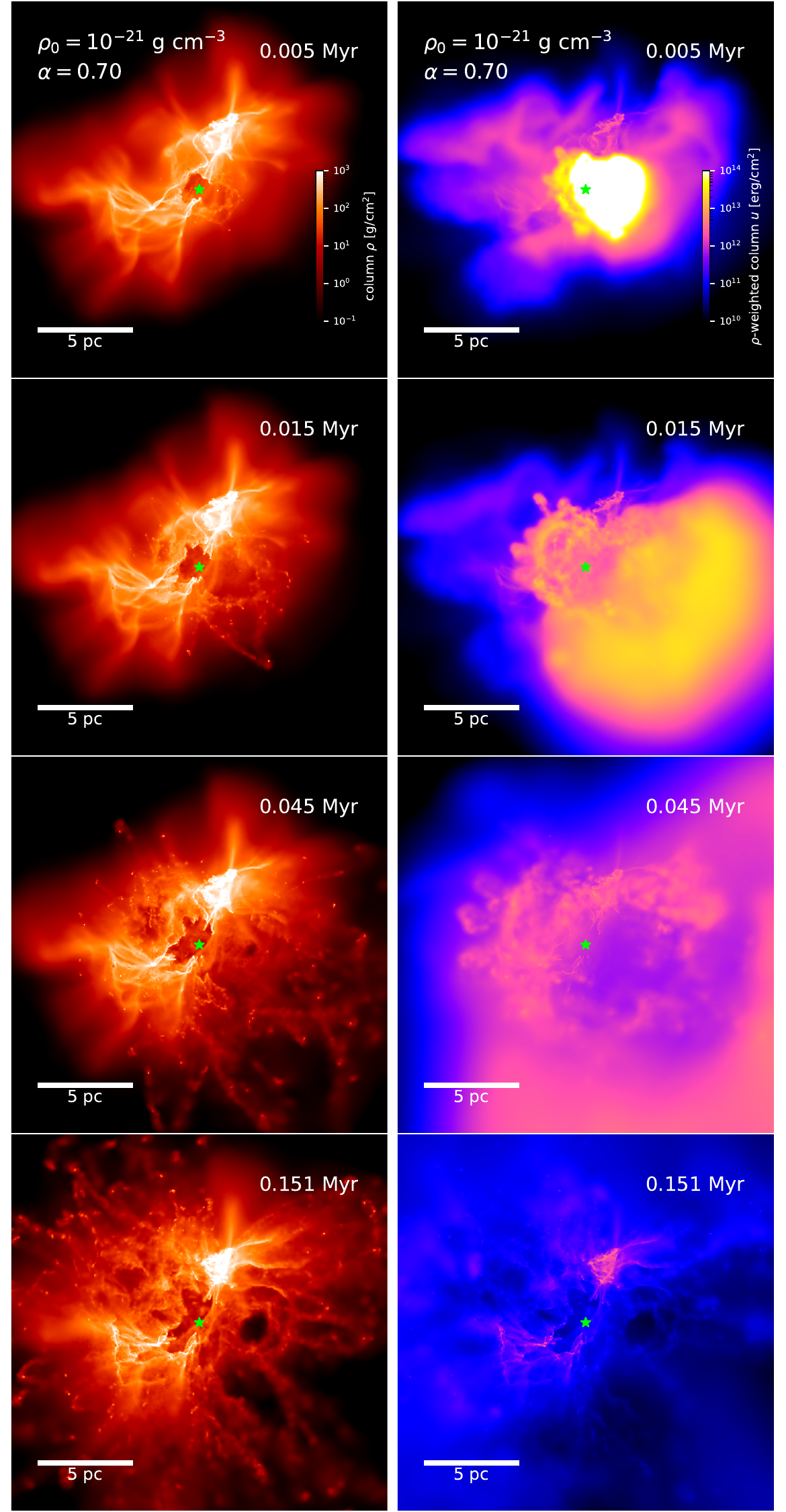}
        \subcaption{Compact environment run}
        \label{fig:radsn_21_07_evol_intermed}
    \end{minipage}
    \begin{minipage}{.48\textwidth}
        \centering
        \includegraphics[width=\linewidth]{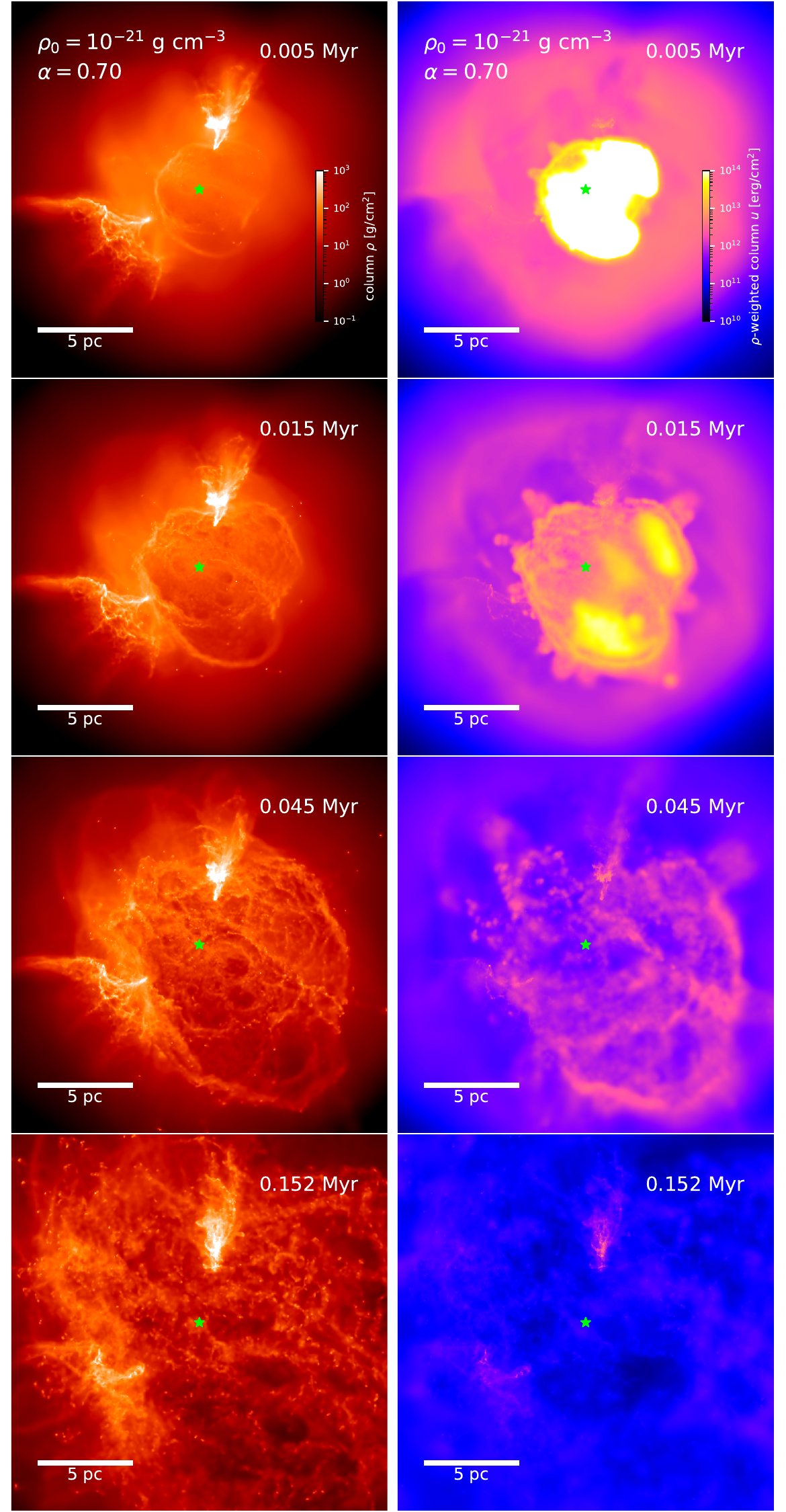}
        \subcaption{Dispersed environment run}
        \label{fig:radsn_21_07_evol_final}
    \end{minipage}
    \caption{Column density and column internal energy evolution of SN in the $10^{-21}\ \mathrm{g\ cm^{-3}}|\alpha=0.7$ GMC, within (a) a compact environment and (b) a dispersed environment.}
\end{figure*}

This appendix presents snapshots of column density and column internal energy of SNR evolution from six other GMC runs in addition to the two displayed in Fig.~\ref{fig:radsn_20_10_evol_intermed} and Fig.~\ref{fig:radsn_20_10_evol_final}. Their initial conditions are detailed in Table~\ref{tab:sims_summary}. Fig.~\ref{fig:radsn_21_10_evol_intermed} and \ref{fig:radsn_21_10_evol_final} show the results for the $10^{-21}\ \mathrm{g\ cm^{-3}}|\alpha=1.0$ compact and dispersed environment runs. Plots from their $\alpha=0.7$ counterpart are shown in Fig.~\ref{fig:radsn_21_07_evol_intermed} and \ref{fig:radsn_21_07_evol_final}. The $10^{-20}\ \mathrm{g\ cm^{-3}}|\alpha=0.7$ runs are presented in Fig.~\ref{fig:radsn_20_07_evol_intermed} and \ref{fig:radsn_20_07_evol_final}. All snapshots are taken at $t_\mathrm{SN}=0.005,0.015,0.045$ and $0.15\ \mathrm{Myr}$.

Multiple outflows are observed in all compact runs, which appear as fuzzy arcs of hot gas with foci centred at the progenitor. Each arc corresponds to a blowout (i.e. a plume), and those seen at larger radii occurred at a slightly earlier time than those nearer to the source. On the other hand, the dispersed runs tend to retain an overall spherical shape, with the only exception being the $10^{-20}\ \mathrm{g\ cm^{-3}}|\alpha=0.7$ run. It has the lowest virial ratio amongst the dispersed runs. All dispersed runs retain significantly more thermal energy in the SNR, thanks to the low-density cavities carved by photoionization.


\bsp	
\label{lastpage}
\end{document}